 \documentclass[review]{elsarticle}

\usepackage{graphicx}
\usepackage{color}
\usepackage[utf8]{inputenc} 
\usepackage[T1]{fontenc}    
\usepackage[colorlinks=false, hidelinks]{hyperref}
\graphicspath{{img/}{../../notes/hcci/}}
\usepackage{subcaption}
\usepackage{caption}
\usepackage{amsmath}
\usepackage{pgffor}
\usepackage{gensymb}
\usepackage{siunitx}
\usepackage{dsfont}
\usepackage{float}
\usepackage{jabbrv}
\usepackage{xcolor}
\usepackage{multirow}
\usepackage[top=.5cm, bottom=.5cm, left=2cm, right=2cm, includehead, includefoot, heightrounded]{geometry}
\journal{Combustion and Flame}

\newcommand{\ignore}[1]{}

\newcommand\rev[1]{\textcolor{black}{#1}}

\begin{document}
\hypersetup{urlcolor=black,citecolor=cyan,linkcolor=black}
\begin{frontmatter}
 \title{A co-kurtosis based dimensionality reduction method for combustion datasets}

 \author[a]{Anirudh Jonnalagadda}
 \author[a]{Shubham P. Kulkarni}
 \author[a]{Akash Rodhiya}
 \author[b]{Hemanth Kolla}
 \author[a]{Konduri Aditya\corref{cor1}}
 \ead{konduriadi@iisc.ac.in}
 \cortext[cor1]{Corresponding Author}
 \address[a]{Department of Computational and Data Sciences, Indian Institute of Science, Bangalore, India}
 \address[b]{Sandia National Laboratories, Livermore, California, USA}
 \begin{abstract}
Principal Component Analysis (PCA) is a dimensionality reduction technique widely used to reduce the computational cost associated with numerical simulations of combustion phenomena. However, PCA, which transforms the thermo-chemical state space based on eigenvectors of co-variance of the data, could fail to capture information regarding important localized chemical dynamics, such as the formation of ignition kernels, appearing as \rev{extreme-valued} samples in a dataset. In this paper, we propose an alternate dimensionality reduction procedure, co-kurtosis PCA (CoK-PCA), wherein the required principal vectors are computed from a high-order joint statistical moment, namely the co-kurtosis tensor, which may better identify directions in the state space that represent stiff dynamics. We first demonstrate the potential of the proposed CoK-PCA method using a synthetically generated dataset that is representative of typical combustion simulations. Thereafter, we characterize and contrast the accuracy of CoK-PCA against PCA for datasets representing spontaneous ignition of premixed ethylene-air in a simple homogeneous reactor and ethanol-fueled homogeneous charged compression ignition (HCCI) engine. Specifically, we compare the low-dimensional manifolds in terms of reconstruction errors of the original thermo-chemical state, and species production and heat release rates computed from the reconstructed state. \rev{The latter -- a comparison of species production and heat release rates -- is a more rigorous assessment of the accuracy of dimensionality reduction.} We find that, even using a simplistic linear reconstruction, the co-kurtosis based reduced manifold represents the original thermo-chemical state more accurately than PCA, 
especially in the regions where chemical reactions are important. 
\end{abstract}


\end{frontmatter}

\section{Introduction} \label{sec:introduction}

Direct numerical simulations (DNS) of turbulent reacting flows, which capture detailed chemical kinetics and their interactions with turbulent structures, demand massive computations and are often performed on large supercomputers. Current state-of-the-art simulations use chemical mechanisms that comprise of tens of species participating in tens or hundreds of reactions \citep{adityaDirectNumericalSimulation2019,savard2019,bergerDNSStudyImpact2020,Nivarti2017,desai2021direct}.
A valuable approach in reducing the associated computational expense involves identifying lower-dimensional manifolds, i.e., a smaller subset of representative thermo-chemical scalars to express, and solve the conservation equations.
\rev{Indeed, based on the modelling strategy being employed, several types of lower-dimensional manifolds can be obtained \cite{pope-challenges-in-turbulent-combustion-2013-proci}.
} 

\rev{The need to reduce dimensionality has long been held imperative for computations of chemically reacting systems. Approaches based on exploiting the dynamics of reacting systems have preceded the more recent data-driven approaches. Intrinsic low-dimensional manifolds (ILDM) \cite{MaasP1992} recognizes the broad range of chemical timescales, characteristic of most reacting systems, and identifies an intrinsic lower-dimensional subspace by decoupling the fastest time scales (equivalent to a local equilibrium assumption) from the slower ones. Eigen decomposition of the chemical Jacobian allows identifying the eigenvalues corresponding to the fastest time scales, and truncating the corresponding eigenvectors basis defines the subspace. Computational Singular Perturbation (CSP) \cite{LamG1989} also exploits the separation of chemical timescales into fast and slow ones, identified from the eigen decomposition of the Jacobian, but primarily to reduce the stiffness of the ODE system rather than to reduce its dimensionality. To account for diffusion and molecular transport effects in addition to chemical reactions, methods such as flame prolongation of ILDM (FPI) \cite{GicquelDT2000} and flamelet-generated manifolds (FGM) \cite{VanoijenLD2001} use one-dimensional laminar flame simulations parameterized by a few controlling variables that describe the full chemical system over a broader range of conditions.
}  

\rev{Among the data-driven approaches, empirical manifolds presented through the application of principal component analysis (PCA) on combustion datasets has received a wide interest, as it provides a rigorous procedure to identify the low-dimensional subspace that consists of orthonormal basis vectors.}
Several flavours of using PCA for dimensionality reduction are reported in the literature.
Sutherland and Parente \cite{sutherland-parante-2009} analysed the application of PCA conditioned on the mixture fraction for DNS datasets, whereas Biglari and Sutherland \cite{biglari-sutherland-2012} considered the application of unconditional PCA with a focus on scaling effects, large eddy simulation (LES) filtering and applicability of multivariate adaptive regression splines (MARS) for the forward and reverse projections.
Thereafter, Yang et al. \cite{yang-pope-chen-2013} applied the analysis presented by Biglari and Sutherland \cite{biglari-sutherland-2012} to high-fidelity DNS datasets with more complex chemical mechanisms, used a more sensitive R$^2$ error metric to evaluate the quality of reconstruction, and presented transport equations for the chosen reduced manifolds for future use.
Echekki and Mirgolbabaei \cite{mirgolbabaei-echekki-ANN-2015} used PCA to tabulate relevant quantities in terms of reduced principal vectors using artificial neural networks (ANNs) which were previously shown to present a better reconstruction as compared to the linear reconstruction \citep{mirgolbabaei-echekki-2014-cnf}.
Ranade and Echekki \cite{ranade-echeckki-2019} used PCA to isolate lower dimensional manifolds to develop closure models for turbulent combustion with the reconstruction being performed using a modified version of the pairwise-mixed stirred reactor model. Malik et al. \cite{malik-obando-coussement-parante-2021} applied the PC-transport approach to 3D LES simulations and used Gaussian Process Regression (GPR) for reconstructing the missing species; the use of GPR reconstruction was motivated by a comparative study performed by Isaac et al. \cite{isaac-thornock-sutherland-smith-parante-2015}.

Apart from the global application of PCA described above, researchers have also applied PCA locally in space/time to capture the non-linear nature of combustion data with greater accuracy.
Particularly, Parente et al. \cite{parante-sutherland-tognotti-smith-2009, parente-sutherland-dally-tognotti-smith-localPCA-MILD-2011} compared and contrasted the effectiveness of two local PCA methods, namely vector quantization PCA (VQPCA) and mixture fraction PCA (FPCA) against global PCA. 
 Coussement et al. \cite{coussement-giquel-parante-2012} conducted investigations for representing the non-linear data better by examining the use of a kernel density method based weighted PCA technique coupled with linear reconstruction.
Thereafter, Mirgolbabaei and Echekki \cite{mirgolbabaei-echekki-2014-cnf} studied the characteristics of an ANN based reconstruction method for the kernel density weighted PCA.

It is clear that while several methods to transform the thermo-chemical scalars to and from low-dimensional representations have been investigated in the literature, the reduced manifold has been predominantly identified using PCA.
\rev{In this work, we investigate the suitability of a co-kurtosis measure based dimensionality reduction approach (called CoK-PCA for brevity) for reduced-order modelling of combustion systems and examine its accuracy, compared to conventional PCA, in representing the thermo-chemical state as well as the chemical kinetics of the system. In particular, we present a detailed and comprehensive analysis by considering reconstruction errors of not only the species mass fractions and temperature but also those of the species production rates and heat release rate. The latter, an analysis of the reconstruction errors of species production and heat release rates, is a more rigorous assessment of accuracy and goes beyond existing PCA literature which typically only report species reconstruction errors. Further, we also compare and contrast the robustness of the PCA and CoK-PCA reduced manifolds in modeling unknown test conditions in the neighbourhood of the original training condition.}

\rev{The rationale behind our line of investigation stems from the observation that, since PCA principal vectors are not very sensitive to the presence of extreme events in a dataset \cite{aditya-anomaly-detection-2019-JCP}, empirically reduced manifolds constructed through PCA may not be able to capture stiff chemical dynamics associated with localized extreme events such as the formation of ignition kernels. In contrast, since the identification of extreme-valued states (`anomalous' events) significantly improves upon using principal vectors from the fourth-order co-kurtosis statistic \cite{aditya-anomaly-detection-2019-JCP}, it can be posited that CoK-PCA reduced manifolds may present improved lower dimensional representations of the local chemical state of the system. We note that the present work focuses on the manner in which the principal vectors are identified and uses the simplest form of reconstruction; an assessment on the performance of the proposed co-kurtosis based dimensionality reduction with non-linear reconstruction methods, which are shown to preserve accuracy better, will be the subject of a future study.}

An outline for the remainder of this paper is as follows.
In Sec.~\ref{sec:config}, we first briefly discuss the specifics of PCA and then describe the calculation of the fourth-order co-kurtosis tensor along with its decomposition to yield the co-kurtosis based principal vectors.
In Sec.~\ref{sec:dr}, we first demonstrate a proof-of-concept evaluation of the CoK-PCA for a synthetically generated dataset \rev{that is representative of combustion datasets}.
Thereafter, in Sec.~\ref{sec:results} we present results relevant to combustion datasets and discuss the merits and demerits of the proposed CoK-PCA relative to the standard PCA. Finally, conclusions from this work are presented in Sec.~\ref{sec:conclusions}.

\section{Fourth-order moment co-kurtosis tensor and decomposition} \label{sec:config}

Consider a dataset $\mathbf{X}\in\mathds{R}^{(n_g\times n_v)}$ with $n_g$ observations each having $n_v$ real-valued variables or features; in terms of the feature space, $\mathbf{X}$ can be represented in terms of column vectors as $\mathbf{X} = \left\lbrace x_i \in \mathds{R}^{(n_g\times 1)} ~\forall ~i \in \lbrace 1, \cdots, n_v\rbrace \right\rbrace$. For PCA, the principal vectors represent directions of maximal data variance as captured by the second order co-variance matrix $\mathbf{C} \in \mathds{R}^{(n_v \times n_v)}$ which is represented using the index notation as:
\begin{equation}
    (\mathbf{C})_{ij} \equiv C_{ij} = \mathds{E}(x_i x_j), \quad i,j \in \lbrace 1, \cdots, n_v\rbrace,
\end{equation}
where $\mathds{E}$ is the expectation operator. The required principal vectors ($\mathbf{A}$) are the eigenvectors of the co-variance matrix obtained through an eigenvalue decomposition, $\mathbf{C} = \mathbf{A} \mathbf{L} \mathbf{A}^T$. It should be noted that the data used in the definition of joint moments is assumed to be centered around the mean.

Similarly, with the higher order moment of interest, i.e., the fourth-order co-kurtosis tensor, the principal vectors represent the directions of maximal kurtosis or peakiness in the data. The co-kurtosis tensor is defined as:
\begin{equation}
    T_{ijkl} = \mathds{E}(x_i x_j x_k x_l), ~ i,j,k,l \in \lbrace 1, \cdots, n_v\rbrace.
\end{equation}
As alluded to previously, Aditya et al. \cite{aditya-anomaly-detection-2019-JCP} demonstrated that the principal vectors of the co-kurtosis tensor can be effective in identifying directions of rare or anomalous \rev{or extreme-valued} samples. The required principal vectors were obtained by decomposing the fourth-order \textit{cumulant} tensor, which is related to the co-kurtosis tensor by means of an analogy derived from independent component analysis (ICA) which is detailed next.

ICA assumes that the observed features ($\mathbf{X}$) are a linear combination of statistically independent non-Gaussian sources ($\mathbf{S}$).
Mathematically, the feature space can be represented as:
\begin{equation}
\label{eq:ICA}
    \mathbf{X} = \mathbf{S} \mathbf{M}^T + \mathcal{N},
\end{equation}
where $\mathbf{M}$ is the mixing matrix and $\mathcal{N}$ is an additive Gaussian noise.
The vectors of the mixing matrix $\bf M$ can be computed by realizing that they maximize the non-Gaussian nature of $\mathbf{X}$ \citep{hyvarinen2000}.
Since excess kurtosis is a good measure of the non-Gaussian nature of a dataset, the fourth-order cumulant tensor, $\mathbf{K}$,
of $\mathbf{X}$ can be factorized in terms of the excess kurtosis of $i$-th source, $\kappa_{s_i}$, and the vectors of $\mathbf{M}$ as \citep{comon2002,anandkumar2014}:
\begin{equation}
   \mathbf{K} = \sum_i \kappa_{s_i} m_i \otimes m_i \otimes m_i \otimes m_i,
\end{equation}
where the cumulant tensor is defined as:  
\begin{equation}
    \label{eq:fourth-order-cumulant}
        K_{ijkl} = T_{ijkl} - C_{ij} C_{kl} - C_{ik} C_{jl}- C_{il} C_{jk}.
\end{equation}
Again, it should be noted that the above equation holds good when the data is centered around the mean.

The mathematical statement of ICA, Eq.~\ref{eq:ICA}, is very similar to PCA in that it prescribes a linear transformation from the original features to a transformed basis.
However, while PCA yields a set of linearly uncorrelated features, ICA yields statistically independent features which is particularly attractive for turbulent reacting flow simulations, especially when the conservation equations are solved for the transformed variables in the context of large-eddy simulations (LES) or Reynolds-averaged Navier-Stokes (RANS) simulations wherein closure models of the joint PDFs of the variables are required to account for sub-grid terms \citep{biglari-sutherland-2012}.

To obtain the required CoK-PCA principal vectors, it is necessary to suitably decompose \rev{ the fourth-order cumulant tensor $\mathbf{K}$}. 
However, as all matrix factorization properties cannot be generalized to higher order tensors, alternate decomposition strategies such as the symmetric canonical polyadic (CP), higher order singular value decomposition (HOSVD), etc. are needed to obtain the required principal vectors and values.
Following \cite{anandkumar2014},  Aditya et al.~\cite{aditya-anomaly-detection-2019-JCP} \textit{recast} the fourth-order cumulant tensor as an $n_v \times n_v^3$ matrix ($\mathbf{T})$, which is then decomposed using SVD as $\mathbf{T} = \mathbf{U}\mathbf{S}\mathbf{V}^T$; the required principal vectors are obtained as the columns of the left singular matrix, i.e., $\mathbf{M} = \mathbf{U}$. Note that, thus computed, the vectors of $\mathbf{M}$ will be ortho-normal to each other, which is an additional constraint not central to ICA.
\section{Dimensionality reduction and data reconstruction}\label{sec:dr}

In a dimensionality reduction exercise,  the most relevant information is retained by choosing the most informative subset of principal vectors that characterize the data, thus representing $\mathbf{X}\in\mathds{R}^{(n_g\times n_v)}$ in a reduced space as $\mathbf{Z}_q \in\mathds{R}^{(n_g\times n_q)}$, with $n_q (< n_v)$ being the number of retained principal vectors.
A simple technique to obtain $\mathbf{Z}_q$ utilizes the linear transformation:
\begin{equation}
    \label{eq:data-reduction}
    \mathbf{Z}_q = \mathbf{X}\mathbf{A}_q,
\end{equation}
where $\mathbf{A}_q \in \mathds{R}^{(n_v \times n_q)}$ comprises the chosen subset of principal
\rev{vectors. The similarities and differences between PCA and CoK-PCA stem from the choice of these principal vectors. For PCA, $\mathbf{A}_q$ correspond to the $n_q$ leading singular vectors of the raw data $\mathbf{X}$, which are the same as the first $n_q$ eigenvectors of its co-variance matrix $\mathbf{C}$. On the other hand, for CoK-PCA the principal vectors are obtained from the HOSVD of the fourth-order cumulant tensor $\mathbf{K}$ (Eq.~\ref{eq:fourth-order-cumulant}), i.e. matricize $\mathbf{K}$ to obtain $\mathbf{T} \in \mathds{R}^{(n_v \times n_v^3)}$ and compute $\mathbf{A}_q$ as the $n_q$ leading singular vectors of $\mathbf{T}$. Because of the connections between the principal vectors of $\mathbf{K}$ and ICA noted in the previous section, CoK-PCA is similar to ICA rather than PCA.}

\begin{figure}[t!]
	\centering
	{\includegraphics[width=5cm]{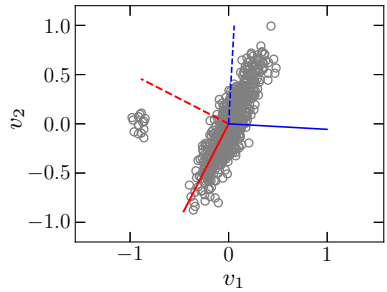}}
	\begin{picture}(0,0)
	\put(-68,-4){\small $v_1$}
	\put(-148,55){\small \rotatebox{90}{$v_2$}}
	\end{picture}
	\caption{A bi-variate synthetic dataset with a few \rev{extreme-valued} samples representing an anomalous event. Red and blue lines are the principal vectors obtained from PCA and CoK-PCA, respectively. Solid and dashed lines indicate the first and the second principal vectors, respectively.}
	\label{fig:synthetic}
\end{figure}

To illustrate the dimensionality reduction procedure and demonstrate the potential of the CoK-PCA method, consider a synthetic dataset, $\mathbf{D} \in \mathds{R}^{( 625 \times 2 )}$, which is constructed by rotating an uncorrelated bi-variate Gaussian distribution by $45\degree$. The two variables, $v_1$ and $v_2$, in the dataset have a zero mean and a variance of $1.0$ and $0.25$, respectively. 
A small, randomly chosen subset of \rev{$15$} samples are isolated from the main cluster and placed \rev{in the neighbourhood of point $(-1,0)$, which is farther away from the normal data, to represent extreme-valued points that characterize} an anomalous event, as shown in Fig.~\ref{fig:synthetic}. To proceed with dimensionality reduction, first the data matrix $\mathbf{D}$ is scaled following the procedure described in \cite{aditya-anomaly-detection-2019-JCP} to obtain the scaled data matrix $\mathbf{X}$; specifically, for each feature, the respective mean is subtracted followed by a normalization with the absolute maximum of the feature.
Thereafter, the PCA and CoK-PCA principal vectors are computed according to the procedure described in Sec.~\ref{sec:config}, and are plotted in Fig.~\ref{fig:synthetic}.
It can be clearly seen that while the first principal vector from PCA (solid-red line) aligns in the direction of maximum spread or variance in the data, the first CoK-PCA vector (solid-blue line) captures the direction of \rev{extreme-valued} samples. The second principal vectors (dashed lines) are, by construction, orthogonal to the respective first principal vectors (solid lines). 

For reducing the dimensionality in the synthetic dataset, we construct two low-dimensional manifolds by retaining the first principal vector from PCA and CoK-PCA, i.e., by choosing $n_q=1$. The two reduced representations of the data are obtained by projecting the original synthetic data onto these manifolds using Eq.~\ref{eq:data-reduction}. Note that a majority of samples in the main cluster are closer, in terms of normal distance, to the first principal PCA vector (solid-red line) and hence, would be represented more accurately in the PCA-reduced manifold.
On the other hand, the \rev{extreme-valued} samples are farther from the first principal PCA vector (solid-red line) when compared to the first principal CoK-PCA vector (solid-blue line). Therefore, these samples would be poorly represented in the PCA based low-dimensional manifold. 

To evaluate the quality of the two reduced manifolds, it is imperative to evaluate the accuracy with which the original state space can be reconstructed. For this, we employ the inverse transformation of Eq.~\ref{eq:data-reduction} to obtain a linear reconstruction,
\begin{equation}
    \label{eq:data-reconstruction}
    \mathbf{X}_q = \mathbf{Z}_q\mathbf{A}_q^T,
\end{equation}
where $\mathbf{X}_q$ is the reconstructed data in the state space, which is the $v_1$-$v_2$ space for the synthetic dataset.
Now, a comparison between $X_q$ and $X$ would reveal the quality of the two manifolds.
It is important to note that standard reconstruction error metrics such as root mean squared, $R^2$, etc. would typically yield smaller errors with PCA as the first PCA principal vector would align well with the bulk of the data, as explained earlier.
Thus, it becomes necessary to identify alternate error metrics which can ascertain the quality of reconstruction for both the \rev{extreme-valued} as well as the bulk samples.
In this work, we use the maximum and average values of the absolute reconstruction error $\left(\varepsilon = |\mathbf{X} - \mathbf{X}_q|\right)$, $\varepsilon_M = \max(\varepsilon)$ and $\varepsilon_A = \mathrm{mean}({\varepsilon})$, respectively, to quantify the accuracy in each reconstructed variable.
Further, we examine the error ratio,

\rev{ \begin{equation}
r_i = \ln\left\lbrace\frac{\varepsilon_i^{\text{PCA}}}{\varepsilon_i^{\text{CoK-PCA}}}\right\rbrace,
    \label{eq:error-ratio}
\end{equation}}
to establish the performance of CoK-PCA against PCA; the subscript $i$ represents either the maximum $(M)$ or the average $(A)$ errors.
Here, we use the simplifying metric $r_i$ as comparisons in terms of absolute errors are cumbersome due to the fact that variables in combustion datasets take values that range of several orders of magnitude.
\rev{Note that the absolute value of $r_i$ represents an improvement in the accuracy of one method relative to the other by a factor of $e^{(r_i)}$ with $r_i$ > 0 indicating that the error from the PCA reduced manifold is larger than that from the CoK-PCA  reduced manifold, and vice-versa for $r_i$ < 0; a value of $r_i$ = 0 indicates that the reconstruction errors from the two methods are equal. }


\begin{figure}[t!]
	\centering
	{\includegraphics[width=5cm]{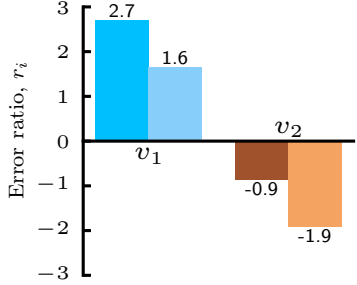}}
	\begin{picture}(0,0)
    \put(-145,42){\footnotesize{\rotatebox{90}{Error ratio, $r_i$}}}
    \end{picture}
	\caption{Error ratio ($r_i$) for variables $v_1$ and $v_2$ in synthetic data. Darker blue and brown colors: error ratio based on maximum error, lighter blue and brown colors: error ratio based on average error.}
	\label{fig:synthetic_ri}
\end{figure}

For the synthetic dataset the values for the error ratio $r_i$ computed based on the maximum and average reconstruction errors are shown in Fig.~\ref{fig:synthetic_ri}.
It can be seen that dimensionality reduction based on CoK-PCA outperforms PCA, both in terms of maximum and average reconstruction errors, in representing $v_1$ data which has significantly greater values for the \rev{extreme-valued} samples. Conversely, PCA performs better than CoK-PCA in reconstructing $v_2$, however, to a lesser extent. Note that the values $v_2$ are not significant for the \rev{extreme-valued} samples.

In the context of combustion, the \rev{extreme-valued} points in the synthetic data \rev{can be considered to represent spatial locations corresponding to} thin reaction zones \rev{associated with the} flame propagation or ignition fronts \rev{which present stiff chemical dynamics}. \rev{The variable} $v_1$ would represent intermediate species which have greater values in the reactions zones and are negligible elsewhere. On the other hand, $v_2$ captures the behaviour of reactant or product species. Accordingly, we can speculate that dynamics in the reaction zone would be better captured by the low-dimensional manifold due to CoK-PCA.
An assessment of this hypothesis is presented in the next section.
\section{Results} \label{sec:results}
\subsection{Homogeneous reactor dataset}
\label{sec:zerod}
To investigate the accuracy of the proposed dimensionality reduction procedure for combustion datasets, we first consider a dataset that captures spontaneous ignition in a simple homogeneous (zero-dimensional) reactor.
The choice of this particular dataset is motivated by the fact that ignition, being a highly nonlinear exponential process, results in a majority of the reactions, along with the production and consumption of intermediate species, to occur over a small fraction of the simulation time.
Thus, the chosen dataset incorporates the kind of  \rev{extreme events that present stiff dynamics which can potentially be well-captured by CoK-PCA}.

The reactants consist of premixed ethylene and air at pressure, temperature, and equivalence ratio of P=\SI{1.72}{atm}, T=\SI{1200}{\K}, and $\phi$=\SI{0.4}{}, respectively, while a 32-species 206-reactions mechanism \cite{luo2012chemical} was used to represent the chemistry.
The homogeneous reactor was evolved for \SI{2.5}{\milli\second} with a time step of \SI{1}{\micro\second} to yield 2501 data samples.
Thus, the \rev{training data matrix $\mathbf{D}$ (dataset used to construct the reduced manifold)} consists of $n_g = 2501$ points and $n_v = 33$ variables, which includes 32-species and temperature.
The data matrix $\mathbf{D}$, as explained in the previous section, is scaled by subtracting with individual feature mean and normalized with feature absolute maximum to obtain the scaled data matrix $\mathbf{X}$.
This matrix is then used to compute the co-variance matrix and the co-kurtosis tensor, which are factorized to obtain the PCA and CoK-PCA principal values and vectors.

\begin{figure}[H]
    \centering
    \includegraphics{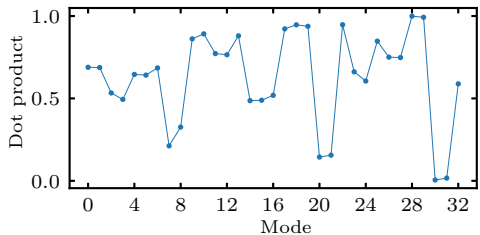}
    \begin{picture}(0,0)
    \put(-93,0){\scriptsize Mode}
    \put(-188,32){\scriptsize{\rotatebox{90}{Dot product}}}
    \end{picture}
    \caption{Alignment between PCA and CoK-PCA principal vectors from homogeneous reactor dataset represented as a dot product for each mode.}
    \label{fig:c2h4-inclination}
\end{figure}

To check whether PCA and CoK-PCA yield different vectors, we first present the dot-product of the two sets of unit vectors in Fig.~\ref{fig:c2h4-inclination}.
It can be seen that due to the inherently different nature of the two methods, the principal vectors are, in general, not aligned and different.
This implies that the two methods identify different low-dimensional sub-spaces.

\begin{figure*}[h!]
    \centering
    {
    \includegraphics[width=7cm]{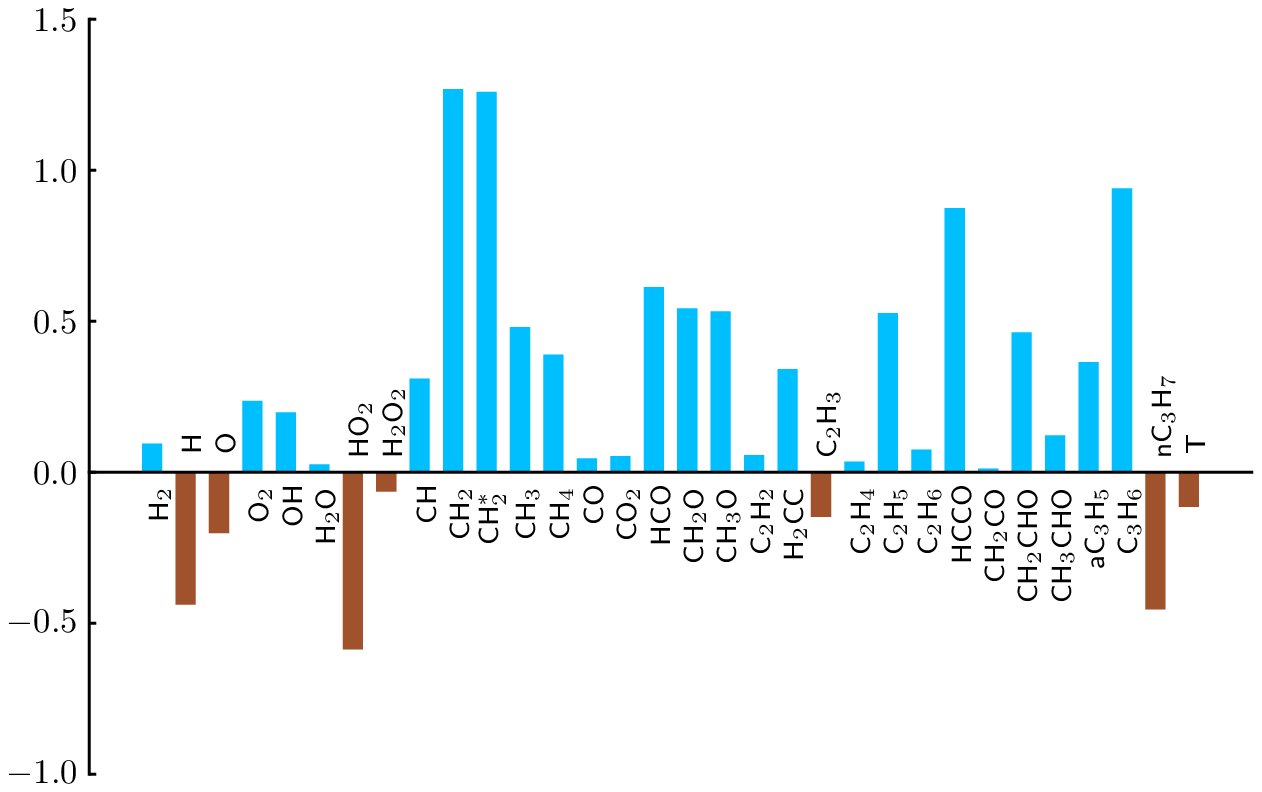}
    \begin{picture}(0,0)
    \put(-180,115){\scriptsize (a) Species mass fractions and temperature}
    \put(-205,45){\scriptsize{\rotatebox{90}{Error ratio $r_M$}}}
    \end{picture}
    } \hspace{0.5cm}
    {
    \includegraphics[width=7cm]{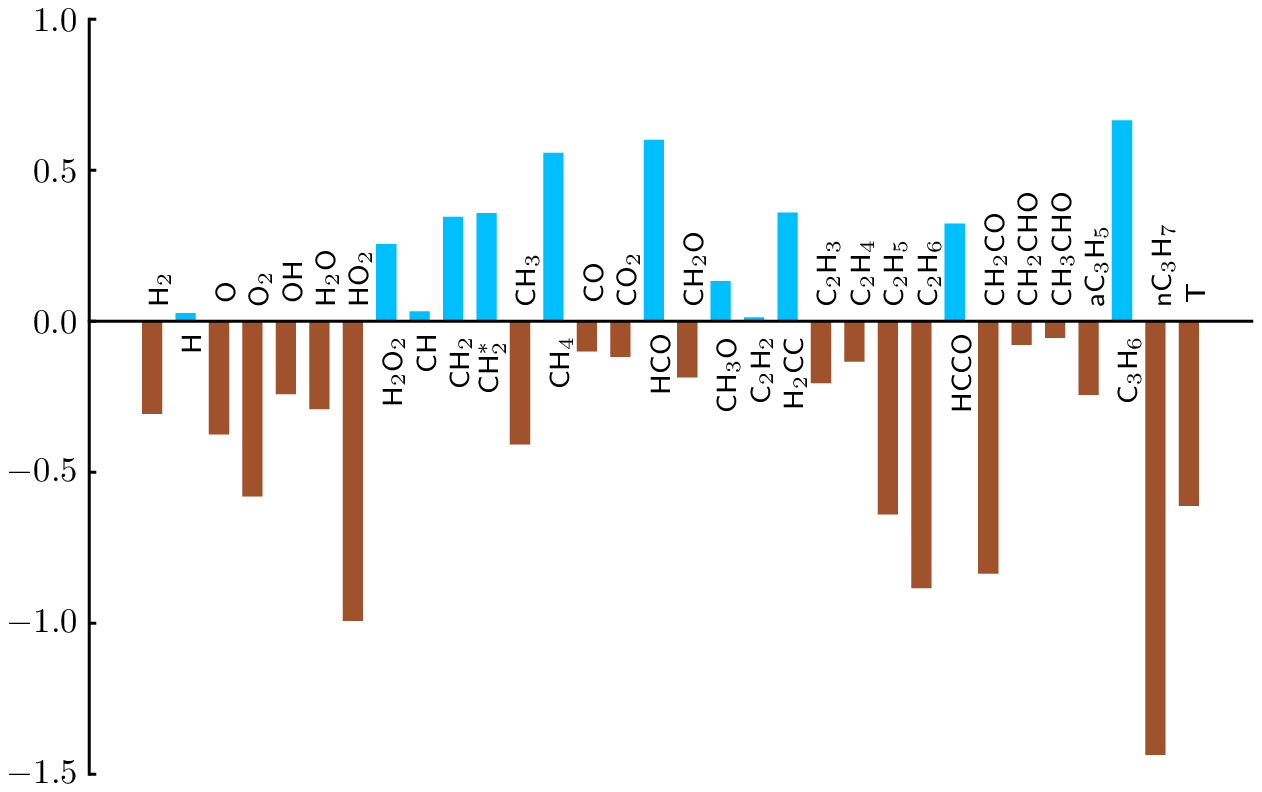}
    \begin{picture}(0,0)
    \put(-180,115){\scriptsize (b) Species mass fractions and temperature}
    \put(-205,45){\scriptsize{\rotatebox{90}{Error ratio $r_A$}}}
    \end{picture}
    }
    
    \vspace{0.4cm}
    {
    \includegraphics[width=7cm]{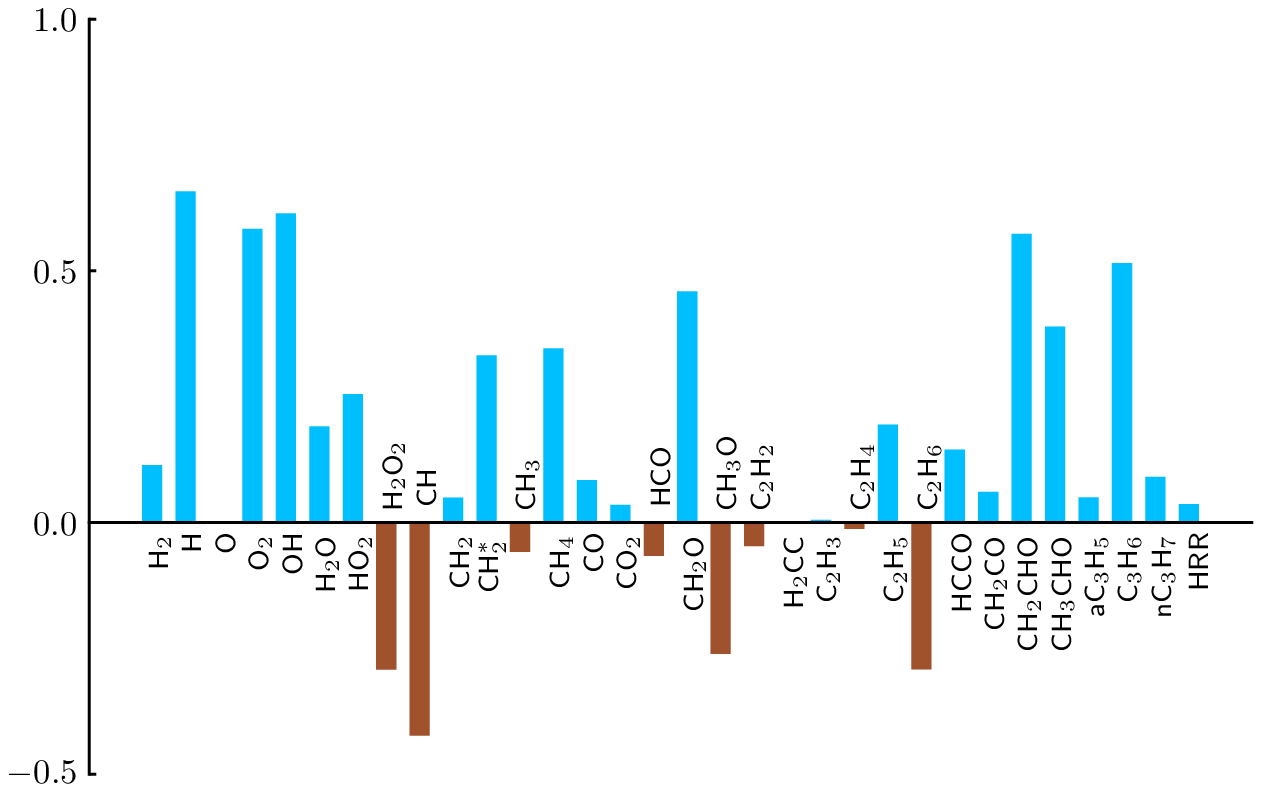}
    \begin{picture}(0,0)
    \put(-180,115){\scriptsize (c) Species production rates and heat release rate}
    \put(-205,45){\scriptsize{\rotatebox{90}{Error ratio $r_M$}}}
    \end{picture}
    }\hspace{0.5cm}
    {
    \includegraphics[width=7cm]{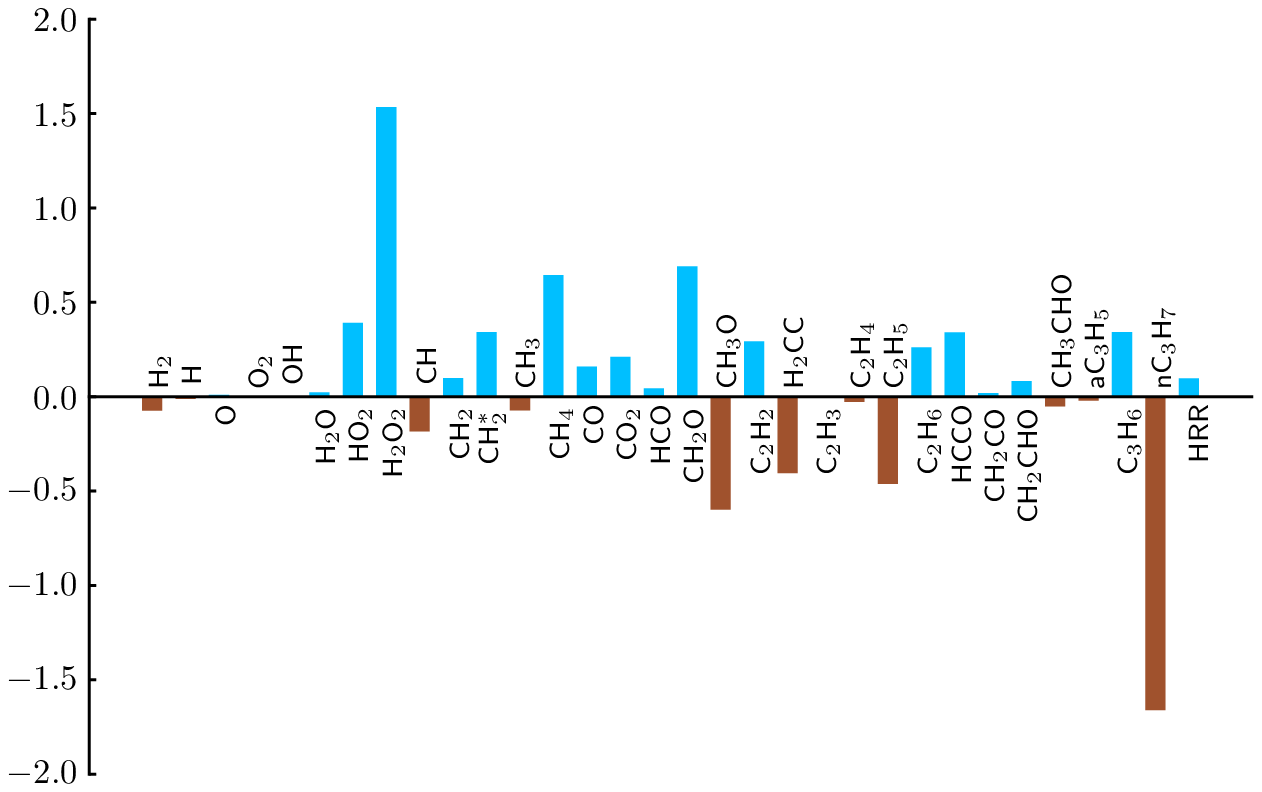}
    \begin{picture}(0,0)
    \put(-180,115){\scriptsize (d) Species production rates and heat release rate}
    \put(-205,45){\scriptsize{\rotatebox{90}{Error ratio $r_A$}}}
    \end{picture}
    }
    \caption{Plots of error ratio $r_i$ ($i\in \{M,A\}$, see Eq.~\ref{eq:error-ratio}) for reconstructed data, $n_q=5$, from homogeneous reactor dataset. Error ratio $r_i$ based on maximum (a) and average (b) errors of species mass fractions and temperature. Error ratio $r_i$ based on maximum (c) and average (d) errors of species production rates and heat release rate.\label{fig:errors-c2h4-q5}}
\end{figure*}


\rev{We now move on to characterize the behaviour of PCA and CoK-PCA reduced manifolds.
First, we assess the quality of reconstruction of the training dataset itself. Here, we construct two reduced manifolds by first adopting an aggressive dimensionality reduction strategy that retains $n_q = 5$ leading principal vectors out of the $n_v = 33$ vectors.
It should be noted that these $n_q = 5$ principal vectors in the PCA and CoK-PCA reduced manifolds correspond to approximately 99\% of the variance and 98\% of the kurtosis in the dataset, respectively.
Thereafter, we consider a less aggressive reduction with $n_q = 15$, i.e., approximately half of the total principal vectors are retained.}

\begin{figure*}[h!]
    \centering
    {
    \includegraphics[width=7cm]{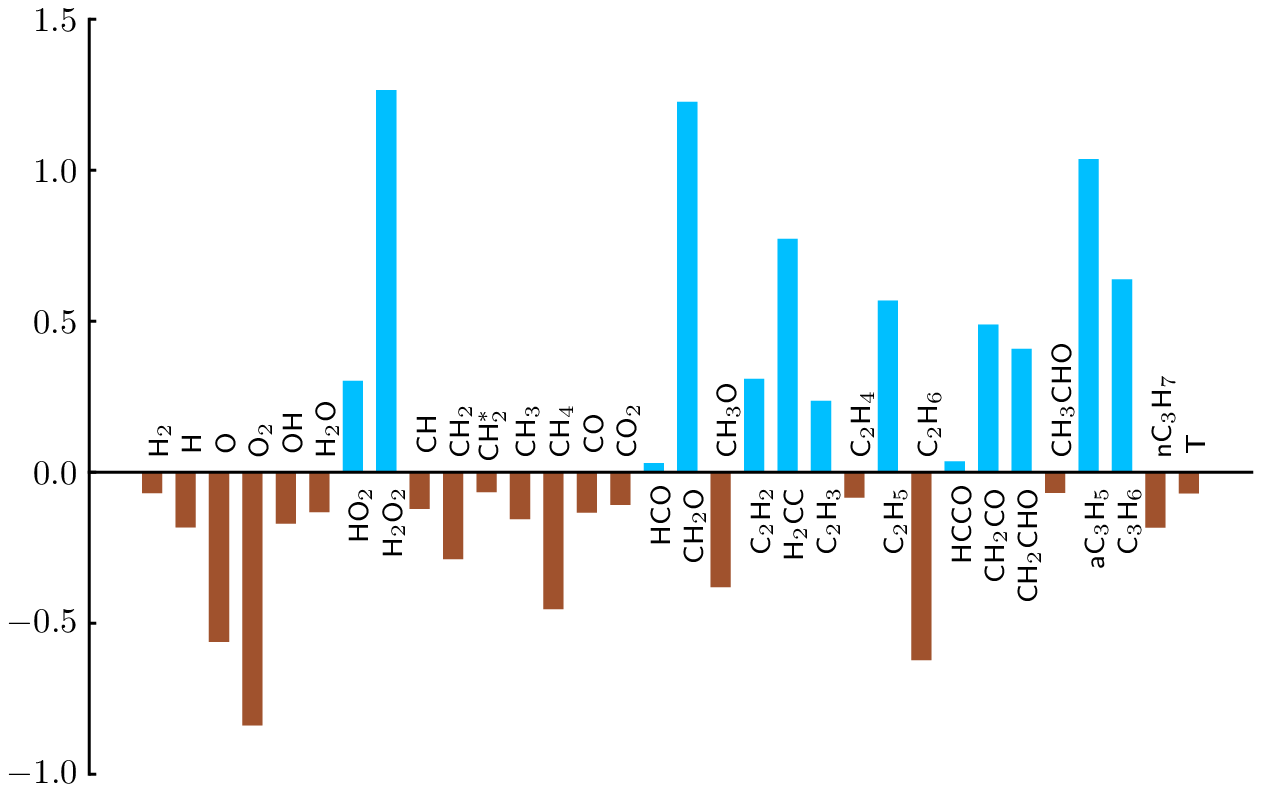}
    \begin{picture}(0,0)
    \put(-180,115){\scriptsize (a) Species mass fractions and temperature}
    \put(-205,45){\scriptsize{\rotatebox{90}{Error ratio $r_M$}}}
    \end{picture}
    } \hspace{0.5cm}
    {
    \includegraphics[width=7cm]{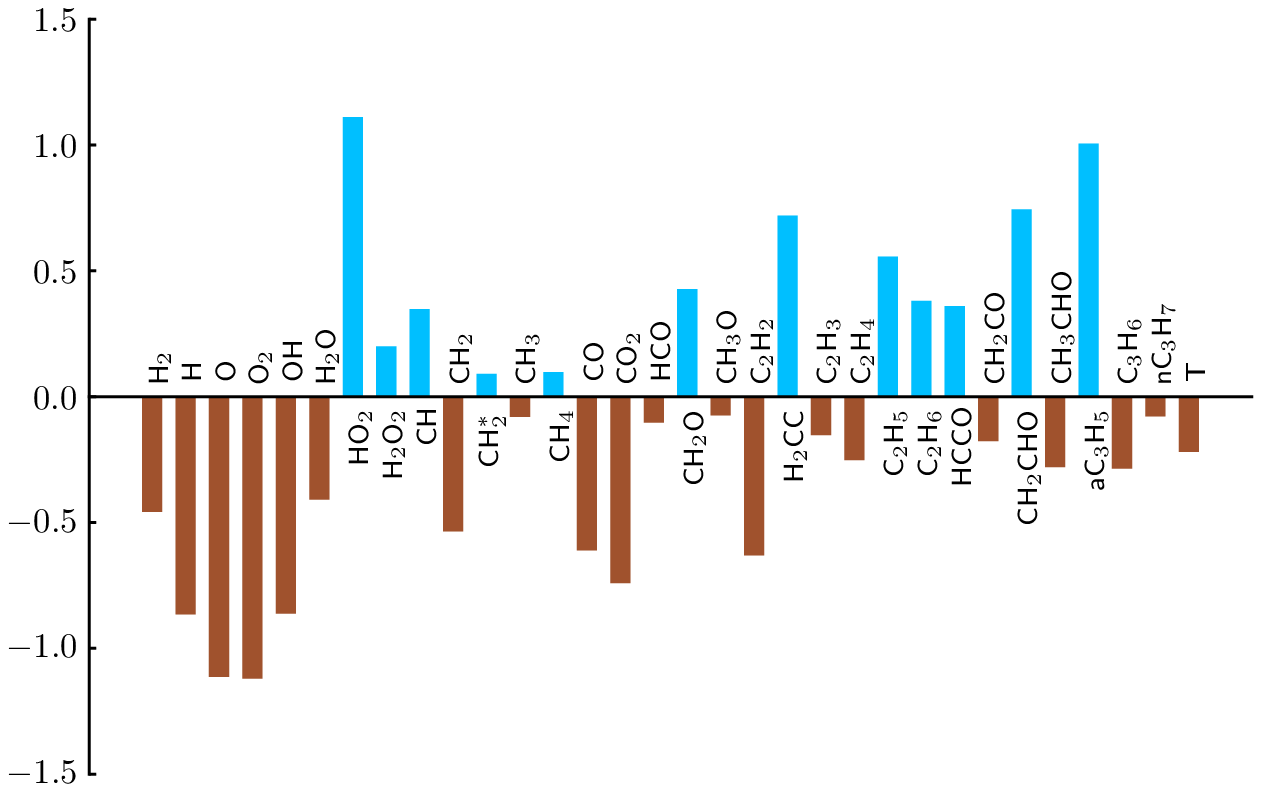}
    \begin{picture}(0,0)
    \put(-180,115){\scriptsize (b) Species mass fractions and temperature}
    \put(-205,45){\scriptsize{\rotatebox{90}{Error ratio $r_A$}}}
    \end{picture}
    }
    
    \vspace{0.4cm}
    {
    \includegraphics[width=7cm]{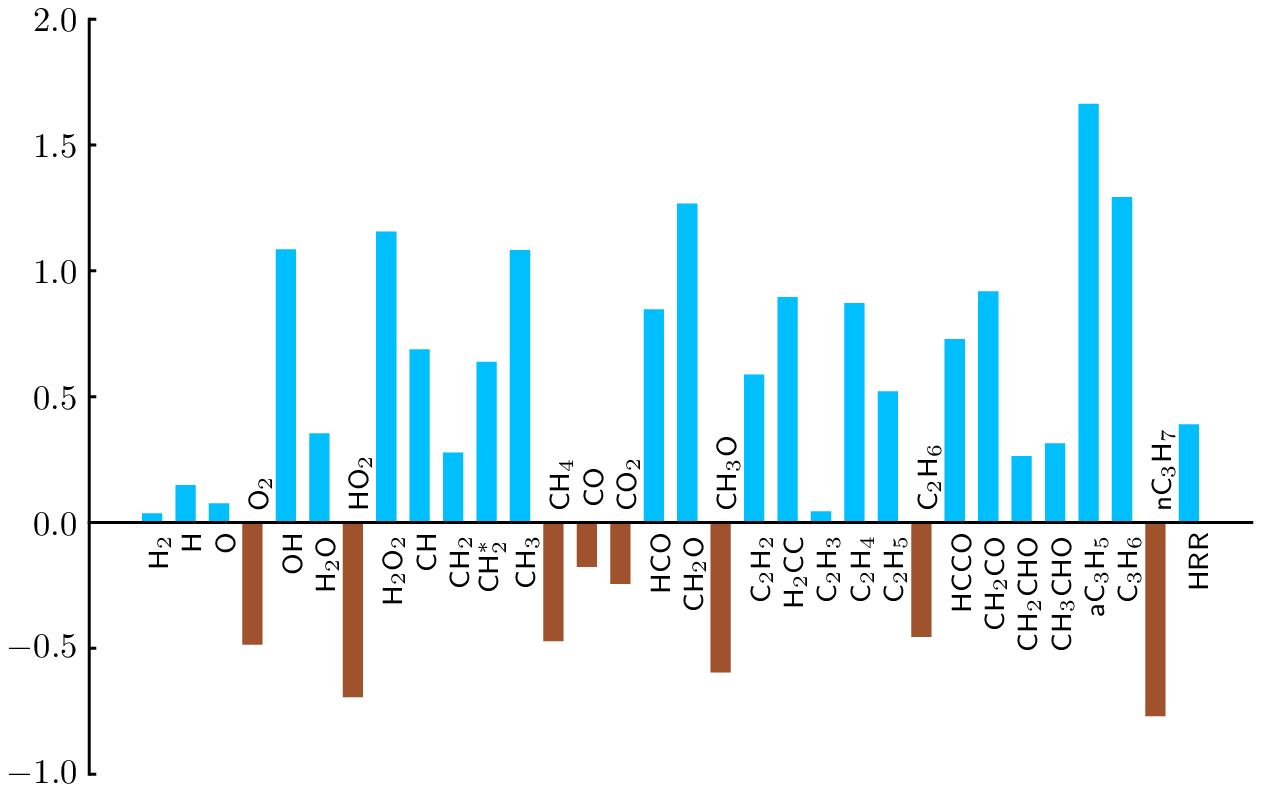}
    \begin{picture}(0,0)
    \put(-180,115){\scriptsize (c) Species production rates and heat release rate}
    \put(-205,45){\scriptsize{\rotatebox{90}{Error ratio $r_M$}}}
    \end{picture}
    }\hspace{0.5cm}
    {
    \includegraphics[width=7cm]{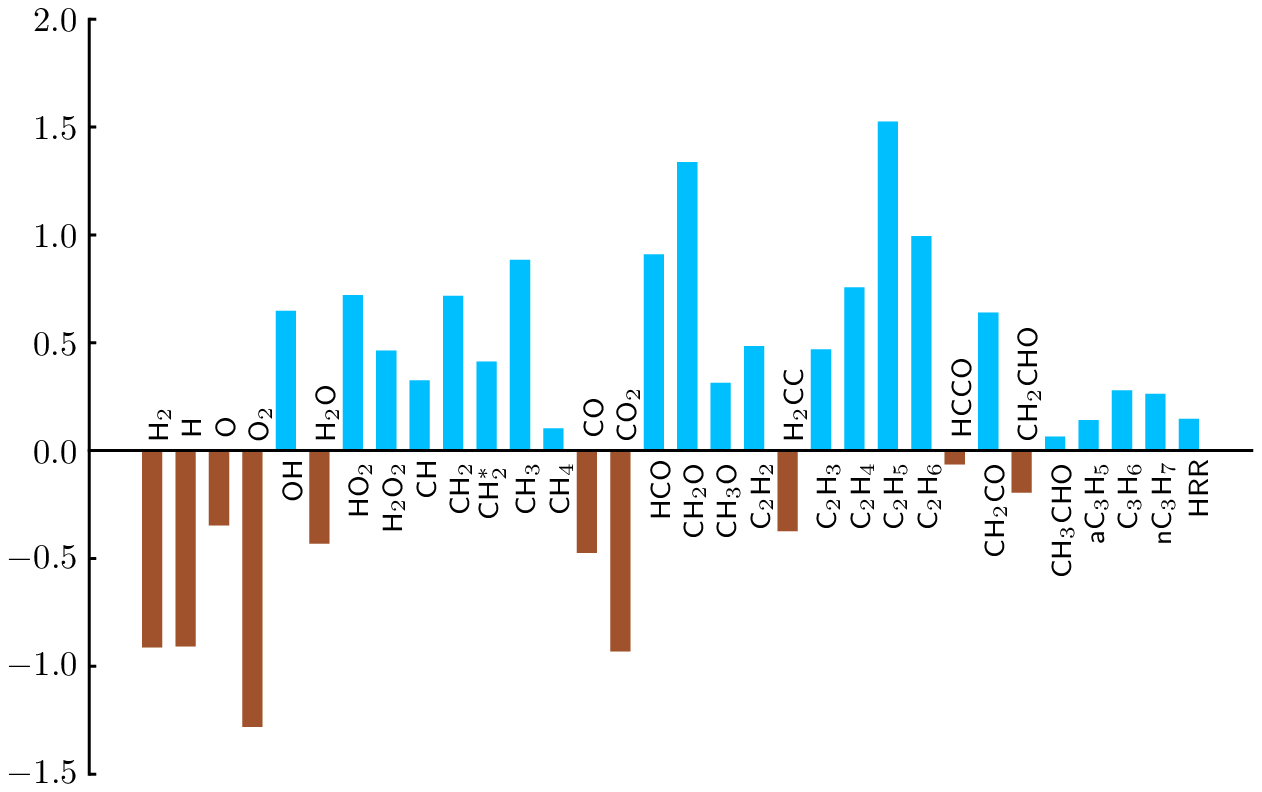}
    \begin{picture}(0,0)
    \put(-180,115){\scriptsize (d) Species production rates and heat release rate}
    \put(-205,45){\scriptsize{\rotatebox{90}{Error ratio $r_A$}}}
    \end{picture}
    }
    \caption{Plots of error ratio $r_i$ ($i\in \{M,A\}$, see Eq.~\ref{eq:error-ratio}) for reconstructed data, $n_q=15$, from homogeneous reactor dataset. Error ratio $r_i$ based on maximum (a) and average (b) errors of species mass fractions and temperature. Error ratio $r_i$ based on maximum (c) and average (d) errors of species production rates and heat release rate.\label{fig:errors-c2h4-q15}}
\end{figure*}

In Fig.~\ref{fig:errors-c2h4-q5}, the accuracy of reduced manifolds with $n_q=5$ is assessed in terms of their ability to represent the full state: thermo-chemical scalars, species production rates, and heat release rate.  Note that the reconstructed thermo-chemical scalars, $\mathbf{D_q}$, are used to compute species production rates and heat release rate, which are compared against those computed with the original state $\mathbf{D}$. Since production rates depend non-linearly on the species concentrations and temperature, reconstruction errors in the latter are likely to be amplified, thereby providing a more stringent test of accuracy.
\rev{Further, apart from having a tangible physical meaning, the reconstruction error associated with the heat release rate also provides an overall assessment of the quality of the reduced manifold since the heat release rate represents an aggregate effect of all the quantities of interest.}
Figure \ref{fig:errors-c2h4-q5} shows the error ratios for these quantities based on the maximum and average errors, $r_M$ and $r_A$, respectively.
$N_2$ being an inert species has not been included here.
Figures~\ref{fig:errors-c2h4-q5} (a) and (b) show that the error due to CoK-PCA reconstruction is smaller than the PCA reconstruction (demonstrated by the blue colored bars) for 25 and 12 out of the 32 variables for $r_M$ and $r_A$ metrics, respectively.
It should be noted that among the 25 species reconstructed better with CoK-PCA in $r_m$ plot, the reconstruction quality for several scalars is much better than that of PCA, as indicated by the taller blue bars.
These results indicate that the PCA reduced manifold, while presenting a better representation of the thermo-chemical scalars on an average, demonstrate significantly larger \textit{maximum} linear reconstruction errors of the thermo-chemical scalars in comparison to the CoK-PCA reduced manifold.
As will be shown later in this section, the samples corresponding to the maximum linear reconstruction error are representative of time instants corresponding to the occurrence of the majority of reactions and, thus, significantly contribute to the overall chemical dynamics of the system.
Figures~\ref{fig:errors-c2h4-q5} (c) and (d), show the reconstruction errors for the species production and heat release rates, which are non-linear functions of the thermo-chemical scalars. It can be clearly seen that CoK-PCA outperforms PCA in predicting the production rates for 23 of the 31 species with the $r_M$ metric and 18 of the 31 species for the $r_A$ metric.
Furthermore, these figures show that the linear reconstruction error in the heat release rates, for both the $r_M$ and $r_A$ metrics, are smaller for the CoK-PCA thus establishing that the CoK-PCA represents the overall chemical state of the system better than PCA.

We next consider the quality of the reduced manifolds for the less aggressive case, $n_q = 15$.
From Figs.~\ref{fig:errors-c2h4-q15} (a) and (b), it can be seen that the error due to CoK-PCA reconstruction is greater than PCA reconstruction for a larger number of variables as signified by the greater count of brown-colored bars: 19 and 20 of 32 variables in terms of the $r_M$ and $r_A$ metrics, respectively.
However, it should be noted that wherever CoK-PCA performs better for the $r_M$ metric, it does so by a larger extent as compared to PCA (signified by the taller blue bars).
Similarly, wherever CoK-PCA performs better in the $r_A$ metric, the performance is just as good as, if not better than, that of PCA.
However, as shown in Figs.~\ref{fig:errors-c2h4-q15} (c) and (d), CoK-PCA significantly outperforms PCA in terms of reconstruction errors for the production rates of 23 and 21 species out of 31 species for the $r_M$ and $r_A$ metrics, respectively.
Also, as in the $n_q = 5$ case, CoK-PCA once again captures the overall chemical state of the system better than PCA as represented by the average and maximum reconstruction errors in the heat release rate shown in Figs.~\ref{fig:errors-c2h4-q15} (c) and (d).

\rev{It should be noted that in order to ensure that the reconstructed thermo-chemical state results in unit sum of species mass fractions, as is the standard practice, all reconstructed species mass fractions which yielded negative values (that were all slightly smaller than zero) were taken to be zero after which any deviation from the sum equalling unity was adjusted for in the non-participating or bath species.}

\begin{figure*}[h!]
    \centering
    {
    \includegraphics[width=6cm]{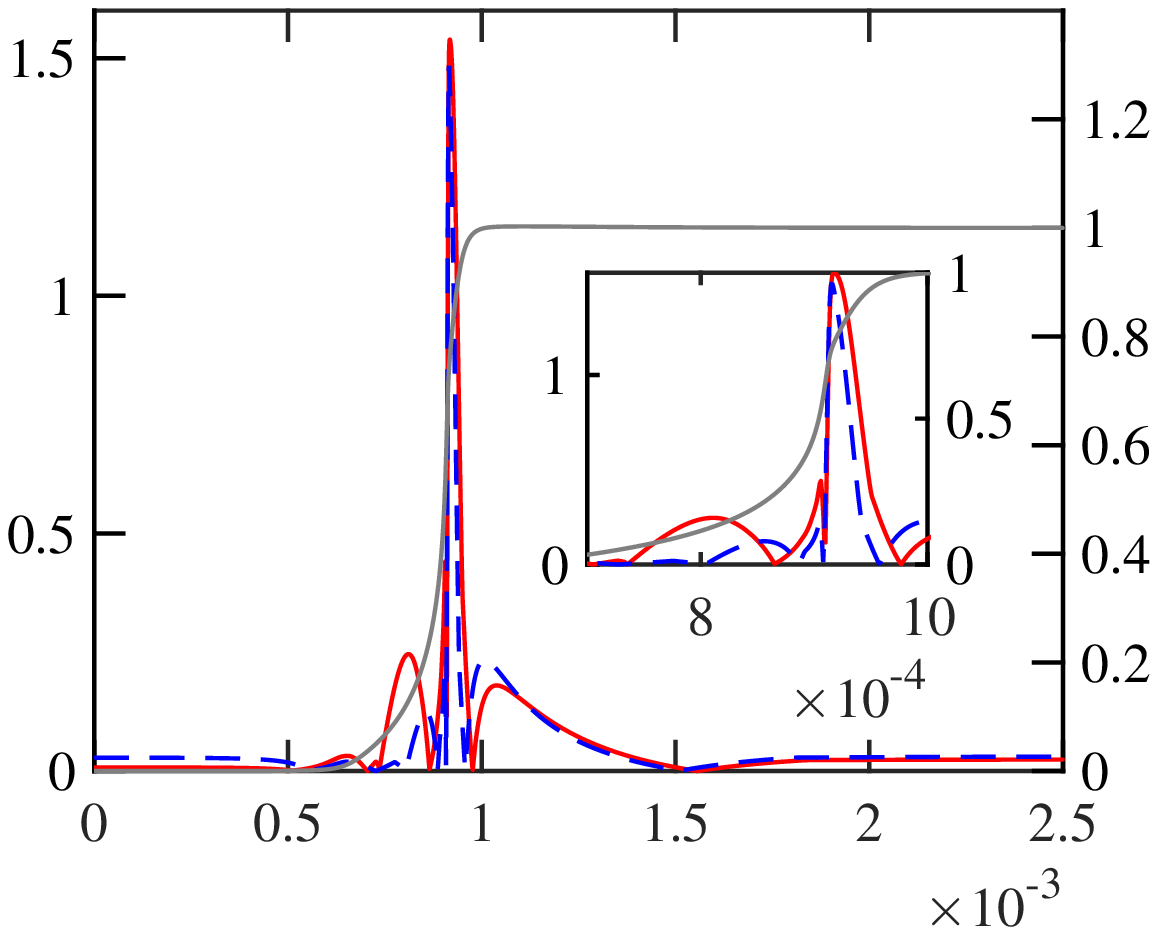}
    \begin{picture}(0,0)
    \put(-143,110){\small (a)}
    \put(-173,35){\scriptsize{\rotatebox{90}{Normalized error in HRR}}}
    \put(-3,95){\scriptsize{\rotatebox{270}{Progress variable}}}
    \put(-100,2){\scriptsize Time (\SI{}{\second})}
    \end{picture}
     }  \hspace{0.6cm}
    {
    \includegraphics[width=6cm]{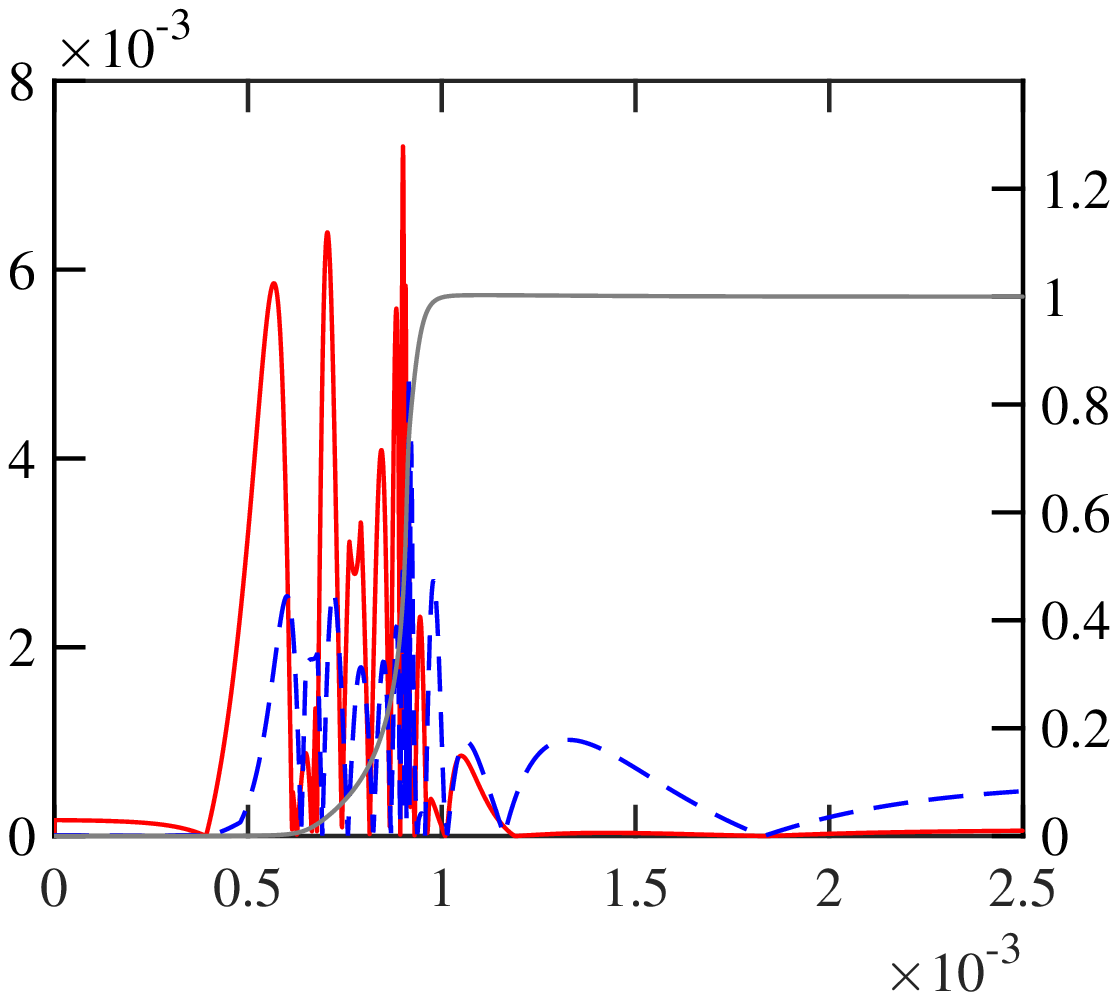}
    \begin{picture}(0,0)
    \put(-143,110){\small (b)}
    \put(-170,35){\scriptsize{\rotatebox{90}{Normalized error in HRR}}}
    \put(-3,95){\scriptsize{\rotatebox{270}{Progress variable}}}
    \put(-100,2){\scriptsize Time (\SI{}{\second})}
    \end{picture}
    }
    \caption{Time evolution of normalised error in heat release rate (HRR) computed from reconstructed data: (a) $n_q=5$ and (b) $n_q=15$. The error values are normalised with peak heat release rate value in the original data. Solid-red and dashed-blue lines represent errors from PCA and CoK-PCA, respectively. The progress variable is plotted in grey for reference.}
    \label{fig:errors-c2h4-hrr}
\end{figure*}

\begin{figure*}[h!]
    \centering
    {
    \includegraphics[width=7cm]{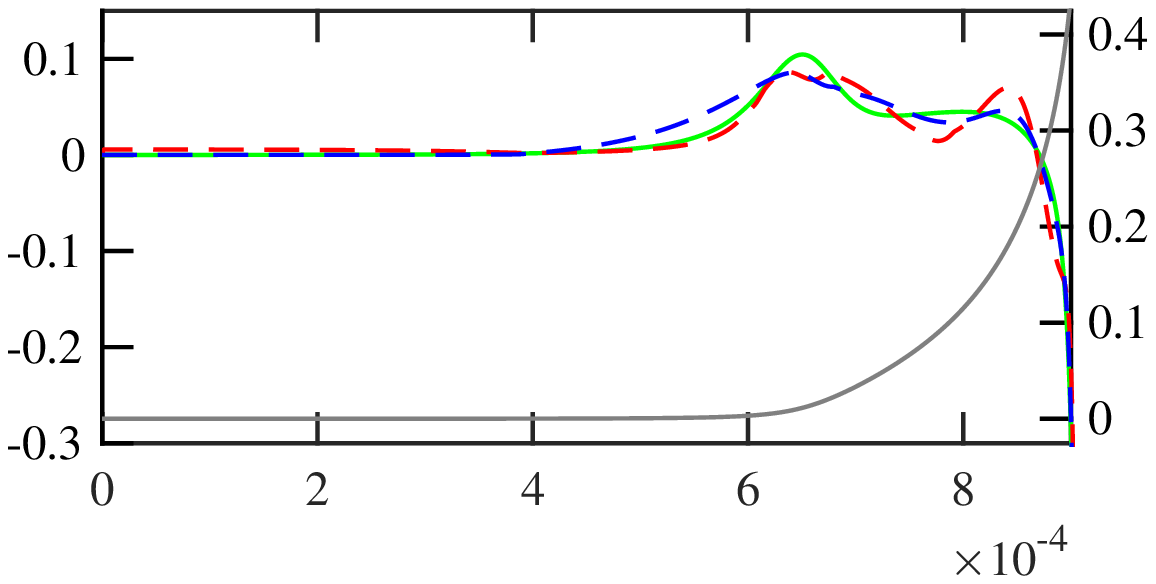}
    \begin{picture}(0,0)
    \put(-167,80){\small (a)}
    \put(-200,25){\scriptsize{\rotatebox{90}{Normalized PR ($HO_2$)}}}
    \put(-6,80){\scriptsize{\rotatebox{270}{Progress variable}}}
    \put(-115,5){\scriptsize Time (\SI{}{\second})}
    \end{picture}
     }  \hspace{0.25cm}
    {
    \includegraphics[width=7cm]{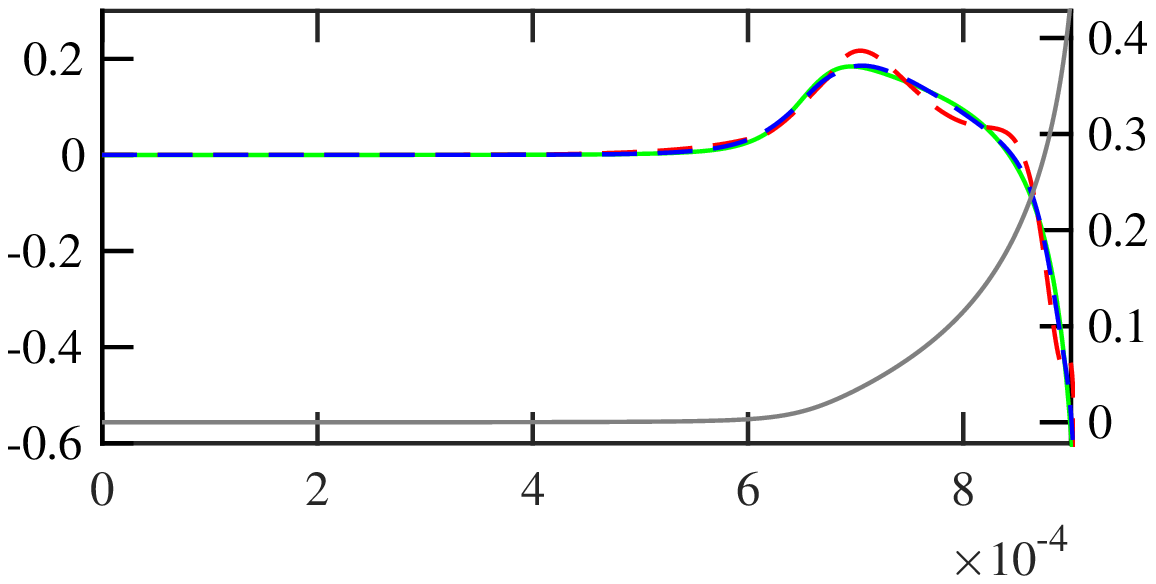}
    \begin{picture}(0,0)
    \put(-167,80){\small (b)}
    \put(-200,20){\scriptsize{\rotatebox{90}{Normalized PR ($CH_2O$)}}}
    \put(-6,80){\scriptsize{\rotatebox{270}{Progress variable}}}
    \put(-115,5){\scriptsize Time (\SI{}{\second})}
    \end{picture}
    }
    \caption{\rev{Time evolution of production rate (PR) of (a) $HO_2$ and (b) $CH_2O$ computed from reconstructed data obtained from the $n_q=15$ reduced manifolds. Solid-green, dashed-red and dashed-blue lines are production rates from original data, PCA and CoK-PCA, respectively. The progress variable is plotted in grey for reference.}}
    \label{fig:errors-c2h4-pr-key-radicals}
\end{figure*}

Up to this point, the characterization of the CoK-PCA and PCA reduced manifolds that have been presented is agnostic of the temporal evolution of the system.
Specifically, it is unclear if the data samples that incur large reconstruction errors for the reduced manifolds with the CoK-PCA and PCA belong to the reactants, the products, or the ignition front.
To relate the results presented earlier to the progress of the combustion process, we present, in Fig.~\ref{fig:errors-c2h4-hrr}, the normalized error in heat release rate obtained through linearly reconstructed scalars from both the reduced manifolds for $n_q=$ $5$ and $15$, along with progress variable.
The normalized error, $\varepsilon_{\text{HRR}}$, is defined as:
\begin{equation}
\varepsilon_{\text{HRR}}= \frac{|\text{HRR}_{rm}-\text{HRR}_{original}|}{max(\text{HRR}_{original})},
\label{eq:normHRR}
\end{equation}
where the subscript ``$rm$'' refers to reduced manifold.
The error profile for $n_q=5$ presented in Fig.~\ref{fig:errors-c2h4-hrr}(a) shows that the HRR obtained from CoK-PCA incurs greater error in reactants and products as compared to PCA.
In contrast, in the reaction zone where significant heat release would be present, CoK-PCA outperforms PCA.
Similarly for the $n_q=15$ reduced manifolds, as evident from Fig.~\ref{fig:errors-c2h4-hrr}(b), CoK-PCA predicts HRR with a greater accuracy in the reaction zone.


The trends in Fig.~\ref{fig:errors-c2h4-hrr} can be explained as follows.
Figs.~\ref{fig:errors-c2h4-q5} (a)-(b) and ~\ref{fig:errors-c2h4-q15} (a)-(b) show that upon reconstructing the thermo-chemical scalars, CoK-PCA and PCA perform better in the $r_M$  and $r_A$  metrics, respectively.
It is noteworthy that while $r_A$ captures the average error, including the errors incurred at all the time steps, $r_M$ captures the peak error.
For PCA, this peak error would usually manifest in the flame front where majority of reactions occur (see Fig.~\ref{fig:errors-c2h4-hrr}), and thus would significantly affect the species production rates and the heat release rate.
Further, as the flame front is represented by only a fraction of all the data samples, and would typically be considered as \rev{extreme-valued samples}, this region is expected to be represented better by the CoK-PCA reduced manifold (again see Fig.~\ref{fig:errors-c2h4-hrr}).
Accordingly, as demonstrated in Figs.~\ref{fig:errors-c2h4-q5} (c)-(d) and ~\ref{fig:errors-c2h4-q15} (c)-(d), the thermo-chemical scalars reconstructed from CoK-PCA reduced manifold yield a more accurate species production rates and overall heat release rate.
Figure~\ref{fig:errors-c2h4-hrr} presents further evidence that the proposed CoK-PCA method predicts the overall chemical kinetics in the reaction zone to a much better extent than PCA.

\begin{table}[h!]
\centering
\begin{tabular}{ccc}
\hline\hline
\multirow{2}{*}{\begin{tabular}[c]{@{}c@{}}Method\end{tabular}} & \multicolumn{2}{c}{Cumulative $\varepsilon_{PR}$} \\
\cline{2-3}
                                                                            & HO$_2$                & CH$_2$O              \\
\hline\hline
PCA                                                                         & 4.9517                & 3.8152               \\
CoK-PCA                                                                     & 4.3663                & 1.1834               \\
\hline\hline
\end{tabular}
\caption{\rev{Cumulative errors in the production rates for the ignition markers in the early reaction zones (up to a progress variable of 0.1) obtained from different reduced manifolds for the homogeneous reactor case.} \label{table:cumulative-pr-errors}}
\end{table}

\rev{Further, in order to compare the relative accuracy of the PCA and CoK-PCA reduced manifolds in very-low progress variable region, which characterizes the thermal runaway that is important in accurately predicting the ignition delay, we examine the production rates of two key radicals in these early reaction zones, namely the hydroperoxy (HO$_2$) and formaldehyde (CH$_2$O) radicals, respectively, as ignition markers. 
For the HO$_2$ radical (Fig.~\ref{fig:errors-c2h4-pr-key-radicals} (a)), we qualitatively observe that the CoK-PCA reduced manifold slightly over-predicts the production rate  between t = 0.4 ms to t = 0.6 ms, while the PCA reduced manifold under-predicts to small extent. After t=0.6 ms, the HO$_2$ production rates from both reduced manifolds oscillate around the true production rate values.
In contrast, from Fig.~\ref{fig:errors-c2h4-pr-key-radicals} (b), the production rates from the CoK-PCA reduced manifold can, qualitatively, be seen to be far better aligned with the true values as compared to those obtained from the PCA reduced manifold.
To make a further quantitative assessment, we evaluate the cumulative normalized error in the production rates of the HO$_2$ and CH$_2$O radicals up to a progress variable value of 0.1. The normalized error values are evaluated as:
\begin{equation}
    \varepsilon_{PR} = \frac{\left|PR_{rm} - PR_{original}\right|}{max\left(\left|PR_{original}\right|\right)}, 
\end{equation}
and the cumulative values are tabulated in Table~\ref{table:cumulative-pr-errors}.
It can be seen that the cumulative errors associated with the production rates of the considered ignition markers obtained with the CoK-PCA reduced manifold are smaller than those from the PCA reduced manifold, thus indicating that the proposed method provides a better estimate of the ignition time delay as compared to PCA.
It should be noted that the present discussion is based on the reconstructed production rates obtained from the $n_q = 15$ reduced manifolds; similar observations also hold true for the $n_q = 5$ reduced manifolds.
}

\begin{figure*}[h!]
    \centering
    {
    {\footnotesize (a) T = 1150K, $\phi = 0.375$}
    
    \includegraphics[width=5.5cm]{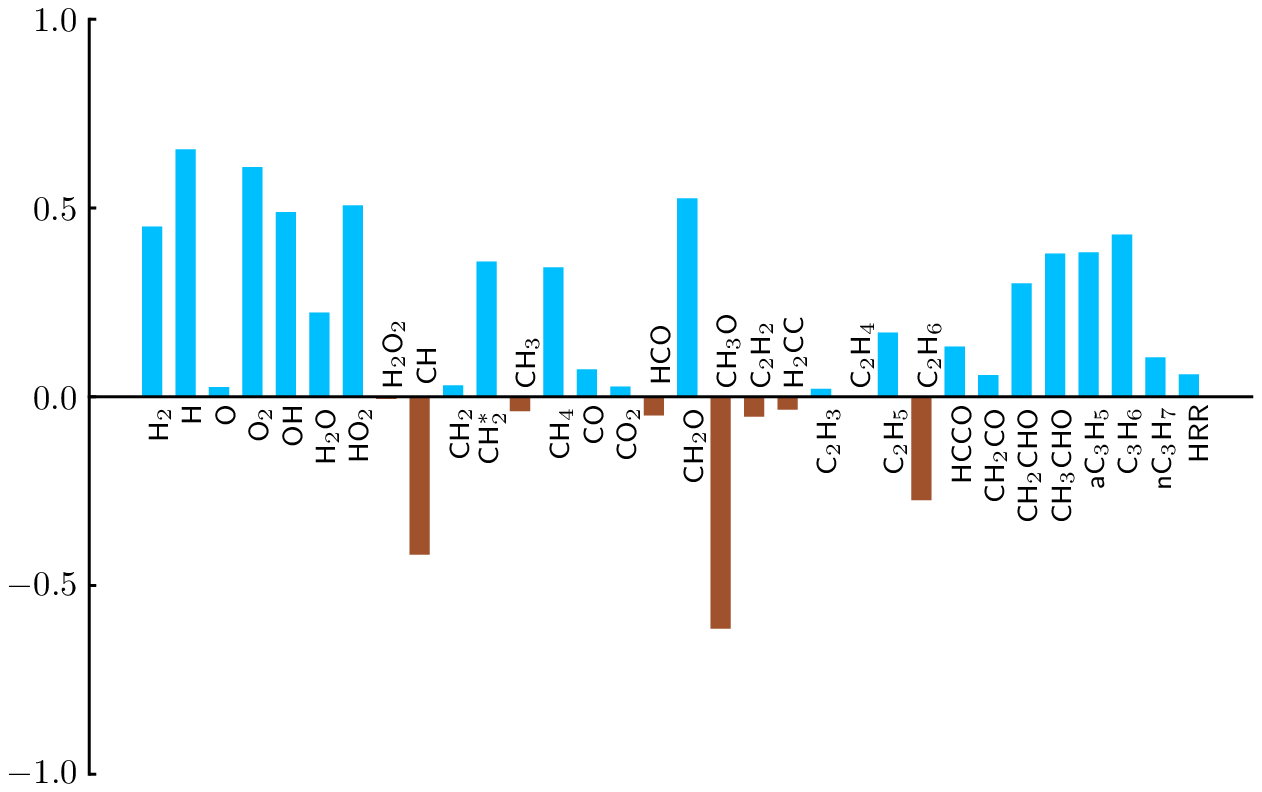}
    \begin{picture}(0,0)
    \put(-165,30){\scriptsize{\rotatebox{90}{Error ratio $r_M$}}}
    \end{picture}
    } \hspace{0.0cm}
    {
    \includegraphics[width=5.5cm]{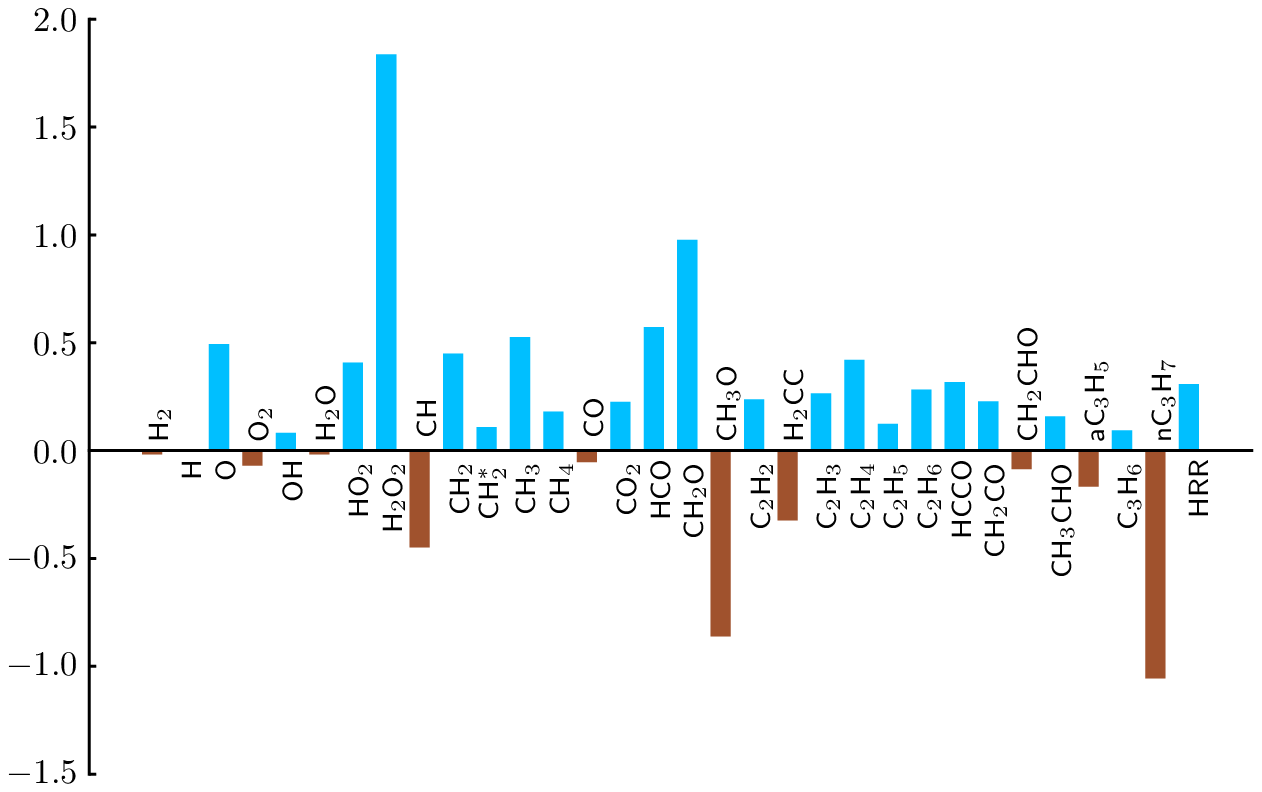}
    \begin{picture}(0,0)
    \put(-165,30){\scriptsize{\rotatebox{90}{Error ratio $r_A$}}}
    \end{picture}
    }
    
    {\footnotesize (b) T = 1150K, $\phi = 0.425$}
    
    \vspace{0.35cm}
    {
    \includegraphics[width=5.5cm]{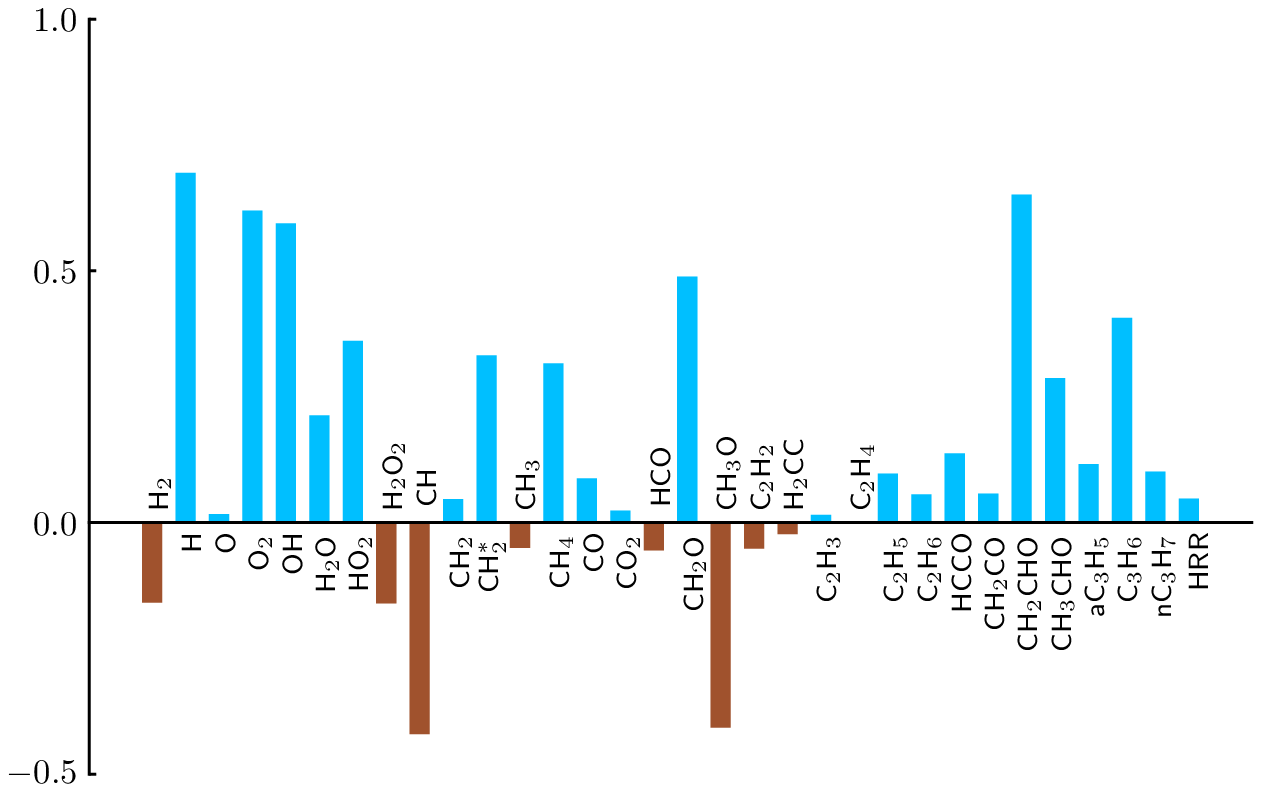}
    \begin{picture}(0,0)
    \put(-165,30){\scriptsize{\rotatebox{90}{Error ratio $r_M$}}}
    \end{picture}
    } \hspace{0.0cm}
    {
    \includegraphics[width=5.5cm]{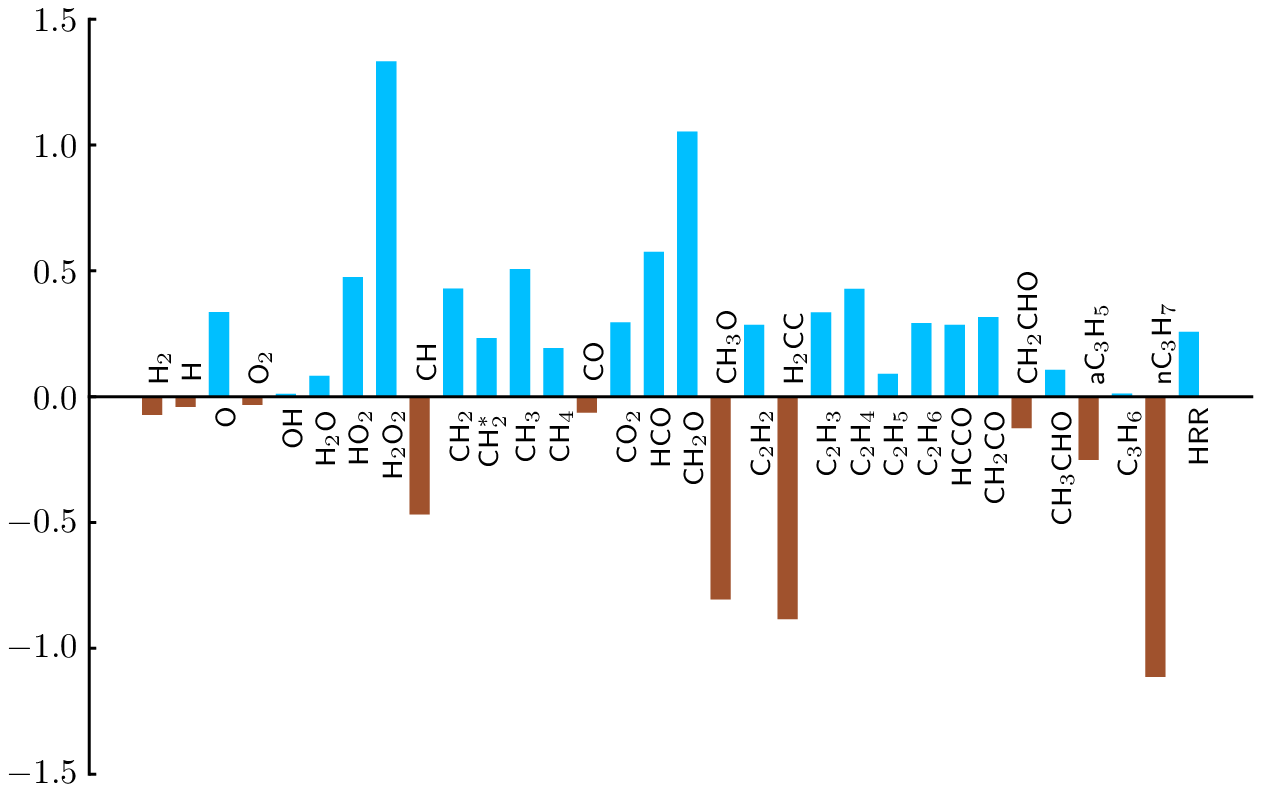}
    \begin{picture}(0,0)
    \put(-165,30){\scriptsize{\rotatebox{90}{Error ratio $r_A$}}}
    \end{picture}
    }
    
    {\footnotesize (c) T = 1250K, $\phi = 0.375$}
    
    {
    \includegraphics[width=5.5cm]{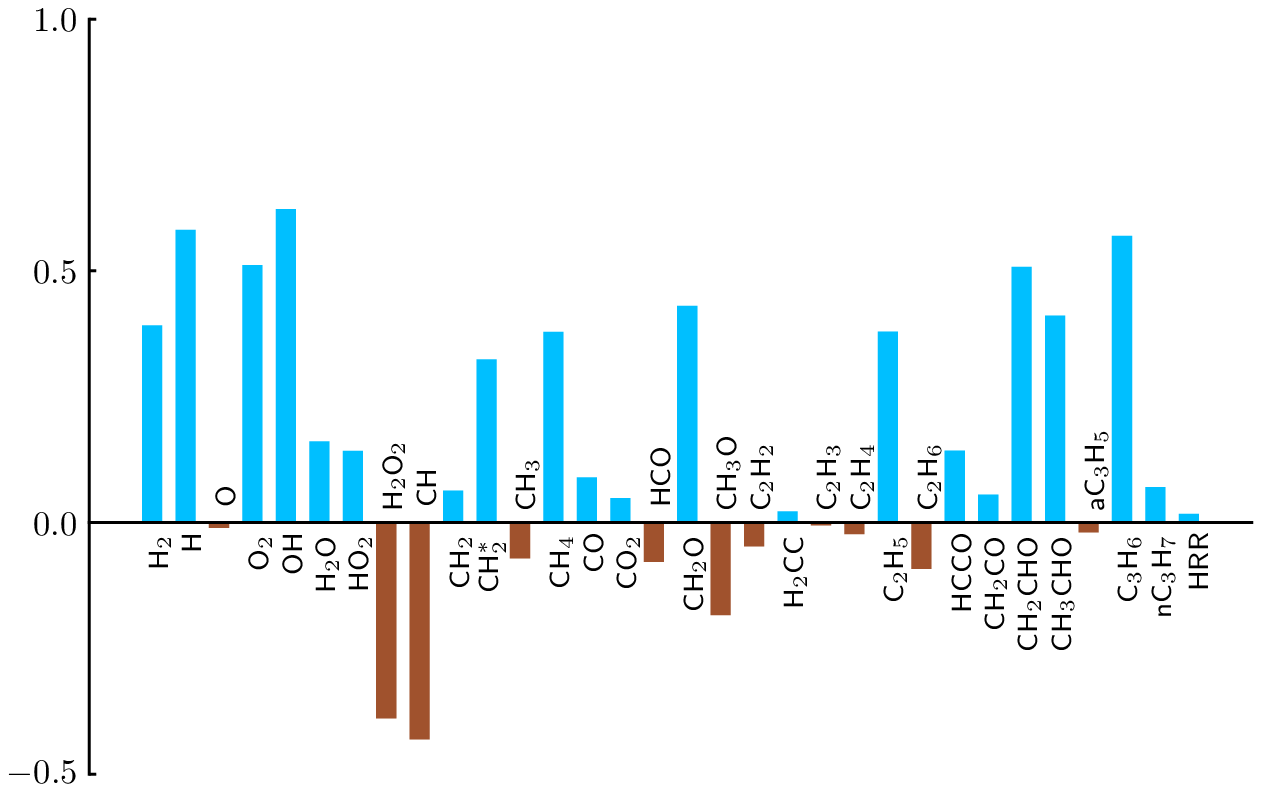}
    \begin{picture}(0,0)
    \put(-165,30){\scriptsize{\rotatebox{90}{Error ratio $r_M$}}}
    \end{picture}
    } \hspace{0.0cm}
    {
    \includegraphics[width=5.5cm]{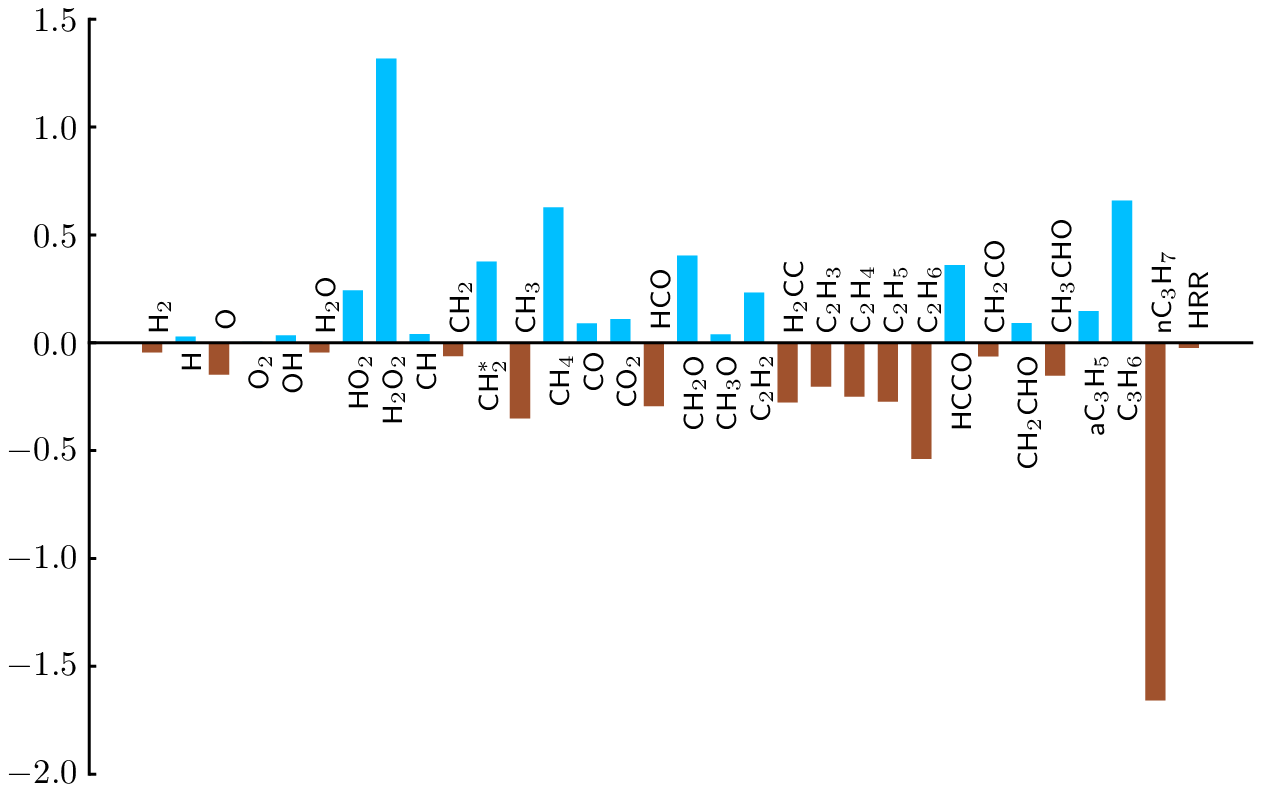}
    \begin{picture}(0,0)
    \put(-165,30){\scriptsize{\rotatebox{90}{Error ratio $r_A$}}}
    \end{picture}
    }
    
    {\footnotesize (d) T = 1250K, $\phi = 0.425$}
    
    {
    \includegraphics[width=5.5cm]{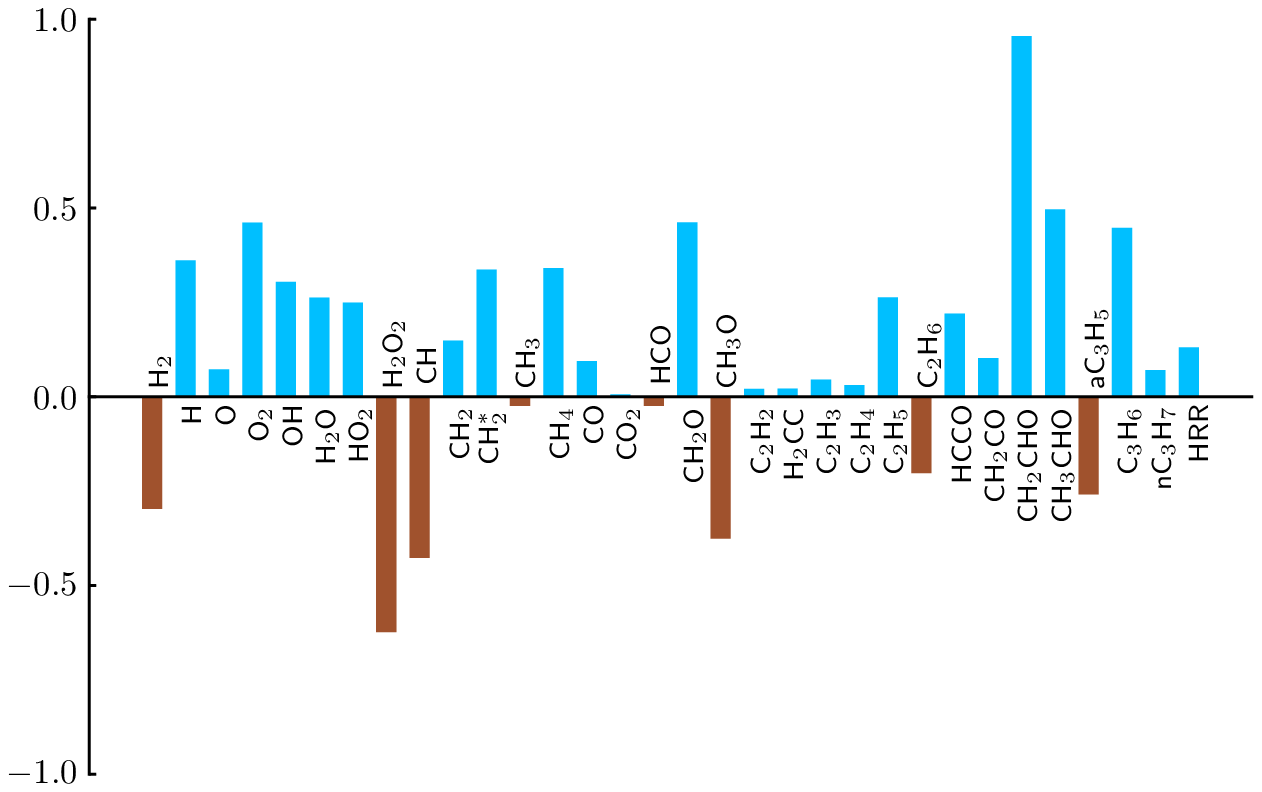}
    \begin{picture}(0,0)
    \put(-165,30){\scriptsize{\rotatebox{90}{Error ratio $r_M$}}}
    \end{picture}
    } \hspace{0.0cm}
    {
    \includegraphics[width=5.5cm]{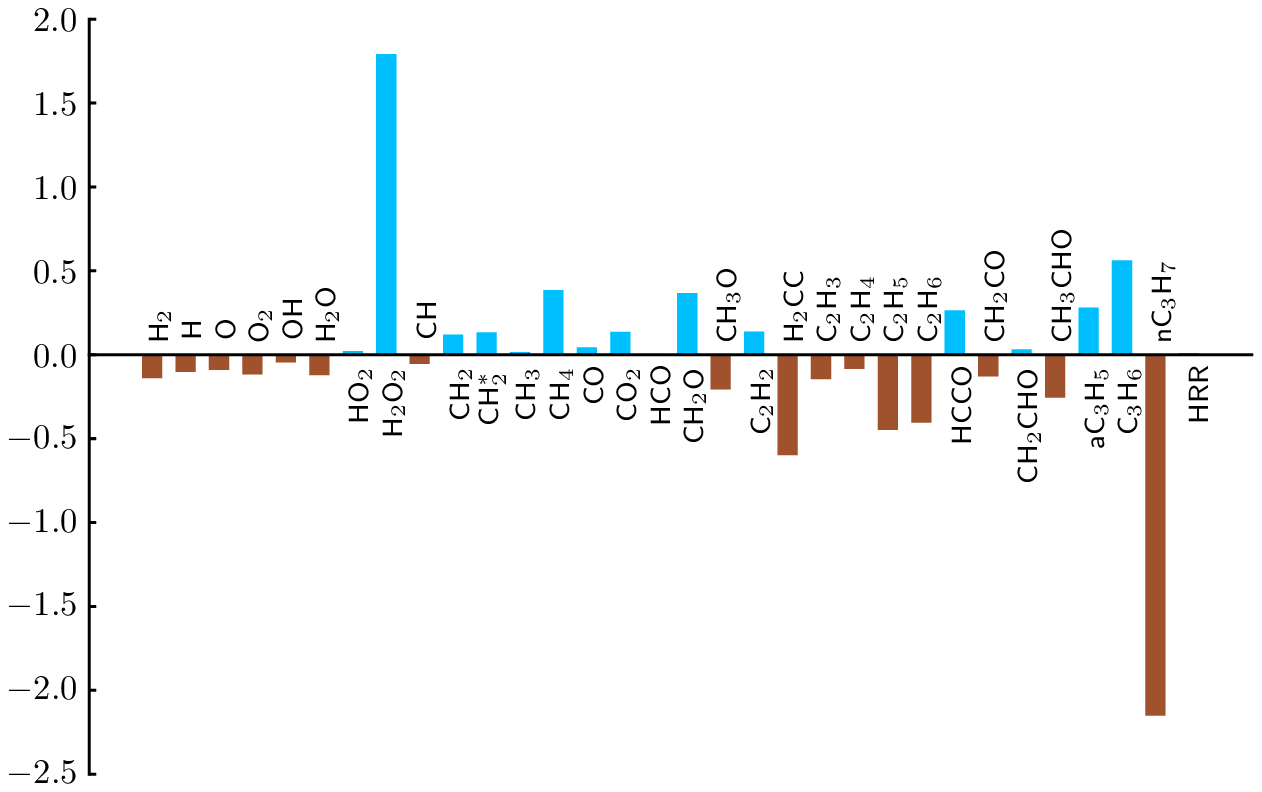}
    \begin{picture}(0,0)
    \put(-165,30){\scriptsize{\rotatebox{90}{Error ratio $r_A$}}}
    \end{picture}
    }

    \caption{\rev{Plots of error ratio $r_i$ ($i\in \{M,A\}$, see Eq.~\ref{eq:error-ratio}) of the species production rates and heat release rate for different test configurations of the homogeneous reactor case. The reconstruction is done using the $n_q=5$ reduced PCA and CoK-PCA manifolds obtained for the training case having initial conditions T=1200K and $\phi=0.4$, respectively. The left and right columns represent the error ratio $r_i$ based on maximum error and  average errors, respectively. \label{fig:0D-testing-data}}}
\end{figure*}

\rev{In order to assess the robustness of the dimensionality reduction procedure we test the PCA and CoK-PCA manifolds at conditions different from that of the training dataset. Using the PCA and CoK-PCA manifolds obtained at the training (baseline) conditions of T$_{\text{training}}$=\SI{1200}{\K} and $\phi_{\text{training}}$=\SI{0.4}{},  we evaluate reconstruction accuracy for datasets at test conditions with different initial temperature and equivalence ratio.
We assess the reconstruction accuracy of the test conditions with the manifolds corresponding to the more aggressive reduction, $n_q = 5$ principal vectors.
Based on previous experimental studies \cite{geipel2021cavity,LIEBER2022cavity}, the test conditions differ from the training conditions by $\Delta$T=$\pm$\SI{50}{\K} and $\Delta\phi=\pm$\SI{0.025}{}, respectively.
Reconstruction errors for the species production rates and heat release rate are presented in Fig.~\ref{fig:0D-testing-data}.
It can be seen that the CoK-PCA manifold yields satisfactory predictions on test conditions in both the $r_M$ and $r_A$ metric, as evidenced by the larger number of blue bars in comparison to the brown bars.
Specifically, when compared to the PCA reduced manifold, the CoK-PCA reduced manifold presents a better reconstruction (averaged across all four test cases) of the species production rates for approximately $73\%$ and $60\%$ of the considered species in the $r_M$ and $r_A$ metrics, respectively.
Further, the CoK-PCA manifold is also able to capture the overall chemical kinetics better than the PCA manifold at the test conditions, as evidenced by the positive or comparable values (for the $r_A$ metric in the T=\SI{1250}{\K}, $\phi=$\SI{0.375}{}, \SI{0.425}{} cases) of the heat release rate.}

\subsection{Homogeneous charge compression ignition (HCCI) dataset}
\label{sec:hcci}

\begin{figure}[h!]
    \centering
    \hspace{0.05cm}
    \includegraphics[trim=0.5cm 0cm 2.35cm 0cm,clip=true,height=3.9cm]{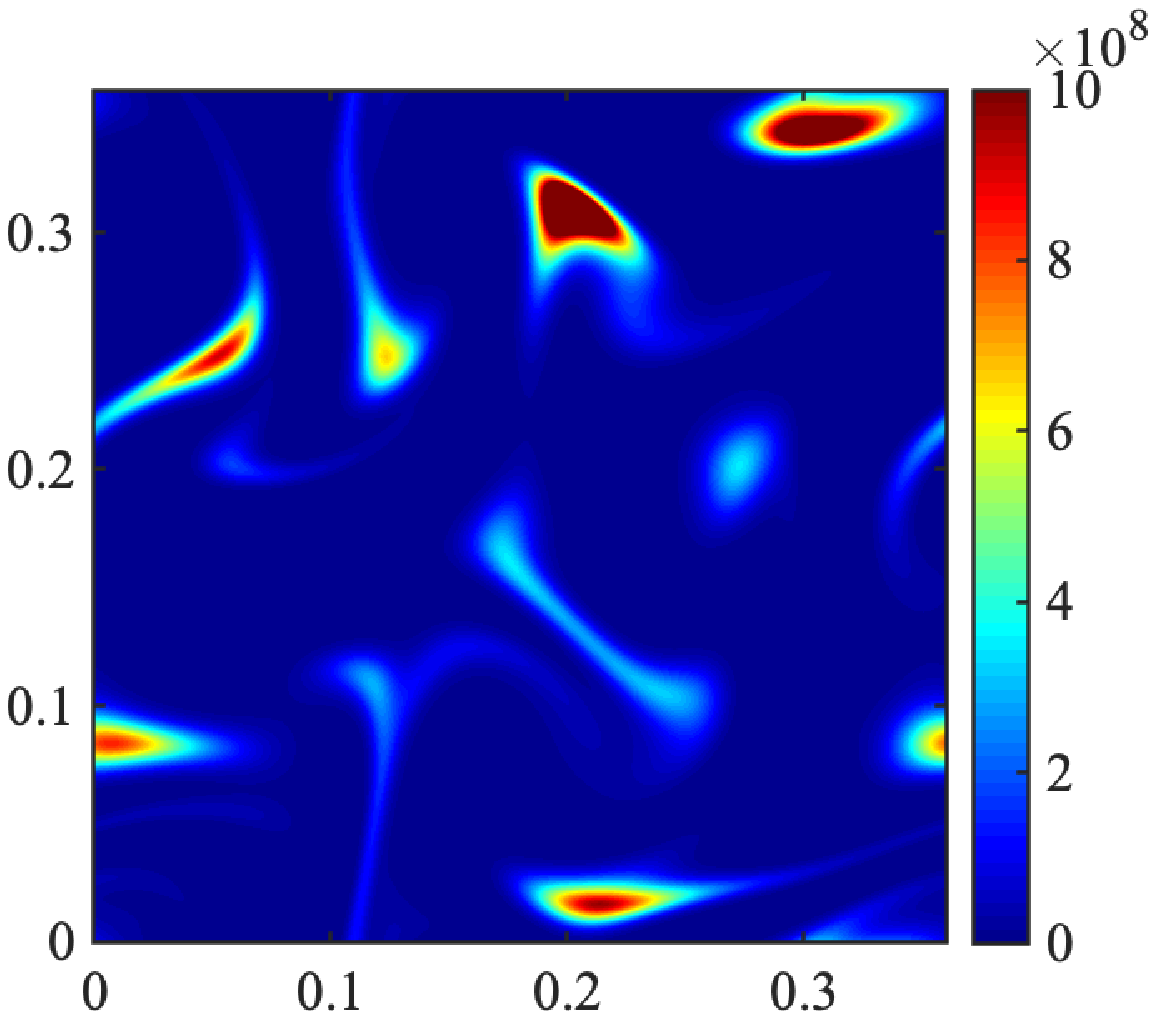}
    \begin{picture}(0,0)
    \put(-60,0){\scriptsize x (\SI{}{\cm})}
    \put(-108,48){\scriptsize{\rotatebox{90}{y (\SI{}{\cm})}}}
    \end{picture}
    \hspace{0.0cm}
    \includegraphics[trim=0.5cm 0cm 0.5cm 0cm,clip=true,height=3.9cm]{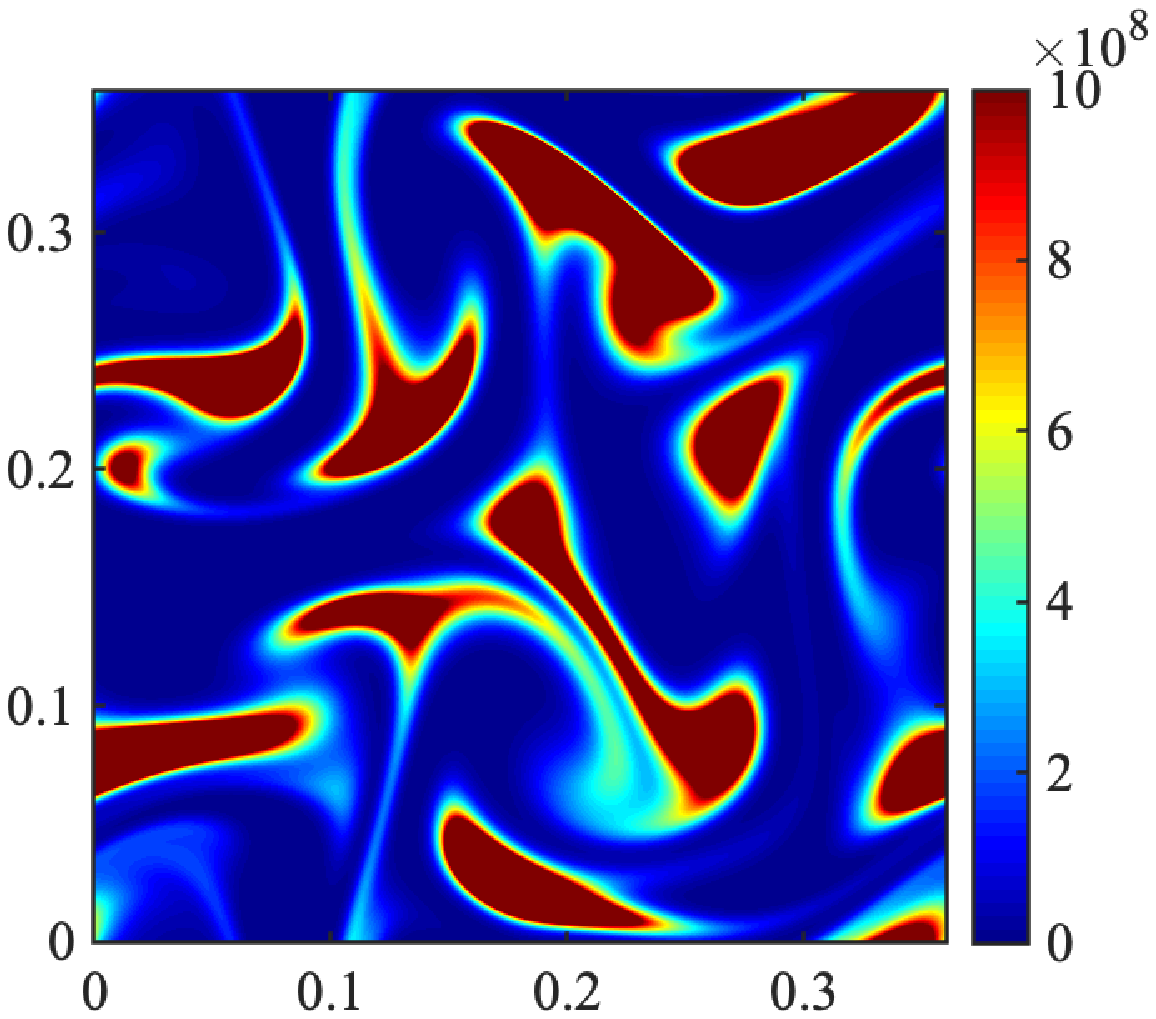}
    \begin{picture}(0,0)
    \put(35,12){\scriptsize x (\SI{}{\cm})}
    \end{picture}
    \caption{Instantaneous contour plots of heat release rates (\SI{}{\joule \meter^{-3} \second^{-1}}) from two-dimensional HCCI dataset at two different times ($t$). (a) $t=$\SI{0.845}{\milli\second}, and (b) $t=$\SI{1.2}{\milli\second}.
    \label{fig:hcci-hrr}}
\end{figure}

Next we consider a dataset that includes the effects of convective and diffusive spatial transport and turbulence, homogeneous charge compression ignition (HCCI) of ethanol representative of internal combustion engines \citep{bhagatwala-hcci-dns}.
The simulation corresponds to high pressure, high temperature auto-ignition of turbulent premixed ethanol-air mixed with combustion products, representing ``exhaust gas recirculation'' (EGR), in a fully periodic domain with a uniform 2D spatial grid comprising 672 $\times$ 672 grid points.
The simulation was initialized with a nominal pressure of \SI{45}{atm} and a mean temperature of \SI{924}{\kelvin}. The equivalence ratio of the reactants was set to 0.4.
To model the uneven mixing caused by the EGR, a spatial temperature fluctuation, along with a separately computed divergence-free turbulent velocity field, was superimposed onto the system.
Additionally, compression heating effects induced due to the motion of the piston were also accounted for in the simulation.
The chemical mechanism associated with ethanol combustion consisted of 28 chemical species.
Thus, at each temporal checkpoint, the dataset $\mathbf{D}$ comprises of $n_g=672\times 672$ samples with $n_v = 29$ thermo-chemical scalars.
The temporal checkpoints of interest correspond to $t=$ 0.845 and 1.2 \SI{}{\milli\second} \citep{aditya-anomaly-detection-2019-JCP}, as these time instances correspond to the inception of auto-ignition kernels and the propagation of the flame front in the bulk of the domain, respectively, as shown in the heat release rate contours in Fig.~\ref{fig:hcci-hrr}.
It should be noted that the data corresponding to $t=\SI{1.2}{\milli\second}$ can be considered as a ``regular'' dataset, whereas the $t=\SI{0.845}{\milli\second}$ dataset can be considered to comprise of outlying events due to the presence of localised ignition kernels.

\begin{figure}[h!]
     \centering
     {
     \includegraphics{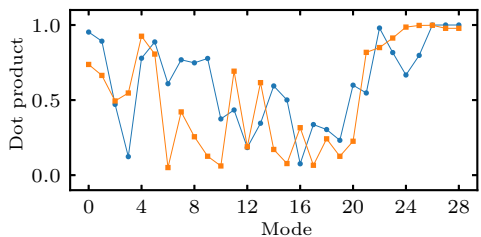}
     \begin{picture}(0,0)
    \put(-93,0){\scriptsize Mode}
    \put(-188,32){\scriptsize{\rotatebox{90}{Dot product}}}
    \end{picture}
     }
     \caption{Alignment between PCA and CoK-PCA principal vectors from HCCI dataset represented as a dot product for each mode. The blue and orange lines represent $t=0.845~ms$ and $t=1.2~ms$, respectively.
     \label{fig:hcci-inclination}}
 \end{figure}

As in the case of the zero-dimensional homogeneous reactor dataset, we obtain the required PCA and CoK-PCA principal vectors from the procedure described in Sec.~\ref{sec:config}. The entire PCA and CoK-PCA analyses were performed  separately on each temporal snapshot, rather than combining them.
In Fig.~\ref{fig:hcci-inclination}, we present the orientation of the principal vectors obtained from the two methods for the considered time instances.
It can be observed that, for both the time instances, the PCA and CoK-PCA manifolds are significantly different from one another and thus identify different lower-dimensional sub-spaces.

\rev{The required reduced manifolds are constructed by retaining $n_q = 5$ out of the $n_v = 29$ principal vectors as these principal vectors correspond to approximately 99\% of the variance and kurtosis in the reduced PCA and CoK-PCA manifolds.}
Thereafter, as in the case of the zero-dimensional homogeneous reactor dataset, the thermo-chemical state is linearly reconstructed.
In the following, we present a comparison of the quality of the PCA and CoK-PCA reduced manifolds in terms of the reconstruction errors of the thermo-chemical scalars, the species production rates and heat release rate.

\begin{figure*}[h!]
    \centering
    {
    \includegraphics[width=7.5cm]{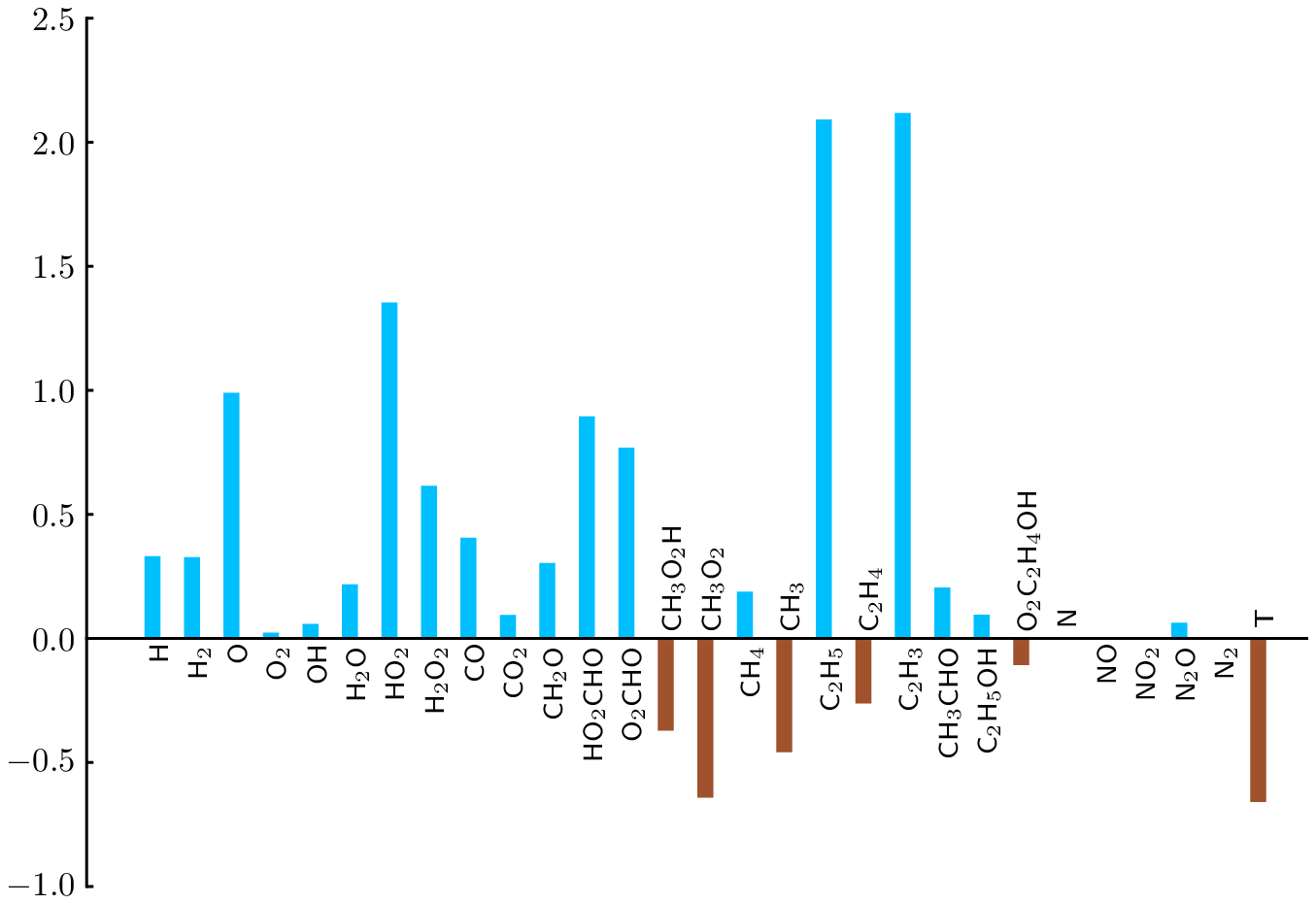}
    \begin{picture}(0,0)
    \put(-190,132){\scriptsize (a) Species mass fractions and temperature}
    \put(-218,50){\scriptsize{\rotatebox{90}{Error ratio $r_M$}}}
    \end{picture}
    } \hspace{0.25cm}
    {
    \includegraphics[width=7.5cm]{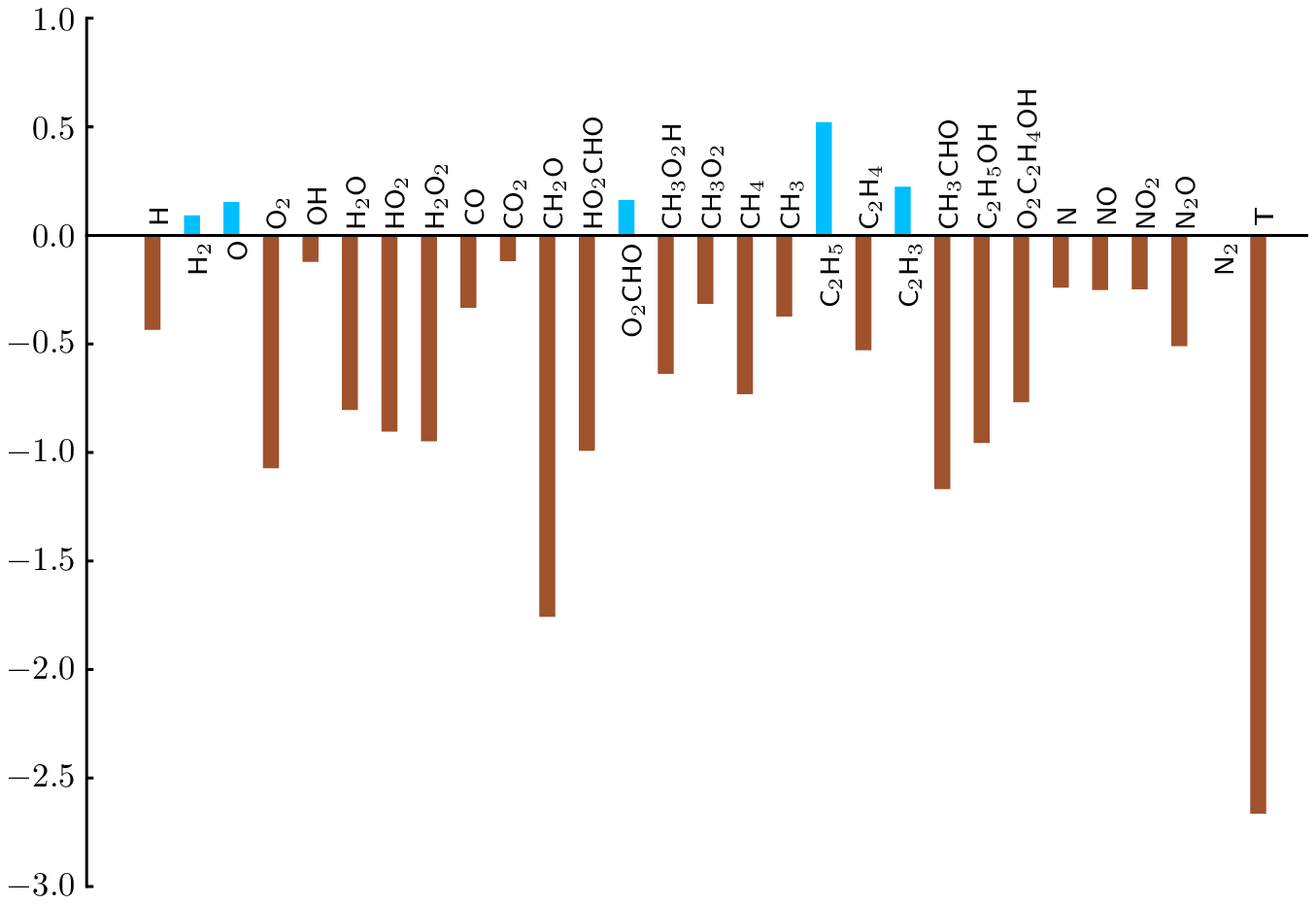}
    \begin{picture}(0,0)
    \put(-190,132){\scriptsize (b) Species mass fractions and temperature}
    \put(-218,50){\scriptsize{\rotatebox{90}{Error ratio $r_A$}}}
    \end{picture}
    }
    
    \vspace{0.1cm}
    {
    \includegraphics[width=7.5cm]{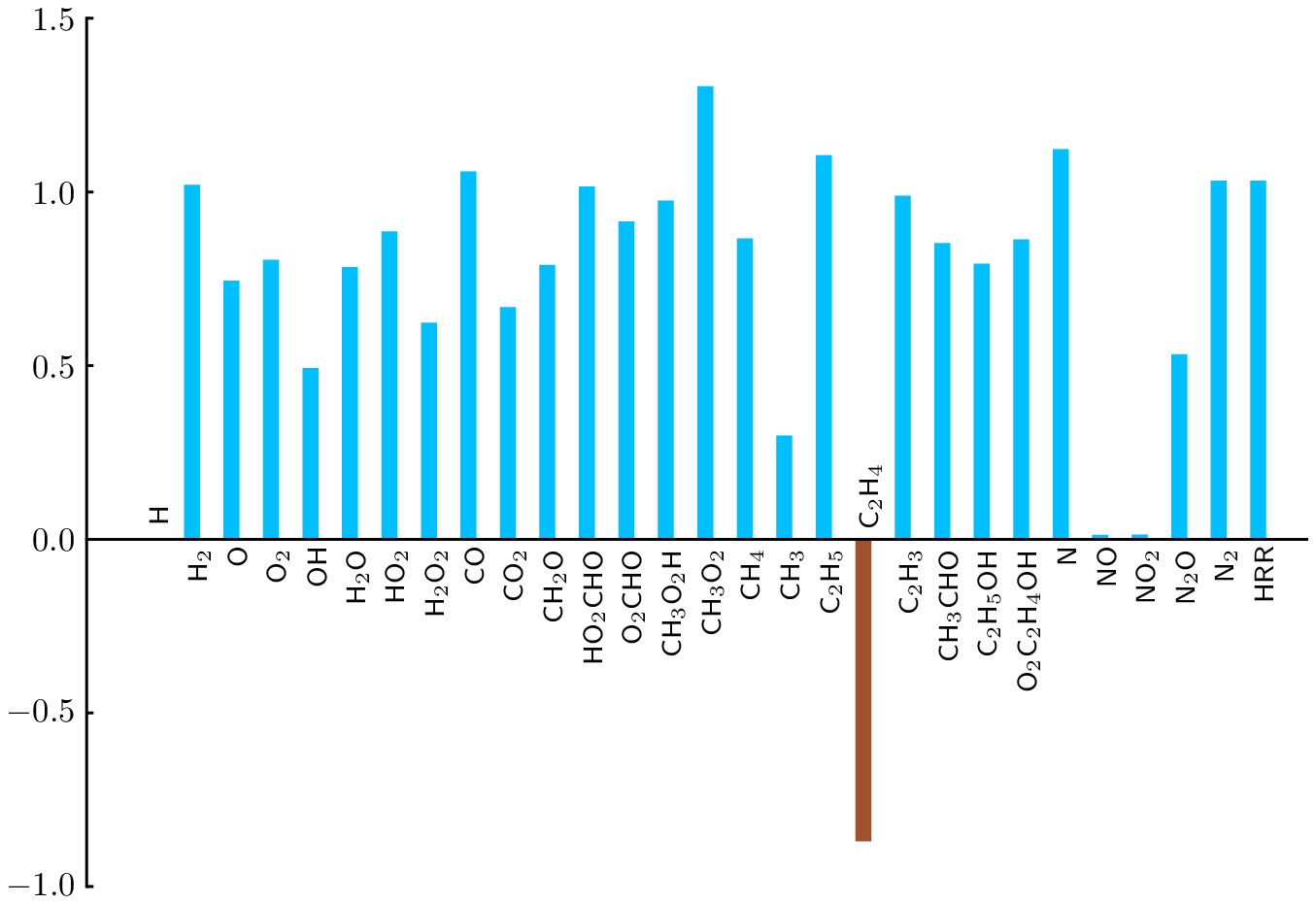}
    \begin{picture}(0,0)
    \put(-190,132){\scriptsize (c) Species production rates and heat release rate}
    \put(-218,50){\scriptsize{\rotatebox{90}{Error ratio $r_M$}}}
    \end{picture}
    }\hspace{0.25cm}
    {
    \includegraphics[width=7.5cm]{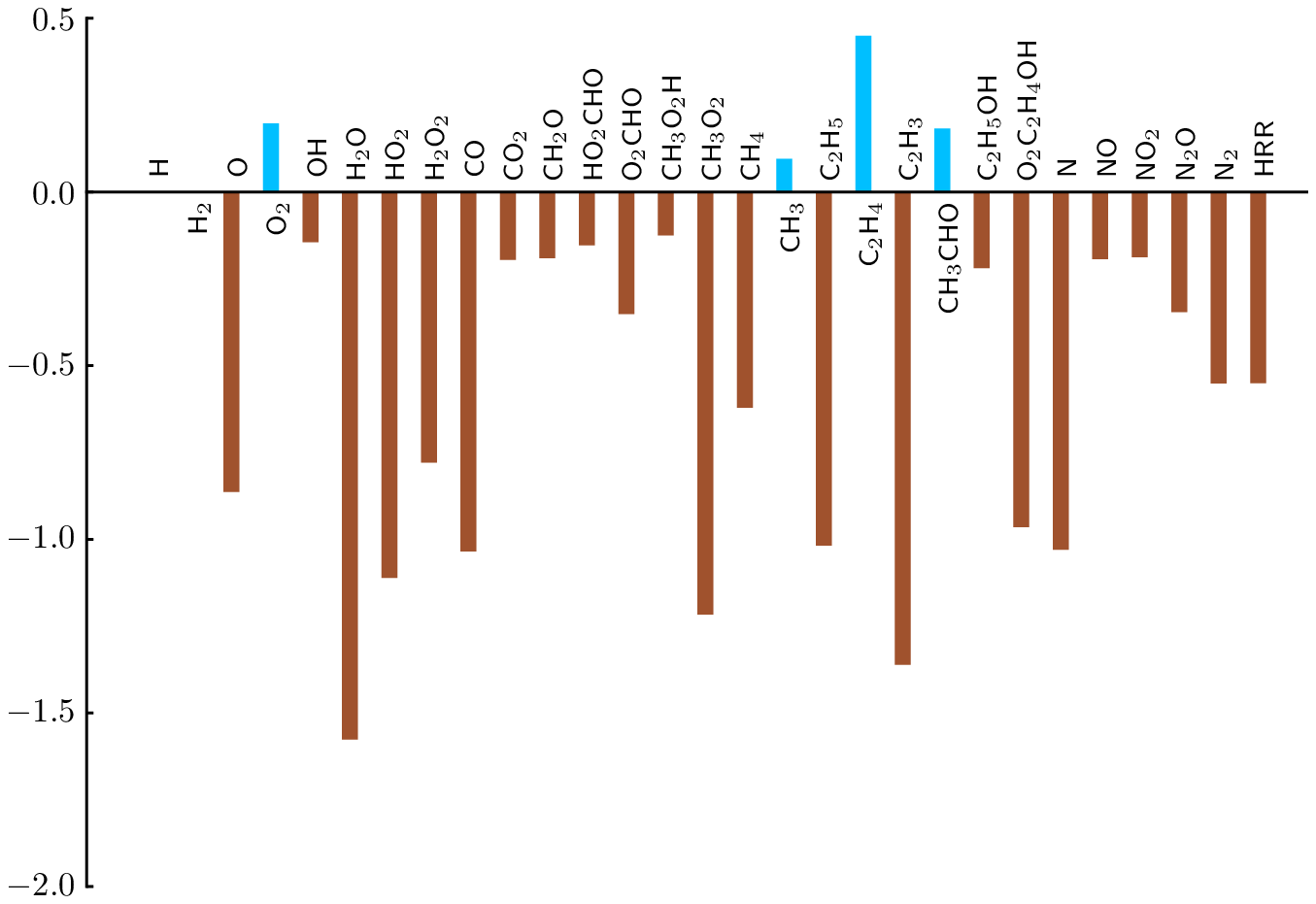}
    \begin{picture}(0,0)
    \put(-190,132){\scriptsize (d) Species production rates and heat release rate}
    \put(-218,50){\scriptsize{\rotatebox{90}{Error ratio $r_A$}}}
    \end{picture}
    }
    \caption{Plots of error ratio $r_i$ ($i\in \{M,A\}$, see Eq.~\ref{eq:error-ratio}) for reconstructed data, $n_q=5$, from HCCI dataset at $t = 0.845~ms$. Error ratio $r_i$ based on maximum (a) and average (b) errors of species mass fractions and temperature. Error ratio $r_i$ based on maximum (c) and average (d) errors of species production rates and heat release rate.\label{fig:errors-hcci-t0}}
\end{figure*}

\begin{figure*}[h!]
    \centering
    {
    \includegraphics[width=7.5cm]{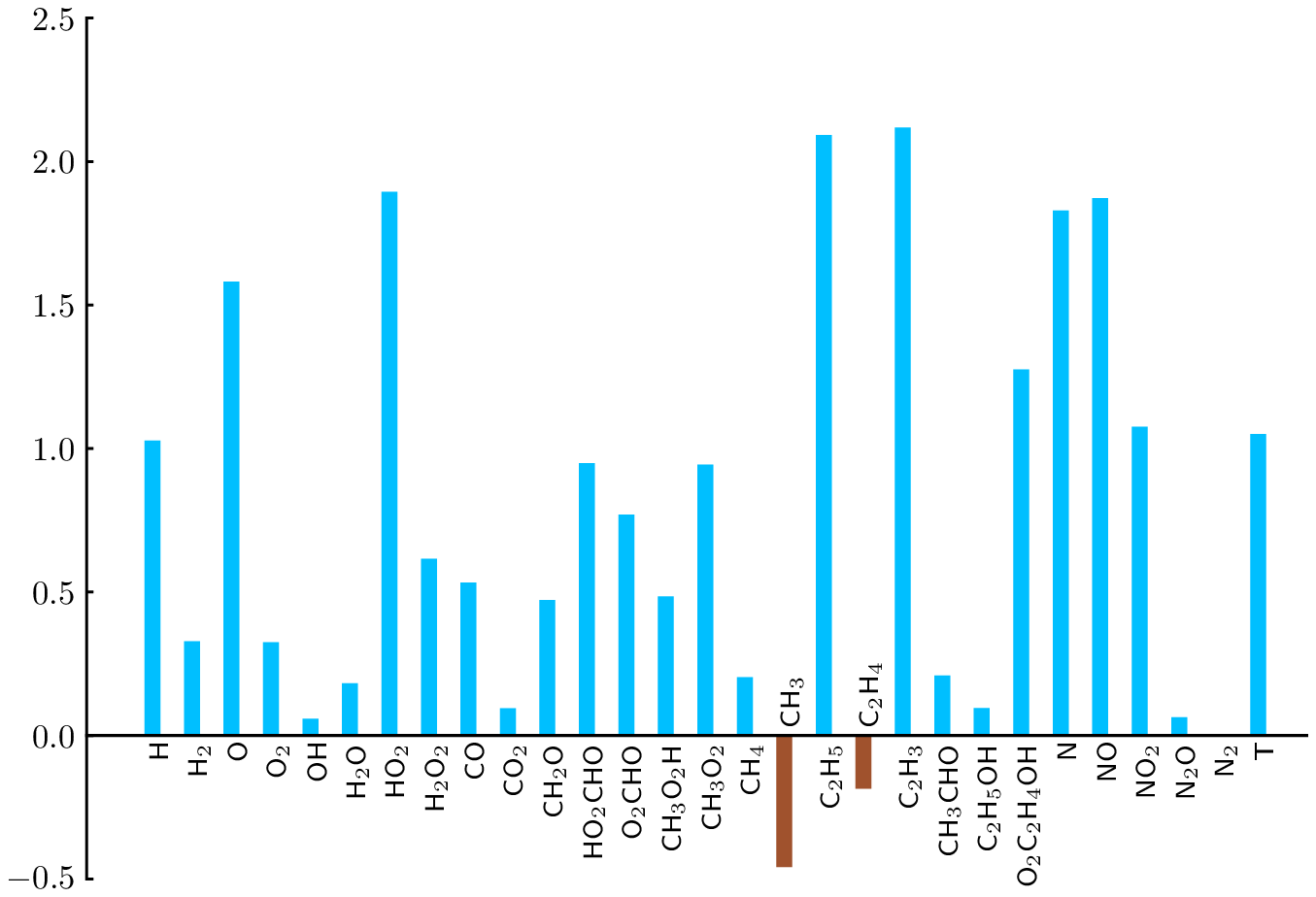}
    \begin{picture}(0,0)
    \put(-190,132){\scriptsize (a) Species mass fractions and temperature}
    \put(-218,50){\scriptsize{\rotatebox{90}{Error ratio $r_M$}}}
    \end{picture}
    } \hspace{0.25cm}
    {
    \includegraphics[width=7.5cm]{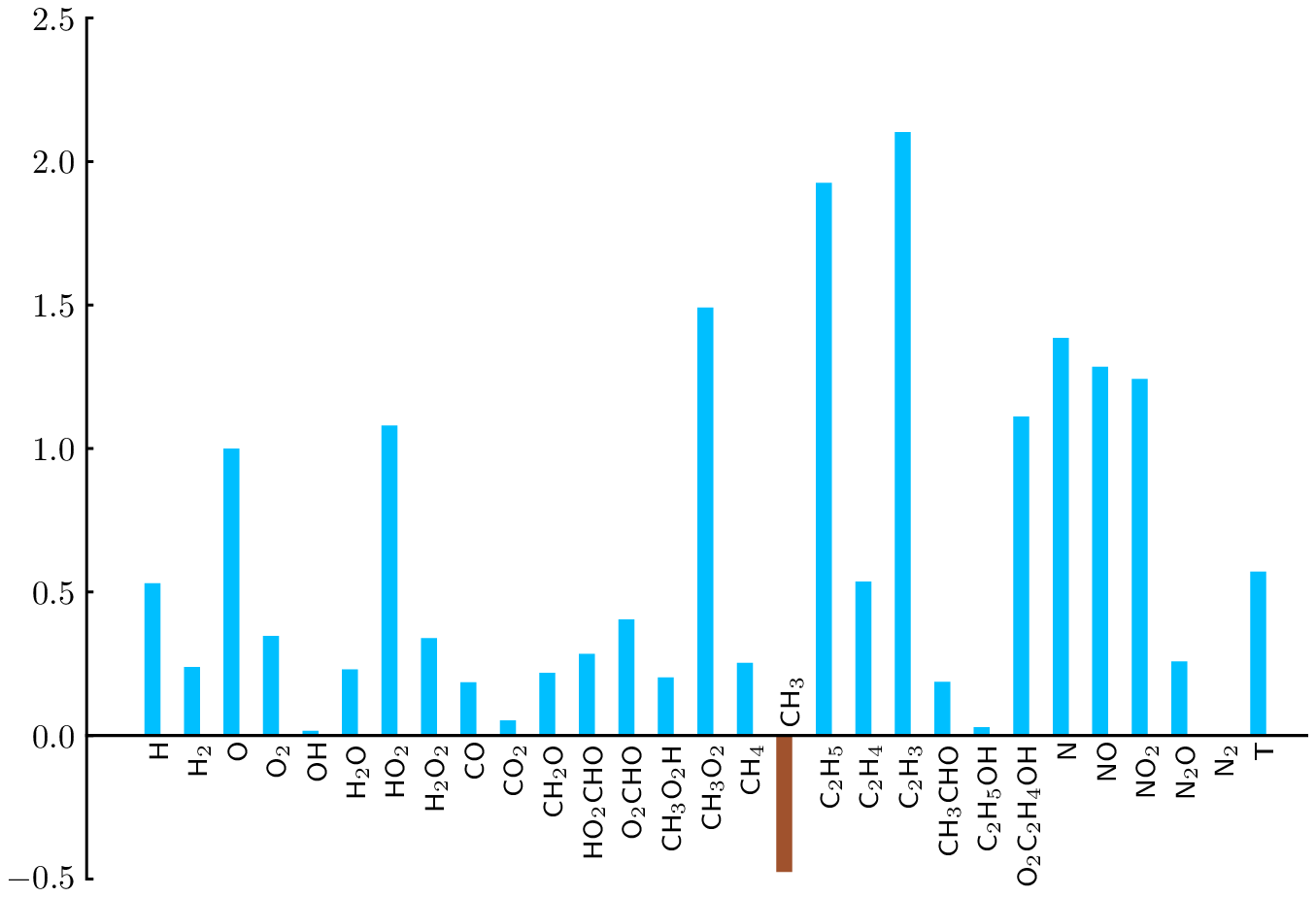}
    \begin{picture}(0,0)
    \put(-190,132){\scriptsize (b) Species mass fractions and temperature}
    \put(-218,50){\scriptsize{\rotatebox{90}{Error ratio $r_A$}}}
    \end{picture}
    }
    
    \vspace{0.1cm}
    {
    \includegraphics[width=7.5cm]{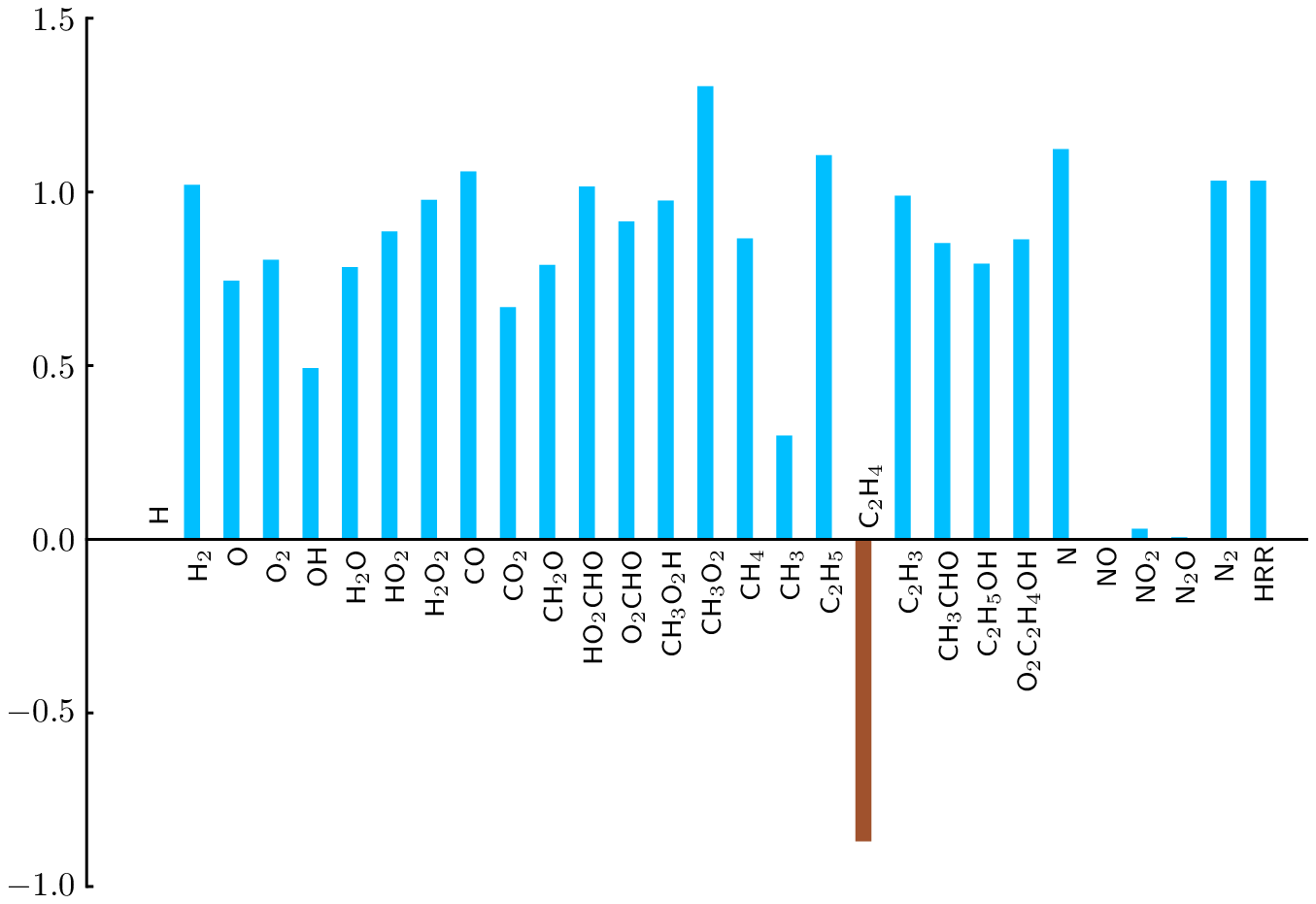}
    \begin{picture}(0,0)
    \put(-190,132){\scriptsize (c) Species production rates and heat release rate}
    \put(-218,50){\scriptsize{\rotatebox{90}{Error ratio $r_M$}}}
    \end{picture}
    }\hspace{0.25cm}
    {
    \includegraphics[width=7.5cm]{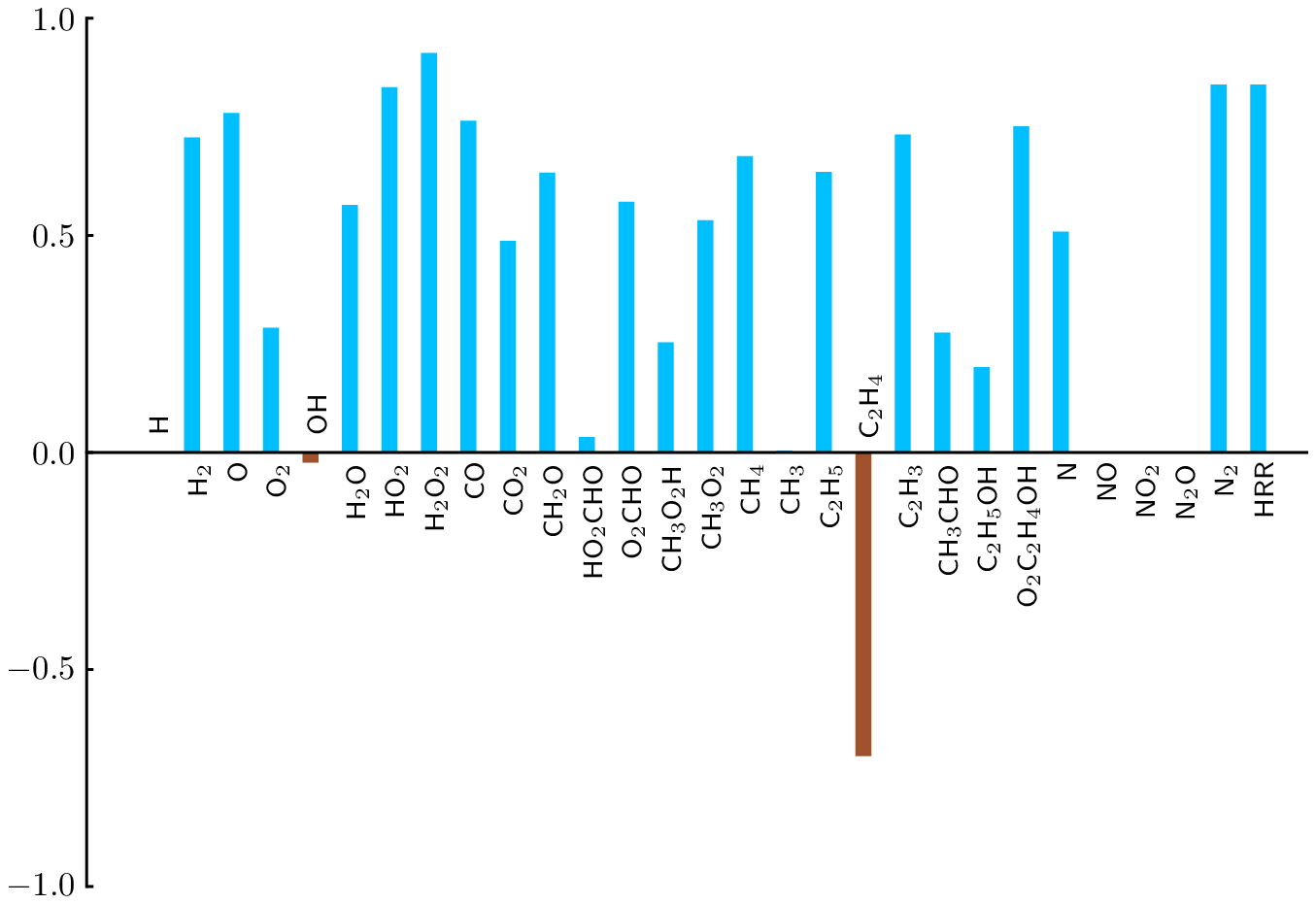}
    \begin{picture}(0,0)
    \put(-190,132){\scriptsize (d) Species production rates and heat release rate}
    \put(-218,50){\scriptsize{\rotatebox{90}{Error ratio $r_A$}}}
    \end{picture}
    }
    \caption{Plots of error ratio $r_i$ ($i\in \{M,A\}$, see Eq.~\ref{eq:error-ratio}) for reconstructed data, $n_q=5$, from HCCI dataset at $t = 0.845~ms$ in the reacting region. Error ratio $r_i$ based on maximum (a) and average (b) errors of species mass fractions and temperature. Error ratio $r_i$ based on maximum (c) and average (d) errors of species production rates and heat release rate.\label{fig:errors-hcci-t0-reacting}}
\end{figure*}

In Figs.~\ref{fig:errors-hcci-t0} (a) and (c), it can be clearly seen that CoK-PCA performs significantly better than PCA for $t = 0.845~ms$ data when reconstructing the thermo-chemical scalars along with the species production rates and heat release rate in terms of the $r_M$ metric.
However, in Fig.~\ref{fig:errors-hcci-t0}(b), we observe that the performance of the CoK-PCA drastically deteriorates in terms of its ability to reconstruct the thermo-chemical scalars on average, i.e., with the $r_A$ metric.
Similarly, in Fig.~\ref{fig:errors-hcci-t0}(d), we observe that the overall chemical kinetics is poorly represented by CoK-PCA as demonstrated by the poor $r_A$ values of the production rates and heat release rate.
Again, it should be recalled that the excess kurtosis is inherently better suited to capture outlier events, which, in the case of the $t=0.845~ms$ dataset, comprises of a small number of samples corresponding to the formation of ignition kernels (see Fig.~\ref{fig:hcci-hrr}(a)). We hypothesize that the chemical manifold identified by CoK-PCA better represents the chemical subspace for the smaller number of ignition samples, rather than the overall domain.
Thus, in the case of the $r_A$ metric, the non-igniting zones of the simulation domain are not represented by CoK-PCA reduced manifold leading to a higher average error.
To test this hypothesis, and to establish an equitable comparison of CoK-PCA against PCA, we evaluate $r_M$ and $r_A$ for the thermo-chemical scalars, species production rates and heat release rate only for those spatial locations that correspond to the reacting zones of the simulation domain.
These reacting zones are identified as spatial points having a heat release rate larger than $7\times10^8$ \SI{}{\joule \meter^{-3} \second^{-1}}, which depict the boundaries of matured ignition kernels. 
For these regions, as shown in Fig.~\ref{fig:errors-hcci-t0-reacting}, it can be seen that the reduced manifold obtained with the CoK-PCA method provides a much better representation of the thermo-chemical state as well as the chemical kinetics of the system when compared to PCA in terms of both the maximum and average errors.

\begin{figure*}[h!]
    \centering
    {
    \includegraphics[width=7.5cm]{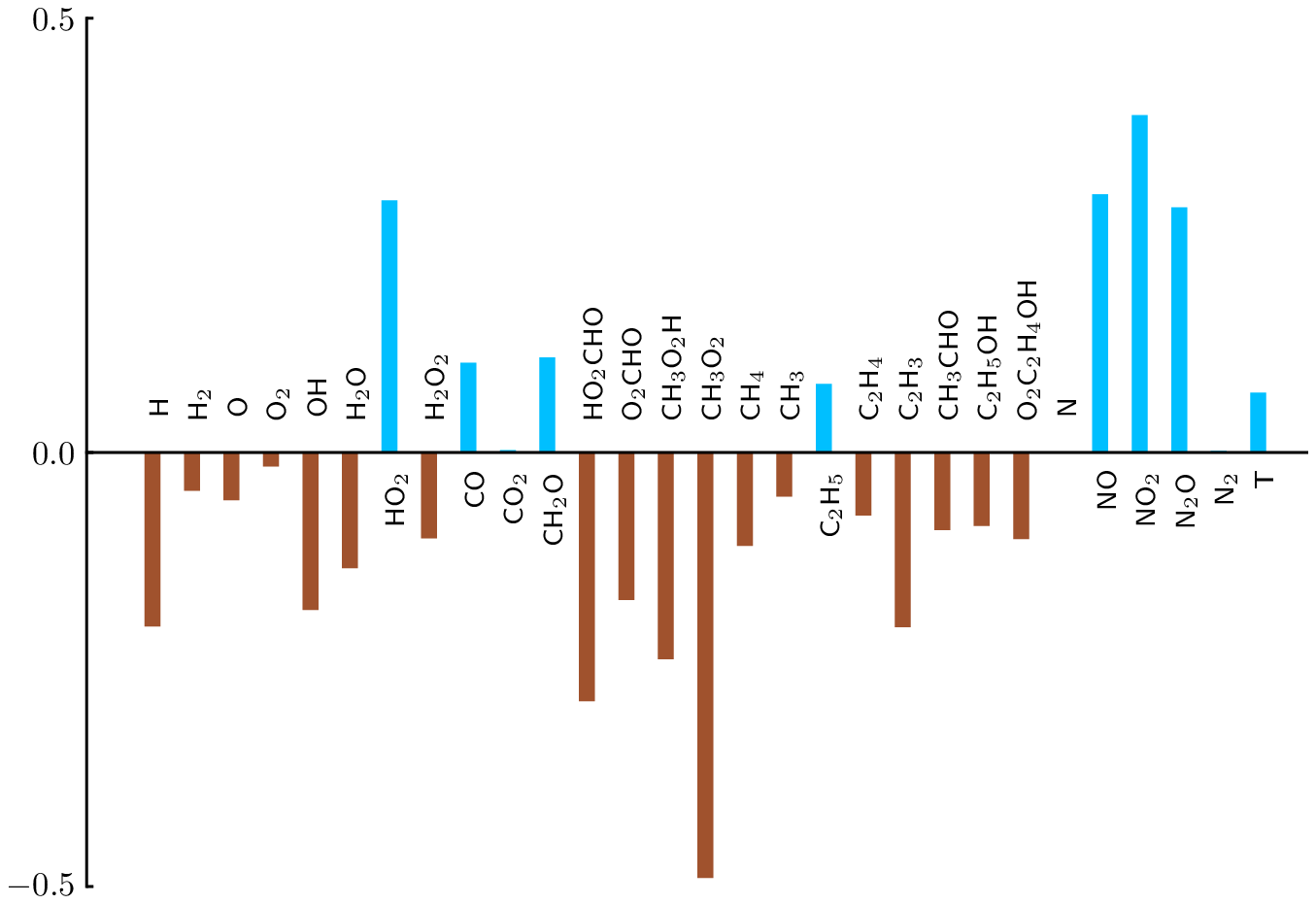}
    \begin{picture}(0,0)
    \put(-190,132){\scriptsize (a) Species mass fractions and temperature}
    \put(-218,50){\scriptsize{\rotatebox{90}{Error ratio $r_M$}}}
    \end{picture}
    } \hspace{0.25cm}
    {
    \includegraphics[width=7.5cm]{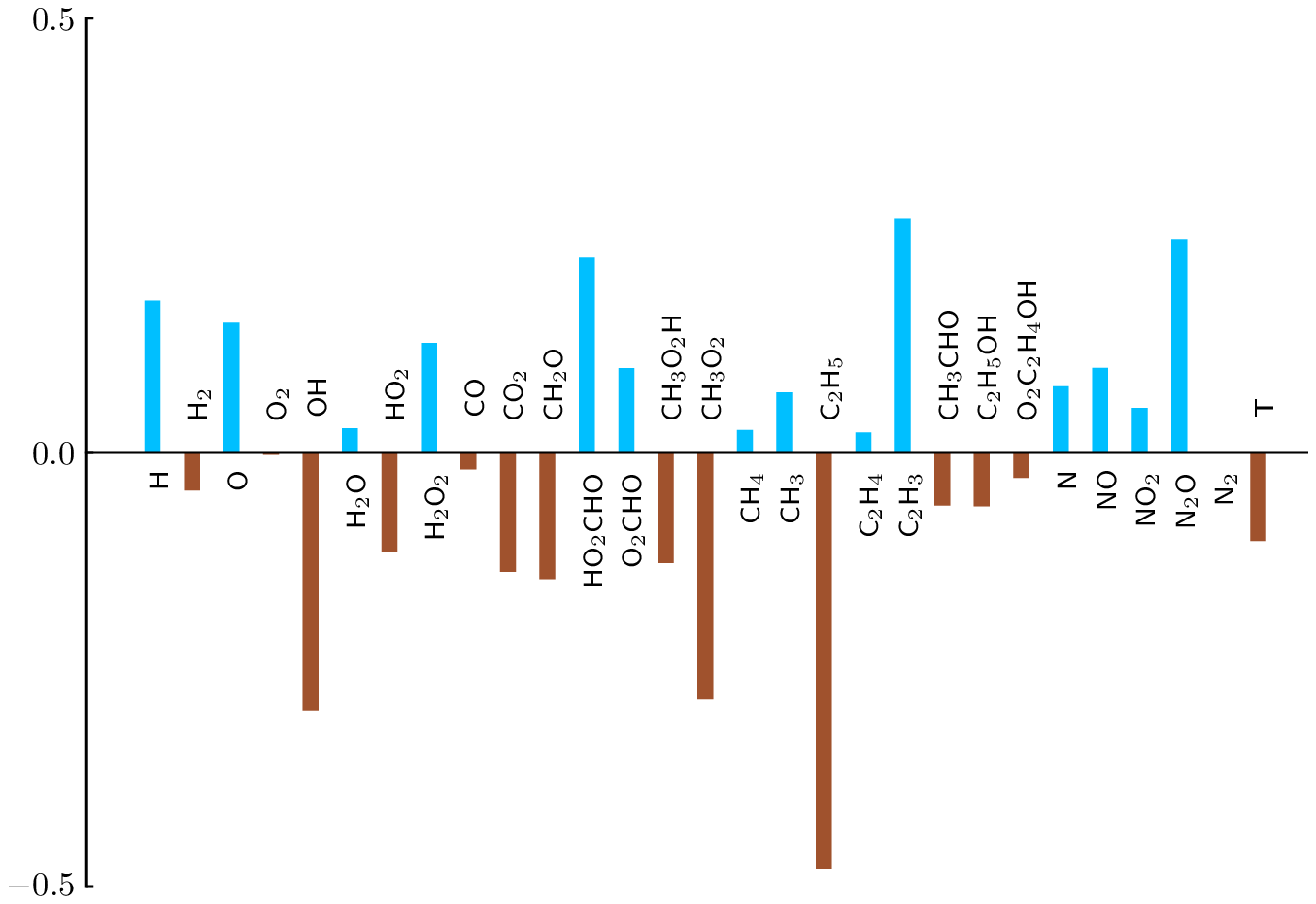}
    \begin{picture}(0,0)
    \put(-190,132){\scriptsize (b) Species mass fractions and temperature}
    \put(-218,50){\scriptsize{\rotatebox{90}{Error ratio $r_A$}}}
    \end{picture}
    }
    
    \vspace{0.1cm}
    {
    \includegraphics[width=7.5cm]{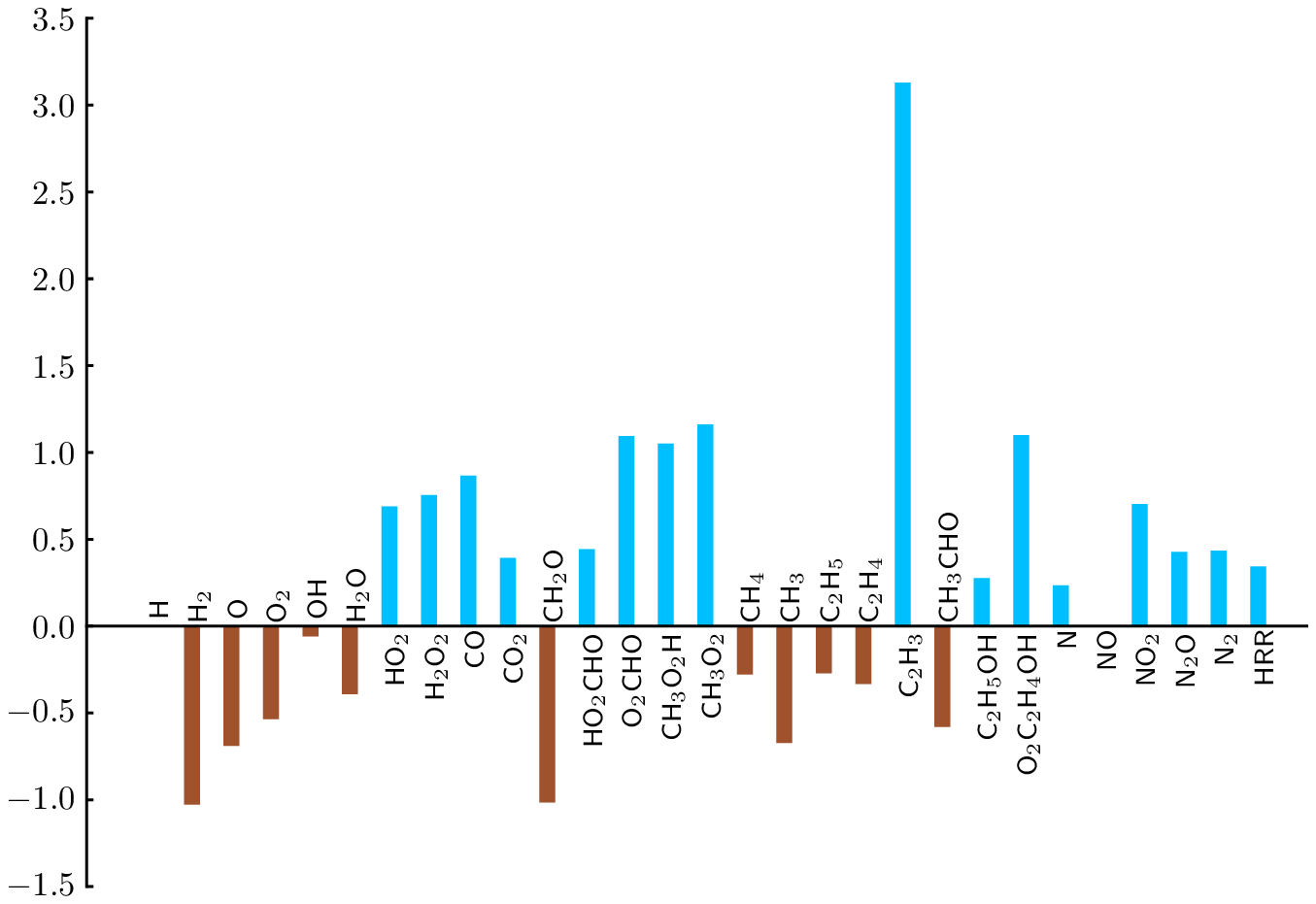}
    \begin{picture}(0,0)
    \put(-190,132){\scriptsize (c) Species production rates and heat release rate}
    \put(-218,50){\scriptsize{\rotatebox{90}{Error ratio $r_M$}}}
    \end{picture}
    }\hspace{0.25cm}
    {
    \includegraphics[width=7.5cm]{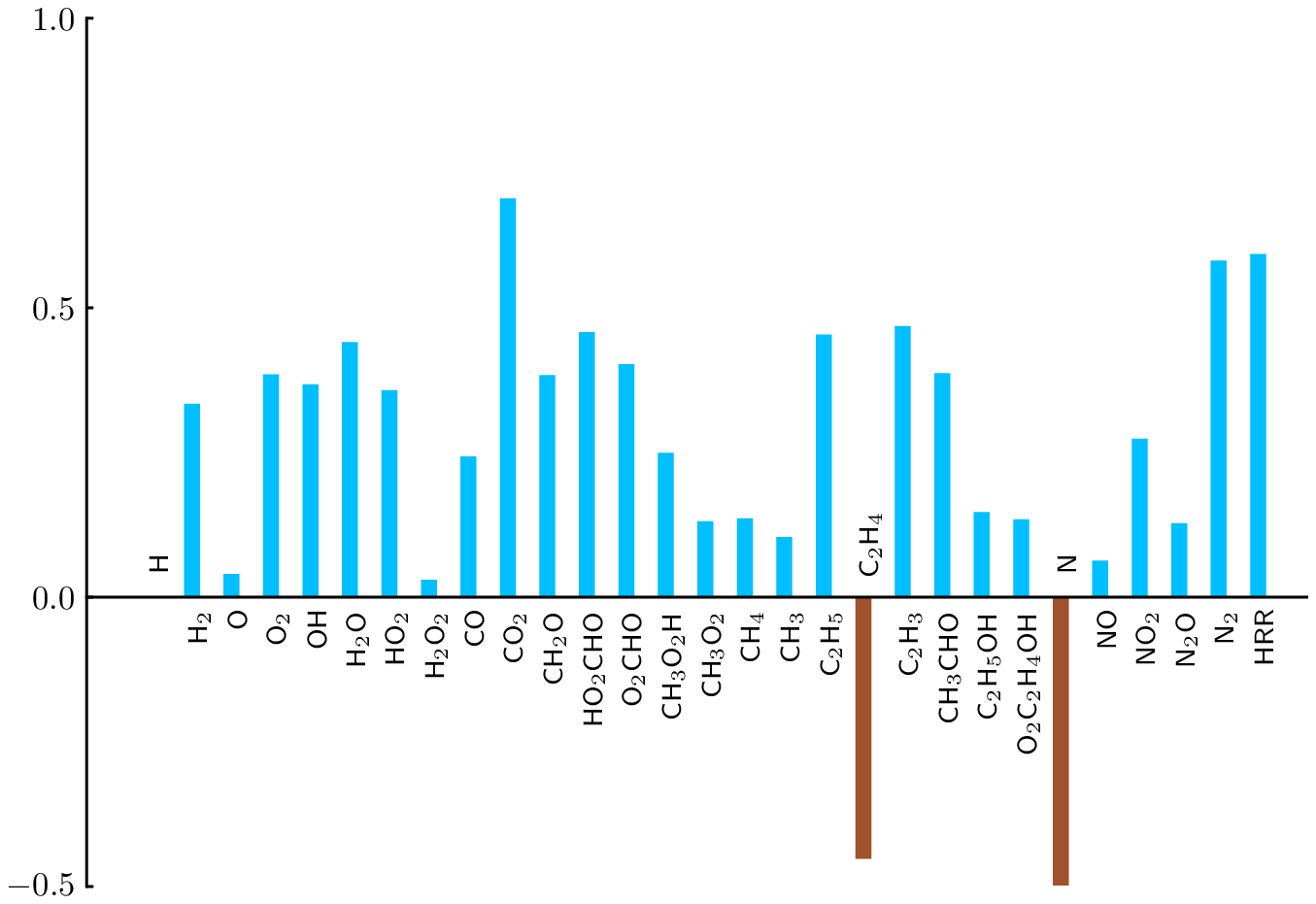}
    \begin{picture}(0,0)
    \put(-190,132){\scriptsize (d) Species production rates and heat release rate}
    \put(-218,50){\scriptsize{\rotatebox{90}{Error ratio $r_A$}}}
    \end{picture}
    }
    \caption{Plots of error ratio $r_i$ ($i\in \{M,A\}$, see Eq.~\ref{eq:error-ratio}) for reconstructed data, $n_q=5$, from HCCI dataset at $t = 1.2~ms$. Error ratio $r_i$ based on maximum (a) and average (b) errors of species mass fractions and temperature. Error ratio $r_i$ based on maximum (c) and average (d) errors of species production rates and heat release rate.\label{fig:errors-hcci-t1}}
\end{figure*}

\begin{figure*}[h!]
    \centering
    {
    \includegraphics[width=7.5cm]{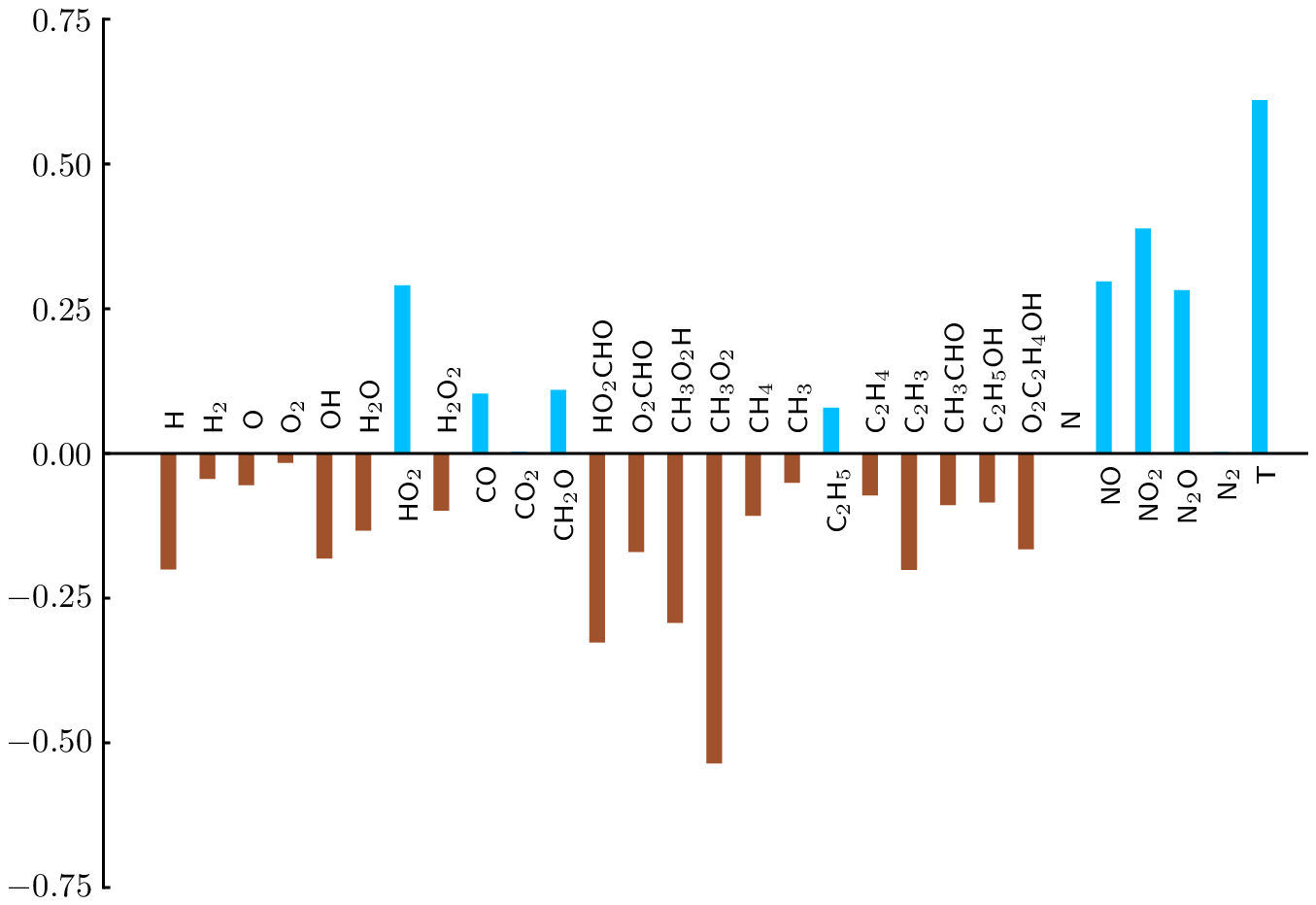}
    \begin{picture}(0,0)
    \put(-190,132){\scriptsize (a) Species mass fractions and temperature}
    \put(-218,50){\scriptsize{\rotatebox{90}{Error ratio $r_M$}}}
    \end{picture}
    } \hspace{0.25cm}
    {
    \includegraphics[width=7.5cm]{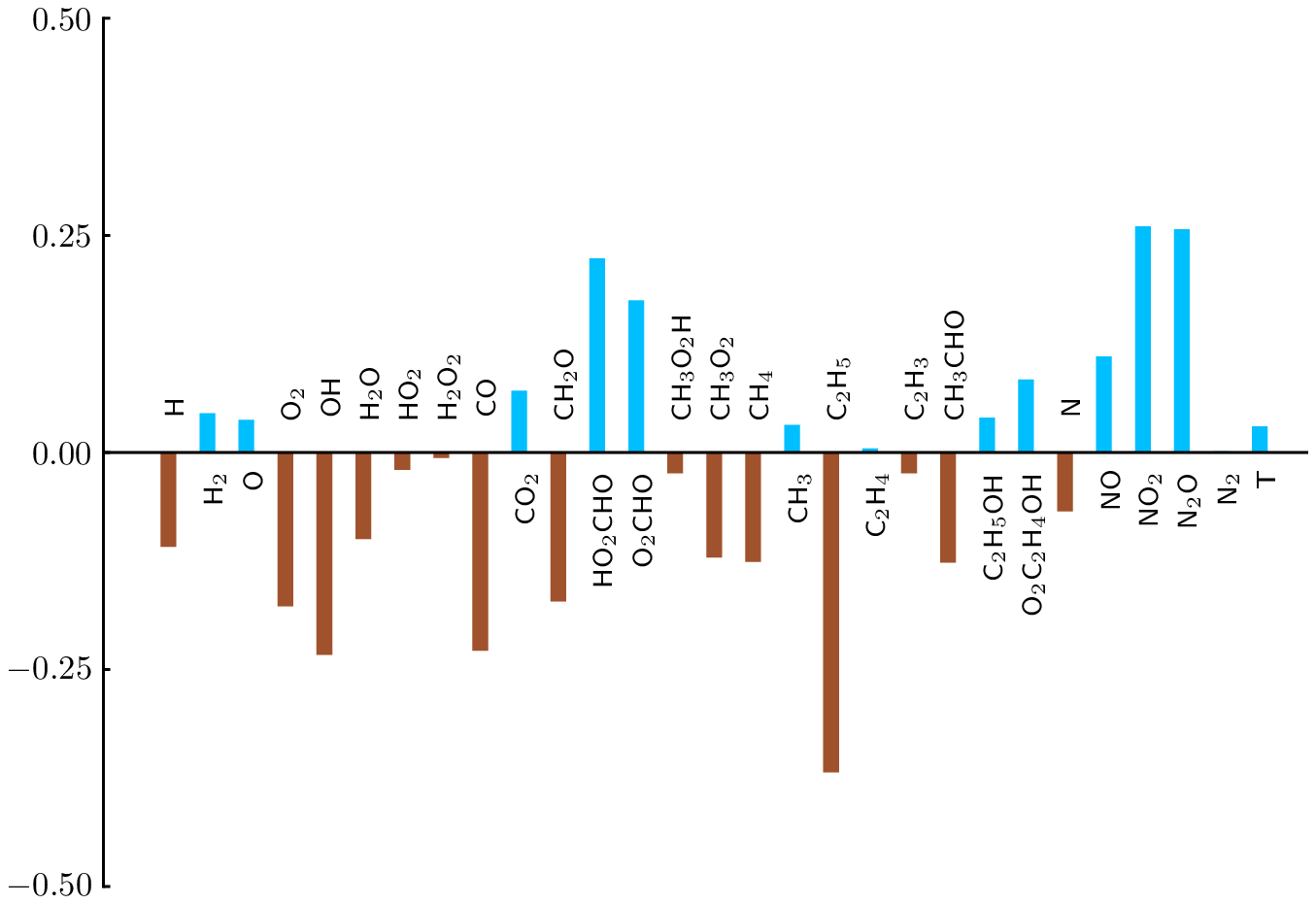}
    \begin{picture}(0,0)
    \put(-190,132){\scriptsize (b) Species mass fractions and temperature}
    \put(-218,50){\scriptsize{\rotatebox{90}{Error ratio $r_A$}}}
    \end{picture}
    }
    
    \vspace{0.1cm}
    {
    \includegraphics[width=7.5cm]{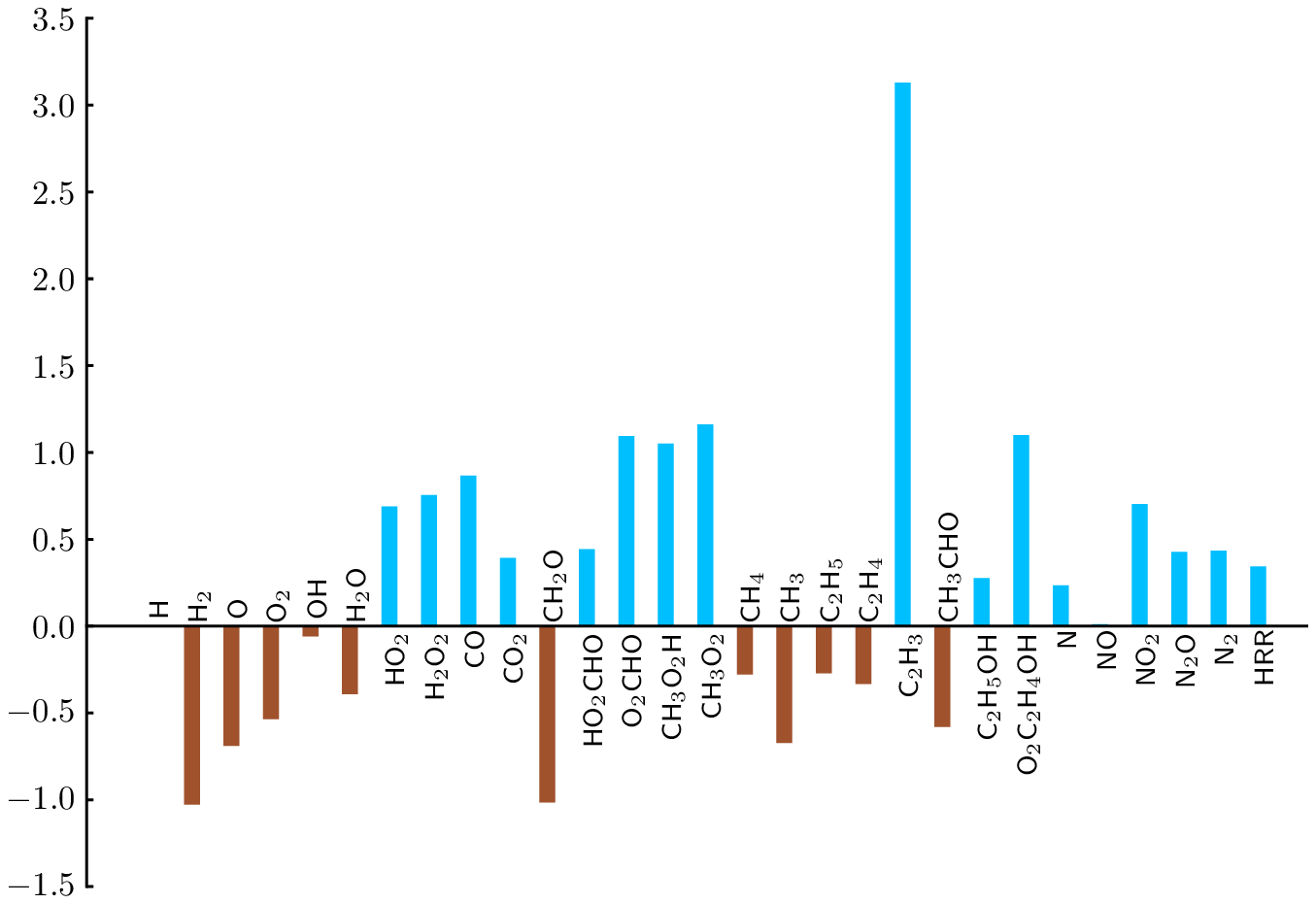}
    \begin{picture}(0,0)
    \put(-190,132){\scriptsize (c) Species production rates and heat release rate}
    \put(-218,50){\scriptsize{\rotatebox{90}{Error ratio $r_M$}}}
    \end{picture}
    }\hspace{0.25cm}
    {
    \includegraphics[width=7.5cm]{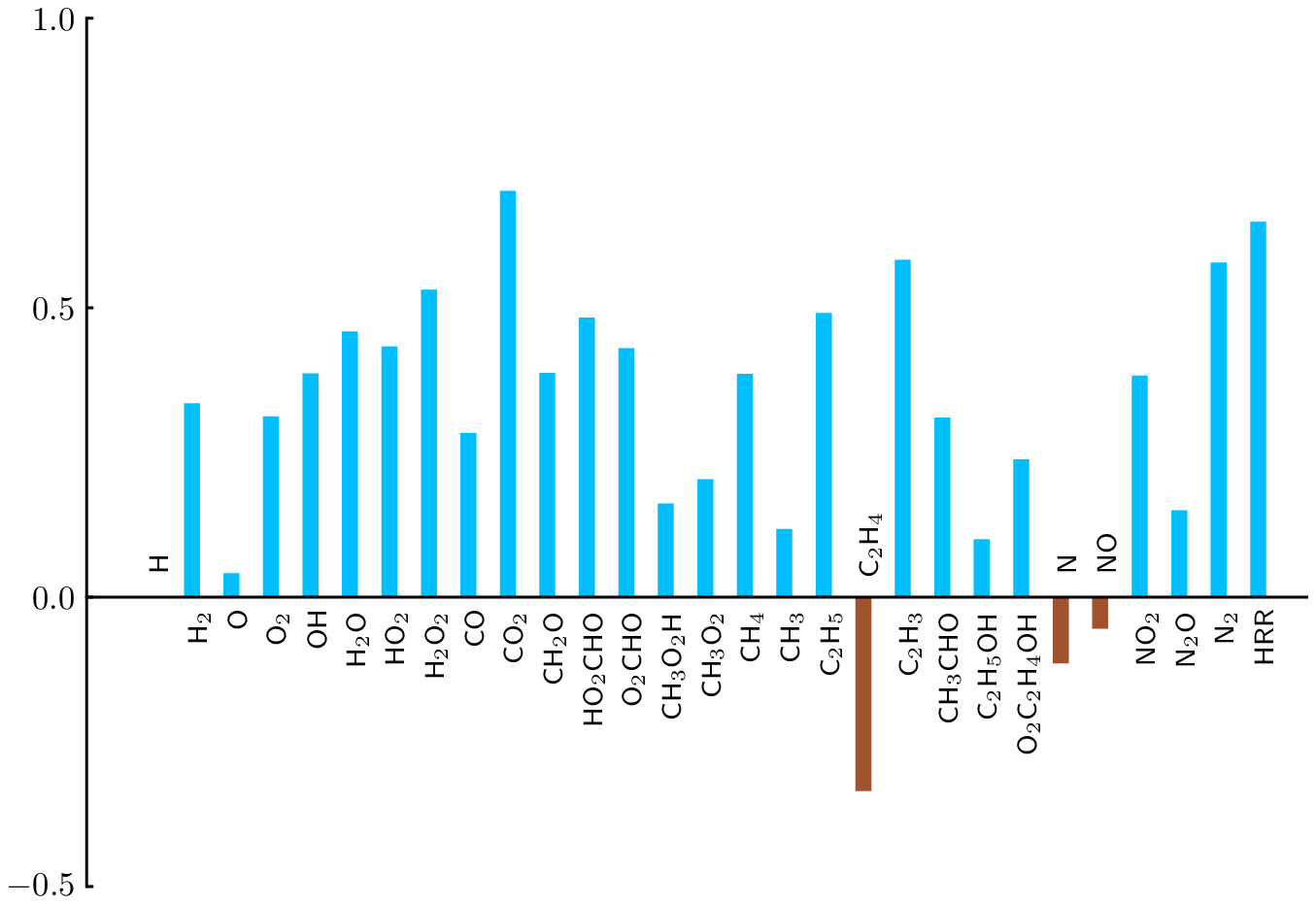}
    \begin{picture}(0,0)
    \put(-190,132){\scriptsize (d) Species production rates and heat release rate}
    \put(-218,50){\scriptsize{\rotatebox{90}{Error ratio $r_A$}}}
    \end{picture}
    }
    \caption{Plots of error ratio $r_i$ ($i\in \{M,A\}$, see Eq.~\ref{eq:error-ratio}) for reconstructed data, $n_q=5$, from HCCI dataset at $t = 1.2~ms$ in the reacting region. Error ratio $r_i$ based on maximum (a) and average (b) errors of species mass fractions and temperature. Error ratio $r_i$ based on maximum (c) and average (d) errors of species production rates and heat release rate.\label{fig:errors-hcci-t1-reacting}}
\end{figure*}

For the $t=1.2~ms$ time instant
it should be noted that the number of samples comprising the reacting and non-reacting zones are comparable (see Fig.~\ref{fig:hcci-hrr}(b)).
Thus, we expect the performance of the PCA and CoK-PCA methods to be largely equivalent to one another in terms of capturing the thermo-chemical scalars.
As shown in Figs.~\ref{fig:errors-hcci-t1} (a) and (b), this is indeed observed to be the case.
However, Figs.~\ref{fig:errors-hcci-t1} (c) and (d) indicate that CoK-PCA captures the species production rates and the heat release rate, both in terms of both the $r_M$ and $r_A$ metrics, much better than PCA.
Similarly, due to the comparable sizes of the reacting and non-reacting zones, the results of the analysis conditioned on the reacting zones (spatial locations with heat release rates larger than $7\times 10^8$ \SI{}{\joule \meter^{-3} \second^{-1}}), as shown in Fig.~\ref{fig:errors-hcci-t1-reacting},  demonstrates similar trends as those for the entire domain.

To summarize, for the HCCI data, the CoK-PCA based lower-dimensional manifold would, at the first glance, appear to present a poor representation of the original chemical state for initial time steps when the number of ignition kernels are small.
However, upon careful inspection, it can be appreciated that the advantages of the CoK-PCA, viz. the ability to effectively represent stiff chemical dynamics in the reaction zones of the simulation domain, remain.
It is important to note that the accurately represented chemical dynamics in the reaction zones would result in smaller errors being propagated as the reduced-order simulations integrate the lower-dimensional system in time.
Further, for the intermediate time instant, CoK-PCA performs just as well as PCA in reconstructing the thermo-chemical scalars due to the equal distribution of the reacting and non-reacting zones. The species production and heat release rates, in this case, are more accurately predicted from CoK-PCA based reconstruction.

\begin{figure*}[h!]
    \centering
    {
    \includegraphics[width=7.5cm]{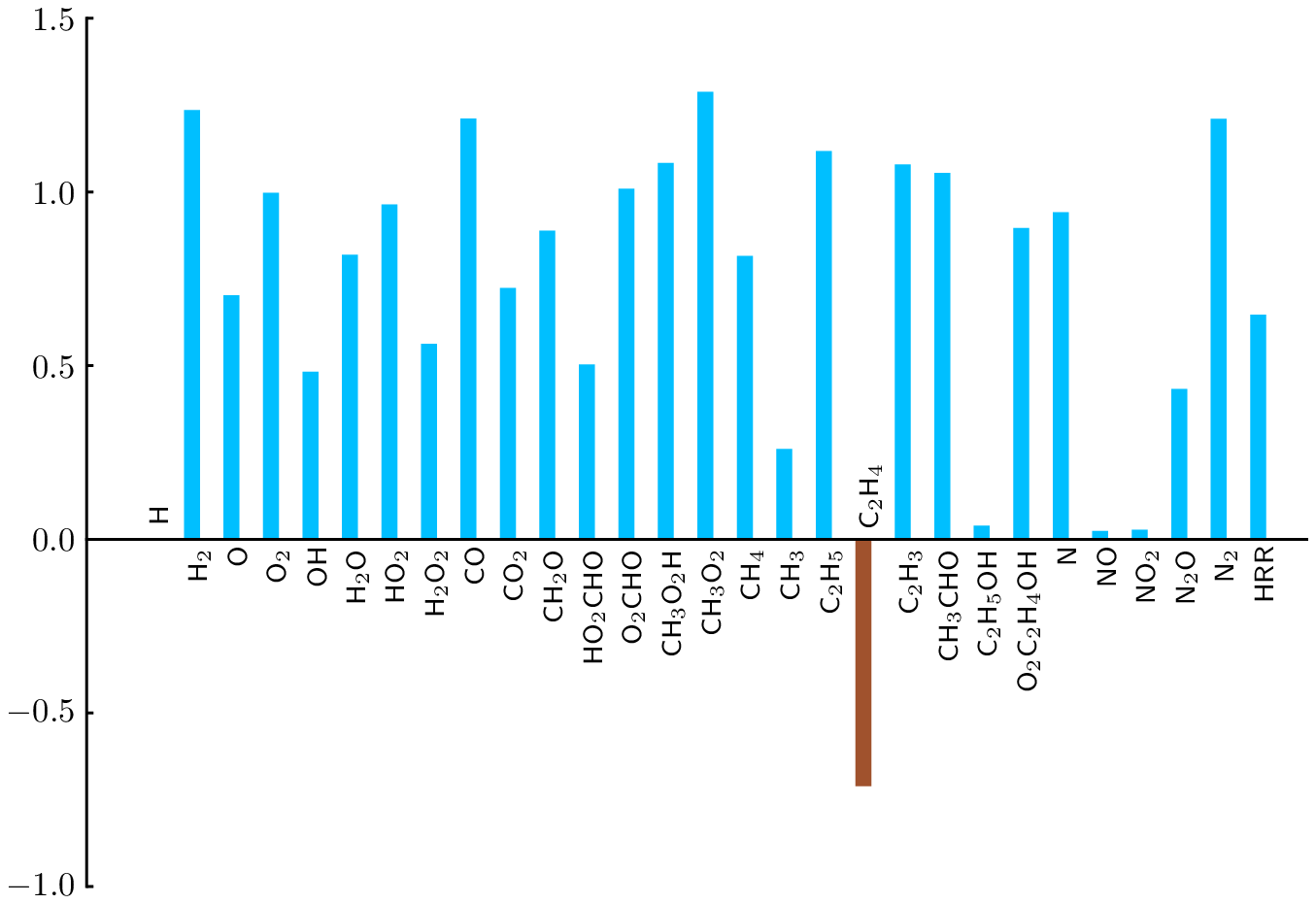}
    \begin{picture}(0,0)
    \put(-190,132){\scriptsize (a) Errors in entire domain}
    \put(-218,50){\scriptsize{\rotatebox{90}{Error ratio $r_M$}}}
    \end{picture}
    } \hspace{0.25cm}
    {
    \includegraphics[width=7.5cm]{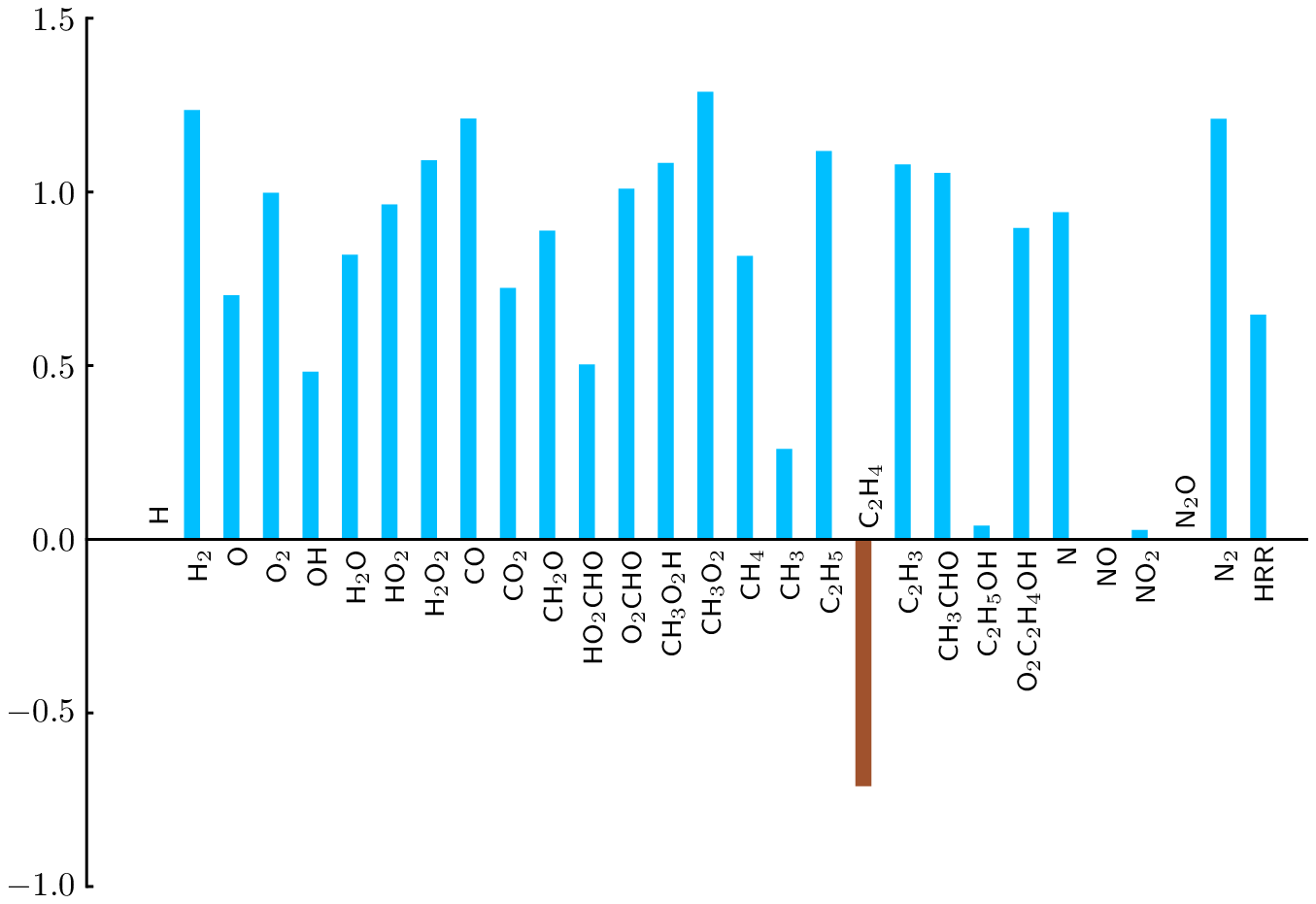}
    \begin{picture}(0,0)
    \put(-190,132){\scriptsize (b) Errors in reacting regions}
    \put(-218,50){\scriptsize{\rotatebox{90}{Error ratio $r_M$}}}
    \end{picture}
    }
    \vspace{0.1cm}
    {
    \includegraphics[width=7.5cm]{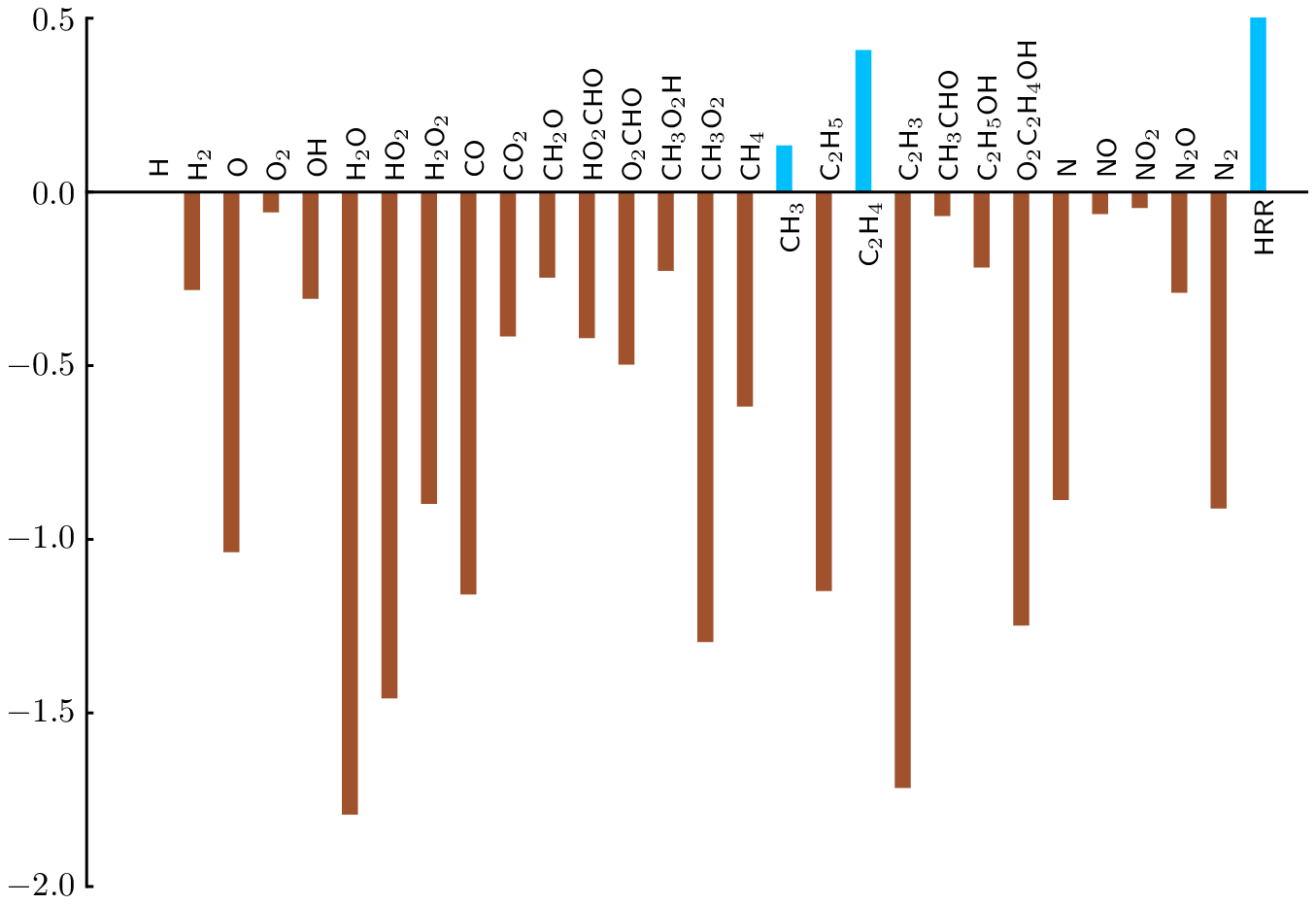}
    \begin{picture}(0,0)
    \put(-190,132){\scriptsize (c) Errors in entire domain}
    \put(-218,50){\scriptsize{\rotatebox{90}{Error ratio $r_A$}}}
    \end{picture}
    }
    \hspace{0.25cm}
    {
    \includegraphics[width=7.5cm]{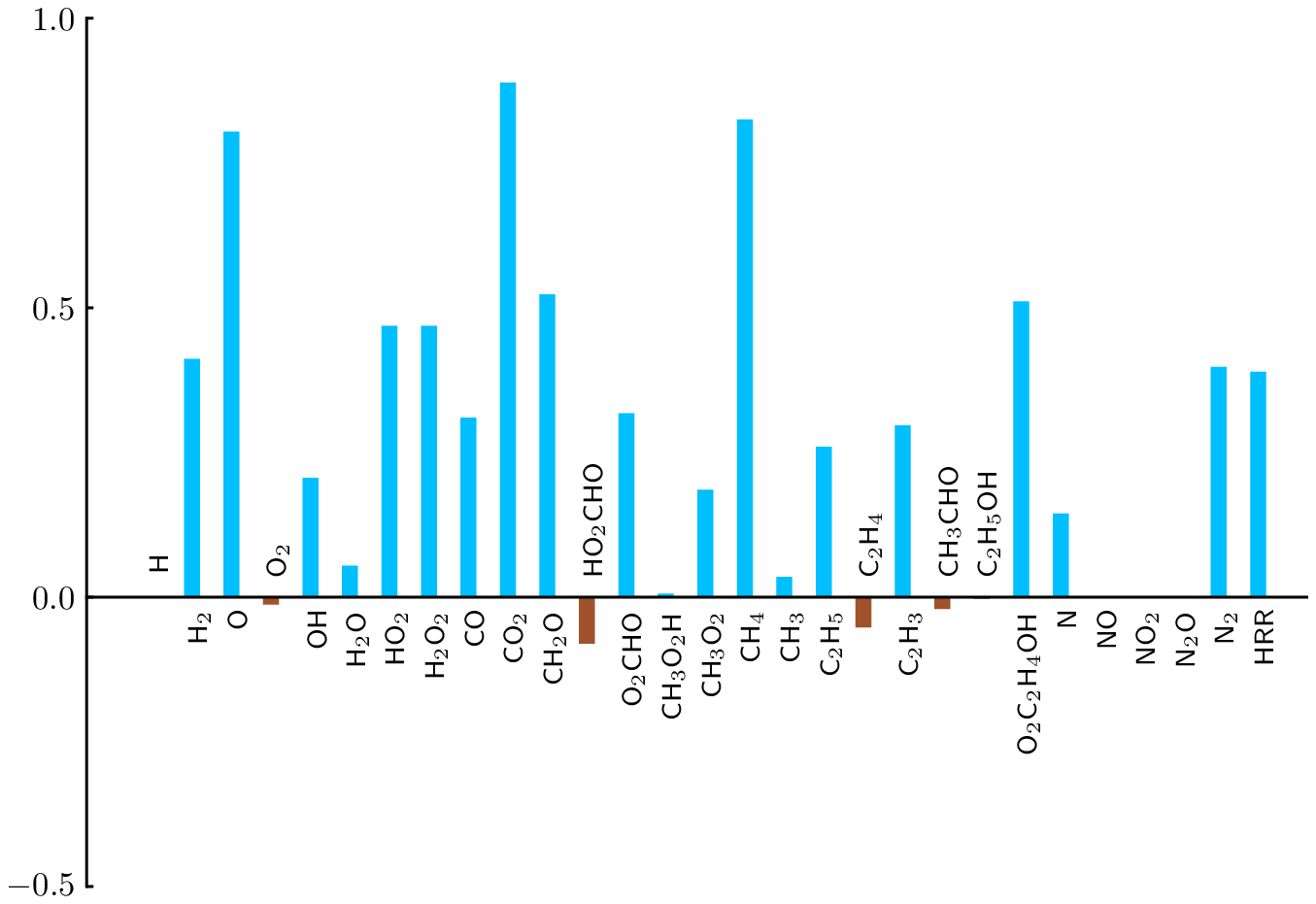}
    \begin{picture}(0,0)
    \put(-190,132){\scriptsize (d) Errors in reacting regions}
    \put(-218,50){\scriptsize{\rotatebox{90}{Error ratio $r_A$}}}
    \end{picture}
    }
    \caption{\rev{Plots of error ratio $r_i$ ($i\in \{M,A\}$, see Eq.~\ref{eq:error-ratio}) for reconstructed data for HCCI dataset at $t = 0.835~ms$. The data is reconstructed by retaining from $n_q=5$ principal vectors from manifolds obtained from the training $t = 0.845~ms$ HCCI dataset.  Error ratio $r_i$ based on maximum (a) and average (b) errors of species production rates and heat release rate in the entire domain. Error ratio $r_i$ based on maximum (c) and average (d) errors of species production rates and heat release rate in reacting regions.}}
    \label{fig:errors-hcci-pr-t1}
\end{figure*}

\begin{figure*}[h!]
    \centering
    {
    \includegraphics[width=7.5cm]{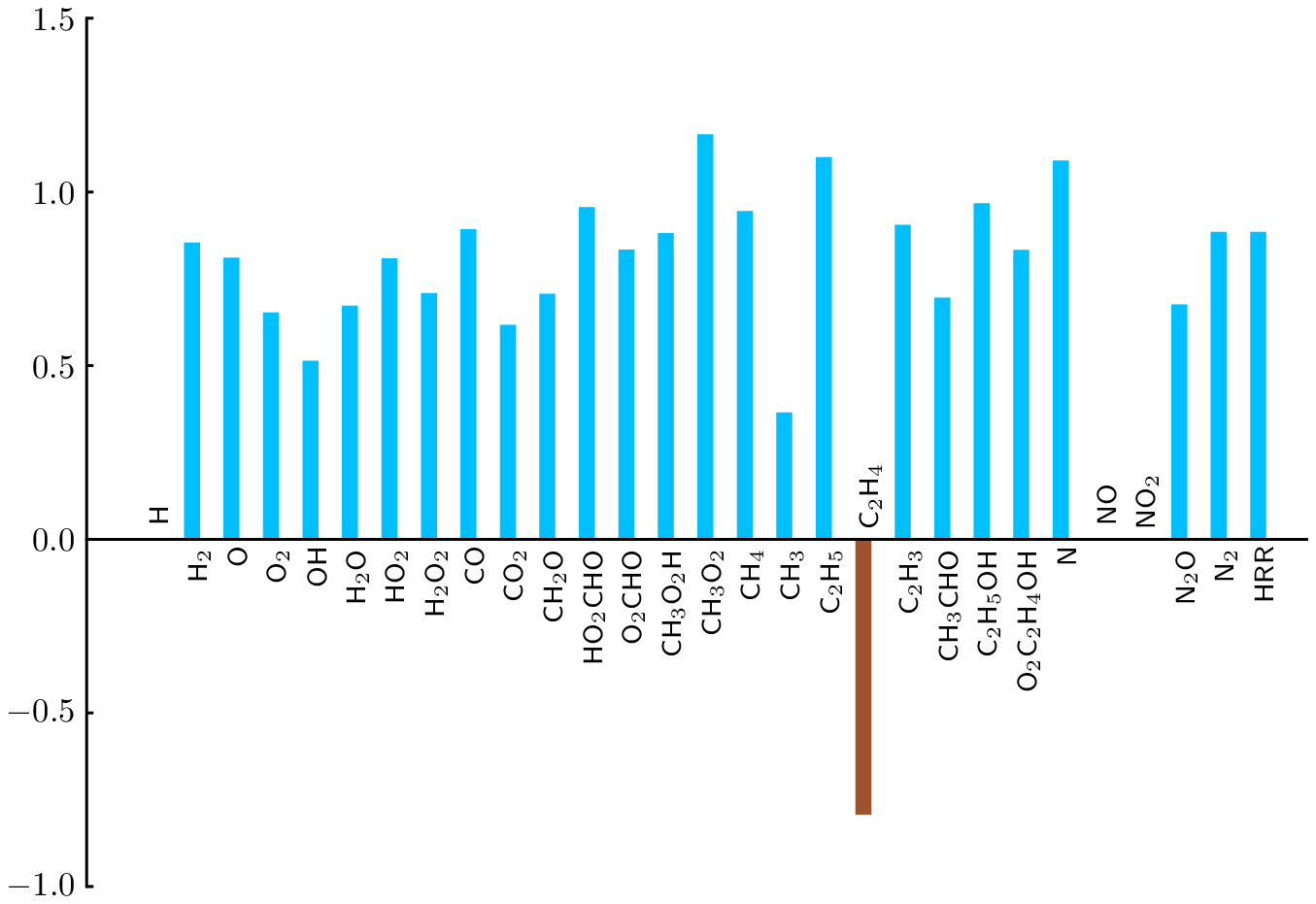}
    \begin{picture}(0,0)
    \put(-190,132){\scriptsize (a) Errors in entire domain}
    \put(-218,50){\scriptsize{\rotatebox{90}{Error ratio $r_M$}}}
    \end{picture}
    } \hspace{0.25cm}
    {
    \includegraphics[width=7.5cm]{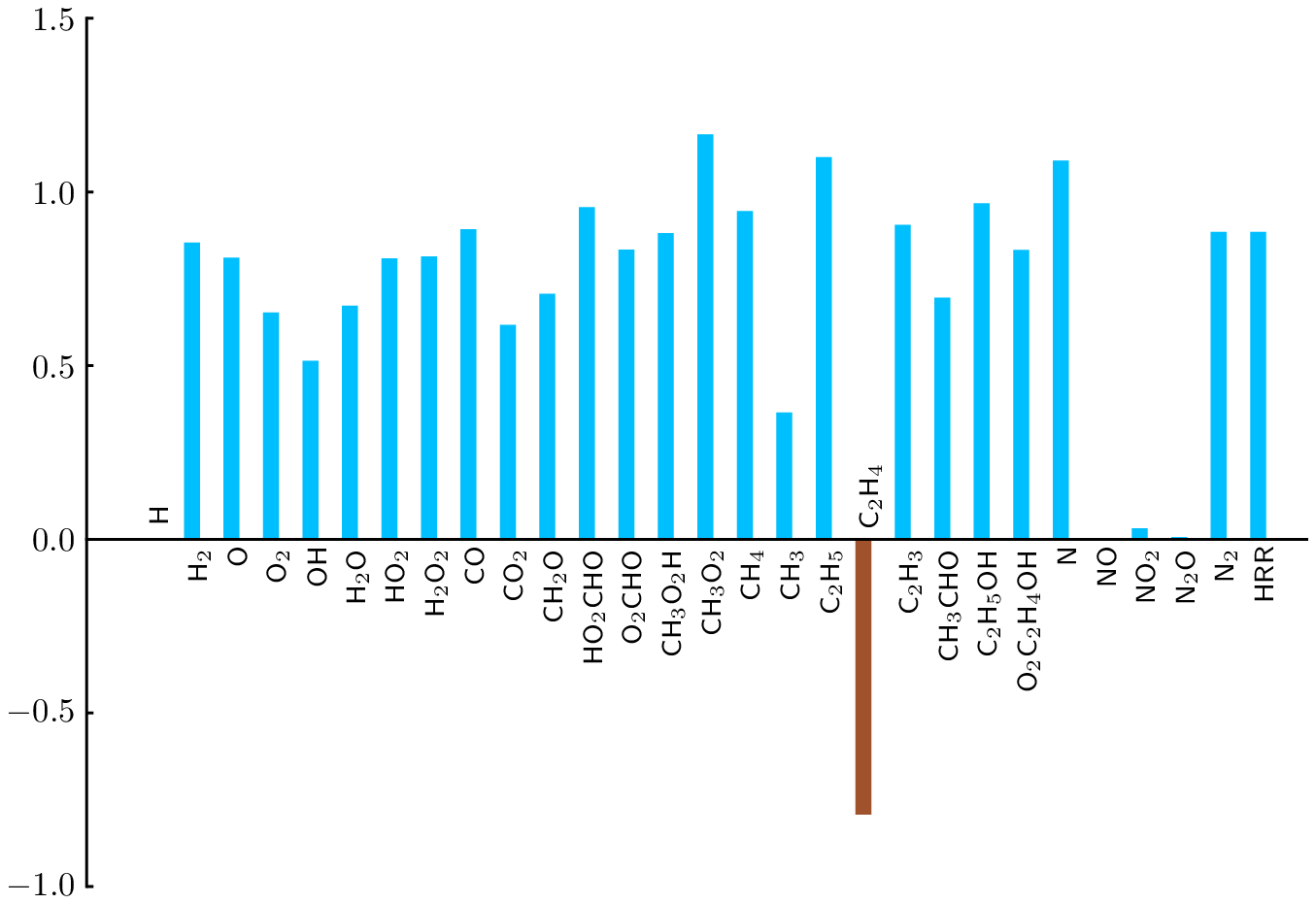}
    \begin{picture}(0,0)
    \put(-190,132){\scriptsize (b) Errors in reacting regions}
    \put(-218,50){\scriptsize{\rotatebox{90}{Error ratio $r_M$}}}
    \end{picture}
    }
    
    \vspace{0.1cm}
    {
    \includegraphics[width=7.5cm]{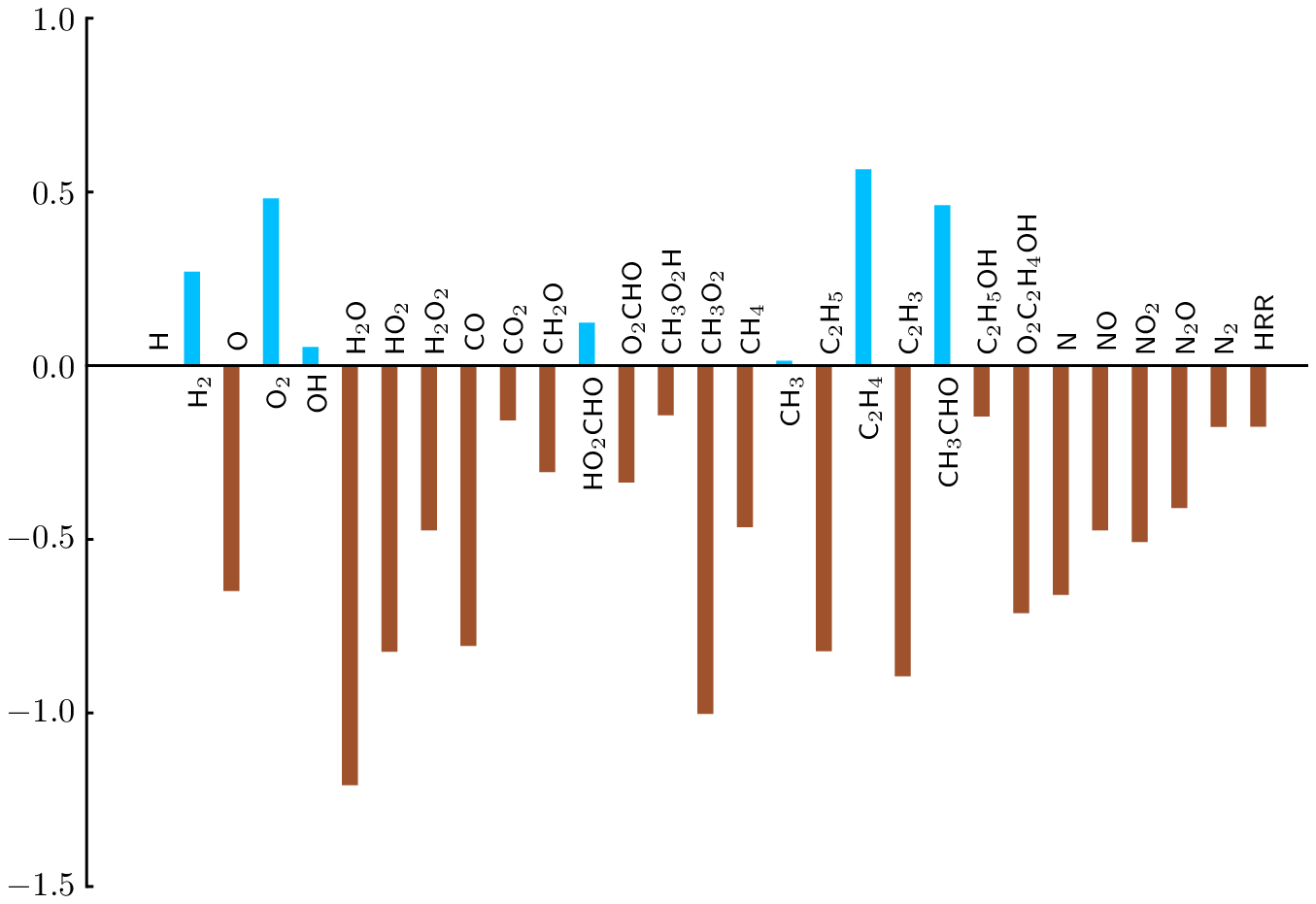}
    \begin{picture}(0,0)
    \put(-190,132){\scriptsize (c) Errors in entire domain}
    \put(-218,50){\scriptsize{\rotatebox{90}{Error ratio $r_A$}}}
    \end{picture}
    }\hspace{0.25cm}
    {
    \includegraphics[width=7.5cm]{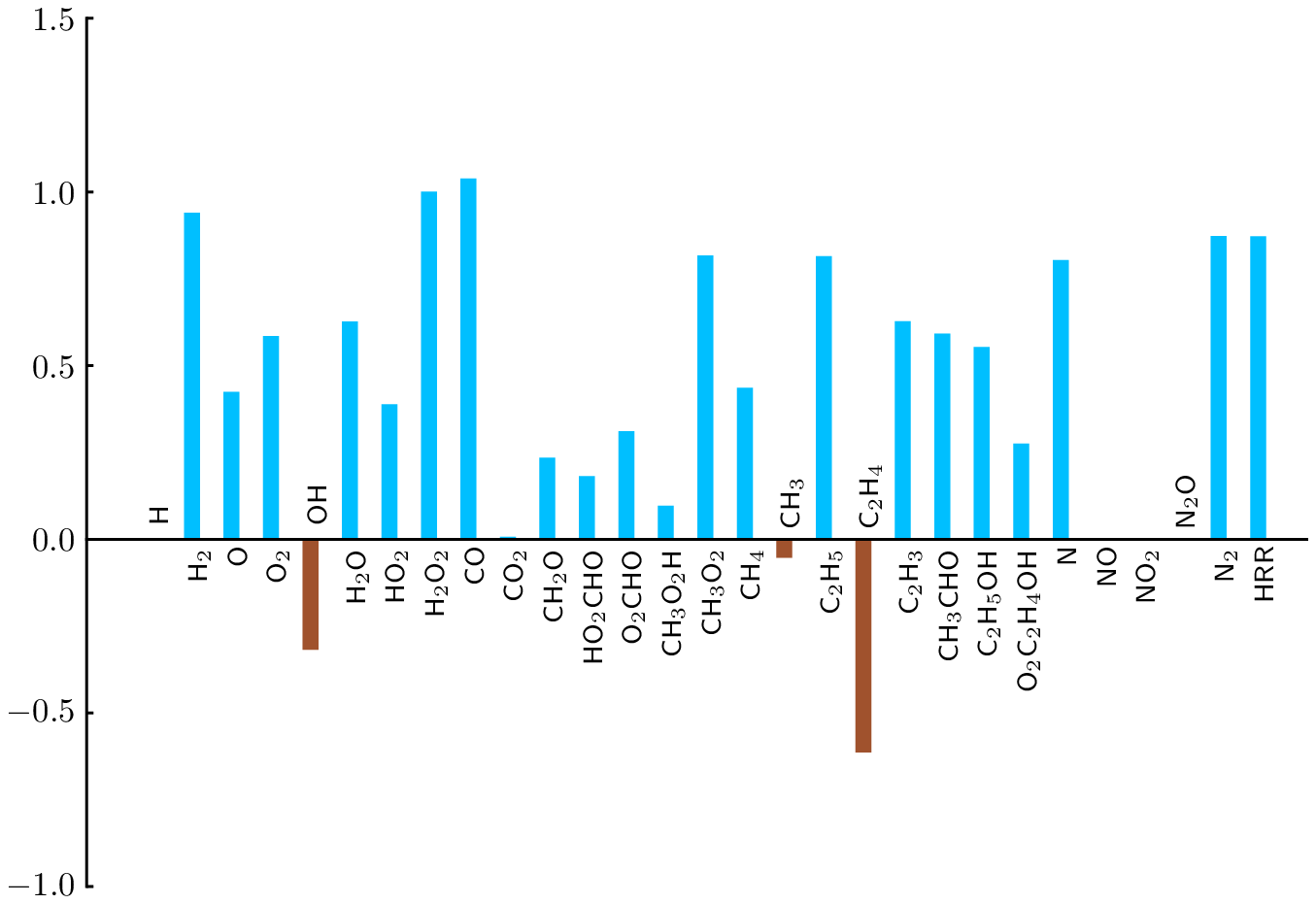}
    \begin{picture}(0,0)
    \put(-190,132){\scriptsize (d) Errors in reacting regions}
    \put(-218,50){\scriptsize{\rotatebox{90}{Error ratio $r_A$}}}
    \end{picture}
    }
    \caption{\rev{Plots of error ratio $r_i$ ($i\in \{M,A\}$, see Eq.~\ref{eq:error-ratio}) for reconstructed data for HCCI dataset at $t = 0.855~ms$. The data is reconstructed by retaining from $n_q=5$ principal vectors from manifolds obtained from the training $t = 0.845~ms$ HCCI dataset. Error ratio $r_i$ based on maximum (a) and average (b) errors of species production rates and heat release rate in the entire domain. Error ratio $r_i$ based on maximum (c) and average (d) errors of species production rates and heat release rate in reacting regions.}}
    \label{fig:errors-hcci-pr-t2}
\end{figure*}

\rev{Next, we assess the robustness with which the PCA and CoK-PCA manifolds reconstruct the state of datasets that were not included in the training exercise.
Here, using the $t = 0.845~ms$ dataset as the training data, we present the reconstruction errors in the production rates and heat release rates obtained from $n_q = 5$ reduced manifolds for two test datasets in the neighbourhood of the training data, namely $t = 0.835~ms$ and $t = 0.855~ms$; as done previously, the analysis comprises a characterization in the whole domain as well as in the reacting zones only. The results are presented in Figs.~\ref{fig:errors-hcci-pr-t1}  and \ref{fig:errors-hcci-pr-t2}.}

\rev{Conforming to the trends observed earlier, in Figs.~\ref{fig:errors-hcci-pr-t1}(a), (b) and ~\ref{fig:errors-hcci-pr-t2}(a), (b) we observe that, for the $r_M$ metric, the CoK-PCA reduced manifold yields a much better reconstruction, in the entire domain as well as the reacting region, for both test instances considered.
Similarly, the results for the $r_A$ metric for both test datasets, shown in Figs.~\ref{fig:errors-hcci-pr-t1}(c), (d) and ~\ref{fig:errors-hcci-pr-t2}(c), (d), also follow the trends observed earlier, in that the PCA and CoK-PCA reduced manifold presents a better reconstruction in the entire domain and the reacting zones, respectively.}

\section{Conclusions and future work} \label{sec:conclusions}

In this paper, we have proposed an alternate dimensionality reduction procedure (CoK-PCA) based on the fourth-order joint moment, i.e., co-kurtosis tensor, as opposed to co-variance which is the basis for the more commonly used Principal Component Analysis (PCA). The rationale for using a higher order joint moment is that chemical reactions represent stiff dynamics and the chemistry-relevant samples in a combustion dataset may typically be extreme-valued (compared to unburnt or fully burnt samples). Accordingly, the directions in thermo-chemical state space that best represent chemical dynamics are captured better by moments higher than co-variance.  
A brief background on the construction and factorization of the co-kurtosis tensor was presented, along with a discussion on the connection of this method to independent component analysis.
The dimensionality reduction (based on PCA and CoK-PCA) and the linear data reconstruction procedures were elucidated by means of a synthetically generated dataset.
The potential of the proposed CoK-PCA method has been evaluated using two datasets: zero-dimensional ignition in a homogeneous reactor, and a two-dimensional DNS of homogeneous charge compression ignition phenomenon.
It has been found that the reduced manifolds obtained from the CoK-PCA method recover the thermo-chemical scalars particularly well, in comparison to PCA, for situations where \rev{extreme-valued} events such as the formation of localized ignition kernels are present, while also accurately capturing the chemical kinetics of the reacting zones of system.
Further, the present study reveals that PCA demonstrates large reconstruction errors in the reaction zones, specifically in terms of the species production and heat release rates. 
In contrast, it is found that CoK-PCA yields reconstruction errors that are more uniformly distributed across all samples and thus presents a qualitatively true representation of the overall reaction dynamics.

There are numerous topics of investigation that have been reserved for future work. We have presented an initial comparison between PCA and CoK-PCA, and to that extent used most straightforward techniques, particularly linear reconstruction of thermo-chemical scalars from the identified principal components. It is well-known that the accuracy of linear reconstruction is limited and other studies have adopted non-linear reconstruction including methods such as regression splines and neural networks. We posit that the reconstruction accuracy of CoK-PCA can similarly be improved using non-linear reconstruction, which should further improve the accuracy of reconstruction of species production rates, and this will be considered in a future study. Similarly, refinements of PCA such as kernel PCA and local PCA that are widely considered in other studies, can be applied to CoK-PCA. Finally, \rev{it should be pointed out that, while it may be unfeasible to use the proposed CoK-PCA method as a drop-in replacement of PCA in every scenario,} the best approach may be a hybrid with certain portions of the dataset represented by PCA manifolds, and others by a CoK-PCA manifold. \rev{Such techniques to consider the PCA and CoK-PCA in conjunction will be examined in future.}
\section*{Acknowledgments}
The work at IISc was supported under a project from the National Supercomputing Mission, India. AJ was funded by a project from Shell Technology Center, Bengaluru, India.
Work by HK was part of the ExaLearn Co-design Center, supported by the
Exascale Computing Project (17-SC-20-SC), a collaborative effort of the U.S. Department of Energy Office of Science and the National Nuclear Security Administration.
Sandia National Laboratories is a multi-mission laboratory managed and operated by National Technology and Engineering Solutions of Sandia, LLC., a wholly owned subsidiary of Honeywell International, Inc., for the U.S. Department of Energy’s National Nuclear Security Administration under contract DE-NA-0003525.
The views expressed in the article do not necessarily represent the views of the U.S. Department of Energy or the United States Government.

\bibliographystyle{elsarticle-num}
\bibliography{dimred.bib}

\end{document}